\documentclass[a4paper,fleqn,usenatbib]{mnras}
\usepackage{newtxtext,newtxmath}
\usepackage[T1]{fontenc}
\usepackage{ae,aecompl}
\usepackage{multicol}
\usepackage{graphicx}	% Including figure files
\usepackage{amsmath}	% Advanced maths commandsfapp
\usepackage{multirow}
\usepackage{longtable}
\usepackage{morefloats}
\usepackage{float}
\usepackage{cleveref}
\usepackage{pdflscape}
\usepackage{placeins}
\usepackage{capt-of}
\usepackage[dvipsnames]{xcolor}

\newcommand{\kms}{{\mathrm{km~s^{-1}}}}

\title[Asteroseismology of the high-amplitude $\delta$ Scuti star V2367 Cygni]
{Asteroseismology of the fast-rotating  high-amplitude $\delta$ Scuti star V2367 Cygni}
\author[Daszy\'nska-Daszkiewicz et al.]{ J. Daszy\'nska-Daszkiewicz\thanks{E-mail:daszynska@astro.uni.wroc.pl},
W. Szewczuk, P. Walczak\\
Instytut Astronomiczny, Uniwersytet Wroc{\l}awski, Kopernika 11, 51-622 Wroc{\l}aw, Poland\\
}

\date{Accepted XXX. Received YYY; in original form ZZZ}

\pubyear{2021}

\hypersetup{draft}
\begin{document}
\label{firstpage}
\pagerange{\pageref{firstpage}--\pageref{lastpage}}
\maketitle

\begin{abstract}
We present the comprehensive analysis of the high-amplitude $\delta$ Sct star V2367 Cygni. 
Firstly, we perform the frequency analysis  for the whole available  set of the {\it Kepler} and {\it TESS} photometry. 
Most of the frequency peaks are harmonics or combinations of the three known independent frequencies with the highest amplitudes, i.e., 
$\nu_1=5.66106$\,d$^{-1}$,  $\nu_2=7.14898$\,d$^{-1}$ and $\nu_3=7.77557$\,d$^{-1}$.
The total number of independent frequencies is 26 and 25 from  the {\it Kepler} and {\it TESS}  light curve, respectively.
Then, using the $UBVRI$ time-series photometry, we unambiguously identify the dominant frequency $\nu_1$ as the radial mode,
whereas in the case of frequencies $\nu_2$ and $\nu_3$  the most probable mode degrees are $\ell=0$ or $\ell=2$.  
However, only the frequency $\nu_2$ can be associated with a radial mode, 
and only if higher order effects of rotation are taken into account.
Including the rotational mode coupling, we constructed complex seismic models of V2367\,Cyg, 
which fit $\nu_1$ and $\nu_2$ as radial modes, and reproduce the amplitude of bolometric flux variations
(the parameter $f$) for the dominant mode. The empirical values of $f$ are derived from the $UBVRI$ amplitudes and phases.
We rely on the Bayesian analysis based on Monte Carlo simulations to derive constraints on evolutionary stage, 
mass, rotation, overshooting from the convective core and efficiency of convective transport in the envelope.
Our seismic analysis clearly indicates that V2367\,Cyg is in a post-MS phase of evolution.
This is the first extensive seismic modelling that takes into account the effect of rotational coupling between pulsation modes.
\end{abstract}


%%%%%%%%%%%%%%%%%%%%%%%%%%%%%%%%%%%%%%%%%%%%%%%%%%
%%%%%%%%%%%%%%%%% BODY OF PAPER %%%%%%%%%%%%%%%%%%

\section{Introduction}
Asteroseismology has already become a mature and well-developed branch of modern astrophysics, 
widely used to obtain constraints on the internal structure of various types of stars.
The most recent excellent reviews by Aerts (2021) and Kurtz (2022) summarize roughly the results obtained to date.
Obviously, there are still many uncertainties in stellar modelling and the uniqueness of a seismic model
depends on the number of observables we have at our disposal. The greatest uncertainties concern processes described 
by free parameters, i.e., primarily the efficiency of convection in the outer layers and all mixing processes like diffusion, 
convective zone boundary mixing or rotationally induced mixing.
Another difficulty arises from the effects of  rotation, both on the equilibrium model as well as on stellar pulsations.
Moreover, in the case of some pulsating stars including $\delta$ Scuti variables, 
the main obstacle to the use of asteroseismic methods is the mode identification.
This is because, with some exceptions \citep{Bedding2020}, we do not observe regular
patterns in the oscillation spectra of $\delta$ Sct pulsators and more sophisticated methods of the mode identification 
must be applied. The most commonly used and effective methods are based on photometric amplitudes and phases,
which were proposed a long time ago \citep[e.g.,][]{Balona1979,Watson1988}.
An additional complication is taking into account the effects of rotation when identifying modes, in particular,
the rotational coupling for close-frequency modes \citep[e.g.,][]{JDD2002}.

A big advantage for seismic modelling is the presence of radial modes, especially two or more. This is because their oscillation spectrum
is sparse and the period ratio of radial modes takes specific values from a very narrow range. There is a subgroup of $\delta$ Sct stars, 
called high-amplitude $\delta$ Scuti (HADS) stars, which often  exhibit two radial modes. HADS stars change their brightness
in the range larger than about 0.3 mag in the Johnson $V$ filter and most of them are already in the post-main sequence phase of evolution.
Asteroseismology of double-mode radial $\delta$ Sct stars provides strong constraints on the parameters of the model and theory
\citep{JDD2020,JDD2022}. Besides, their seismic models are extremely sensitive to adopted opacity data \citep{JDD2023}.
%and only the OPAL seismic models are caught within the observed error box in the HR diagram 

As pointed out by \citet{Breger2000}, HADS stars are rather slow rotators with $V_{\rm rot}\sin i \lesssim40~\kms$.
This paper concerns the HADS star V2367 Cygni which is suspected of having a surprisingly high rotation \citep{Balona2012}.
%This high rotation has been suggested both by some spectroscopic determinations as well as by the unusual period ratio 
%if two of the observed frequencies do correspond to radial modes. 
The effect of rotation on the period ratio
of radial modes was studied by \citet{Suarez2006}. Then, \citet{Suarez2007} included the effect of near-degeneracy.
In both papers, the authors limited their computation to the rotation up to  50\,$\kms$. 
Here, we consider the rotational velocity up to the half of critical velocity.
 
The paper is organized as follows. Sect.\,2  gathers the information on V2367 Cygni.
In Sect.\,3, we present the frequency analysis of the whole light curves from the two space missions: {\it Kepler} and {\it TESS}.
Sect.\,4 is devoted to the mode identification for the three main frequencies using the multi-colour photometry. % $UBVRI$ of \citet{Ulusoy2013}. 
In Sect.\,5, we study the effects of rotation on radial modes, including the effect of mode coupling. 
Extensive asteroseismic modelling of V2367 Cyg with Monte Carlo–based Bayesian analysis, taking into account the rotational mode coupling, 
is presented in Sect.\,6. A summary is given in the last section.

\section{V2367 Cygni}
V2367 Cygni is a high-amplitude $\delta$ Scuti star with the average visual brightness of $V=11.48$\,mag.
It was identified for the first time as a $\delta$ Scuti variable from the ROTSE-I All-Sky Surveys \citep{Akerlof2000}.
Subsequently, \citet{Jin2003} classified V2367 Cyg as HADS based on the two-passband $VI$ photometry. One periodicity 
$P=0.176617$\,d was detected.
This classification was confirmed by \citet{Pigulski2009} who analysed the ASAS data.

The Gaia DR3 parallax of V2367 Cyg amounts to $\pi= 0.7859\pm0.0147$\,mas, which corresponds to the distance 
$d=1272.43\pm23.80$\,pc (Gaia Collaboration, 2020).
According to \citet{Frasca2016}, the spectral type of this HADS is F0III and their classification was based on the low-resolution  
spectra from LAMOST  (The Large Sky Area Multi-Object Fibre Spectroscopic Telescope).
Using the medium-resolution spectra from NOT (the Nordic Optical Telescope), \citet{Niemczura2017} assigned a spectral type F0IVs+?,
where $s$ indicates narrow (sharp) absorption lines. These authors mentioned also that V2367 Cyg is a member of binary system
but they did not provide any justification for this statement. Besides, the Gaia DR3 RUWE parameter equal to 0.948 does not reveal 
that V2367\,Cyg is part of a binary system.

The star was also observed in the framework of the {\it Kepler} and {\it TESS} missions.
\citet{Uytterhoeven2011} analysed the first two quarters of Kepler data, i.e., Q0 (short cadence, SC) and Q1 (long cadence, LC). 
They detected the main frequency $\nu_1=5.661$\,d$^{-1}$ and suggested many more frequency peaks
in the range  (4.3,\,80.0)\,d$^{-1}$. Moreover, using the Str\"omgren photometry,  \citet{Uytterhoeven2011} derived
the effective temperature $T_{\rm eff}=6810\pm130$\,K and the surface gravity $\log g=3.80\pm0.19$\,dex.

Subsequently, \citet{Balona2012} used a longer set of {\it Kepler} data, i.e.,  Q0, Q6, Q7 (SC) and Q1, Q2, Q5 (LC),  and extracted 
three main frequencies  $\nu_1=5.6611$\,d$^{-1}$,  $\nu_2=7.1490$\,d$^{-1}$ and  $\nu_3=7.7756$\,d$^{-1}$.
They also detected a dozen other frequencies, but most of the signals could be explained 
by harmonics and combinations of the three main frequencies. 
Moreover,  using  a single, low resolution ($R\sim 5000$) spectrum, \citet{Balona2012}
derived basic parameters, $T_{\rm eff} = 7300 \pm 150$\,K, $\log g = 3.5 \pm 0.1$ (in cgs), and concluded that V2367 Cyg
has a solar abundance. Besides, the authors announced an unusual fast rotation with the projected value 
of $V_{\rm rot}\sin i=100\pm 10\,\kms$, what made V2367 Cyg the fastest rotating HADS (more than twice that of other HADS stars).  

\citet{Balona2012} attempted also to make seismic modelling of V2367\,Cyg assuming that $\nu_1$ and $\nu_2$ 
are both radial modes: fundamental \& first overtone or first overtone \& second overtone. However, the frequency ratio 
$\nu_1/\nu_2=0.79187$ did not fit to any of these cases and the final conclusion was that higher order effects of rotation, including mode coupling, must be taken into account if $\nu_2$ were to be a radial mode.

The fast rotation of V2367 Cyg was confirmed by \citet{Ulusoy2013} on the basis of  higher resolution spectra ($R\sim 16000$).
However, the most recent determination by \citet{Niemczura2017},  based on the spectra with $R\sim 25000$ and 46000,
indicates rather a low projected rotational velocity $V_{\rm rot}\sin i=16\pm 1\,\kms$. \citet{Niemczura2017} also determined
abundances of 29 chemical elements, which are similar to the solar abundances within the errors.
%, e.g., the abundance of iron is $\log \epsilon(Fe)= 7.53\pm0.13$.
Using their abundances, we determined the bulk metallicity by mass in the range $Z\in [0.0134,~0.0356]$.
% [m/H]$\approx 0.0$      [Fe/H]$=0.08 \pm0.15$  \citet{Frasca2016}  $\log \epsilon(Fe)= 7.53\pm0.13$   \citet{Niemczura2017}
 
\citet{Ulusoy2013}  also obtained time-series photometry in the $UBVRI$  bands.  Their aim was to make the mode identification 
from the amplitudes and phases for the three main frequencies. Unfortunately, their identification was not successful. 
\begin{figure}
	\includegraphics[width=\columnwidth,clip]{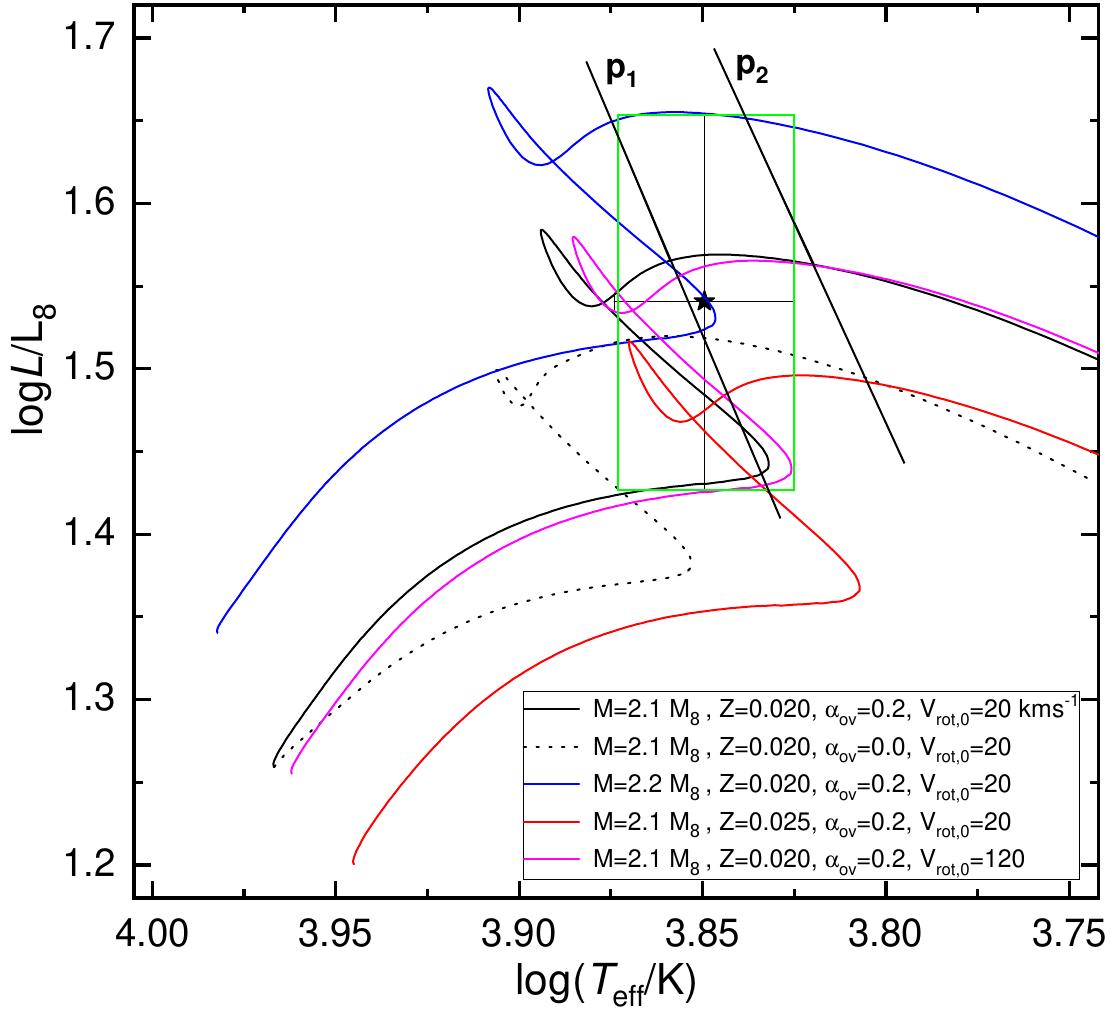}
	\caption{The position of V2367 Cyg in the HR diagram. Evolutionary tracks were computed with the OPAL opacity tables, the solar mixture
		from AGSS09, initial hydrogen abundance $X_0=0.70$ and the MLT parameter $\alpha_{\rm MLT}=0.5$. There is shown the effect of mass $M$, 
		metallicity $Z$, overshooting from the convective core $\alpha_{\rm ov}$ and the initial value of rotation.
		For $Z=0.020,~\alpha_{\rm ov}=0.2, V_{\rm rot,0}=20~\kms$, the line of constant frequency $\nu_1=5.66106$\,d$^{-1}$ are indicated
		for the radial fundamental ($p_1$) and first overtone mode ($p_2$) radial modes.}
	\label{HR}
\end{figure}
 
In our analysis, we adopted the whole range of effective temperature from the above mentioned literature, i.e.,
$T_{\rm eff} \in (6680, 7484)$\,K, which corresponds to $\log (T_{\rm eff}/{\rm K})=3.8495 \pm 0.0246$.
To derive the luminosity $L$, we adopted a distance determined on the basis of Starhorse2 model \citep{Anders2022}, 
using the Gaia DR3 observations (Gaia Collaboration et al., 2022). The bolometric corrections were taken from Kurucz 
model atmospheres for the microturbulent velocity $\xi_t=2,~4\,\kms$  and metallicity [m/H]$=0.0, +0.1,+0.3$.
Finally, we got $\log L/{\rm L}_\odot = 1.5407 \pm 0.1138$. The error in luminosity also includes errors in the effective temperature and bolometric correction. Therefore, the relative error of $L$ is larger than the relative error of the distance $d$.

In Fig.\,\ref{HR}, we show the position of V2367\,Cyg on the Hertzsprung-Russell (HR) diagram. A few evolutionary tracks were depicted, presenting the effect of a mass $M$, metallicity $Z$, the overshooting form convective core $\alpha_{\rm ov}$ 
and the initial rotation $V_{\rm rot,0}$. The track were computed with Warsaw-New Jersey code for the OPAL opacity tables \citep{Iglesias1996} 
and the solar chemical mixture of \citet{Asplund2009}, hereafter AGSS09. We depicted also lines of the constant frequency $\nu_1=5.66106$\,d$^{-1}$ for the fundamental ($p_1$) and first overtone ($p_2$) radial modes, 
for $Z=0.020,~\alpha_{\rm ov}=0.2, V_{\rm rot,0}=20~\kms$. 
In this case, the value of $\nu_1$ as $p_1$ or $p_2$ is reached only in the hydrogen-shell burning (HSB) phase.

\section{Analysis of space photometry}% from Kepler and TESS}
%\subsection{Frequencies from the Kepler data}
%\subsection{Frequencies from the TESS data}
%
\begin{figure}
	\includegraphics[angle=270, width=1\columnwidth]{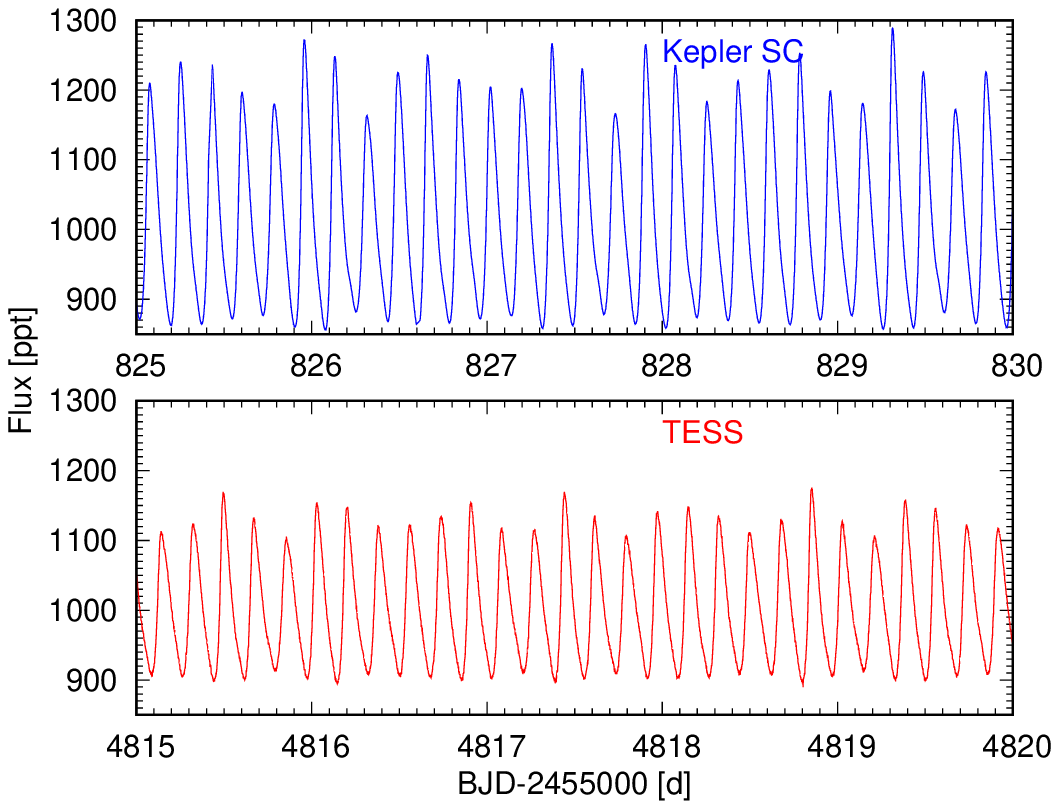}
	\caption{Five-days cuts of {\it Kepler} (top panel) and {\it TESS} (bottom panel) light curves of V2367 Cyg.}
	\label{fig:V2367Cyg_lc}
\end{figure}
\setlength{\tabcolsep}{5pt}
\begin{table*}
	\centering
	\caption{The standard deviation in all available parts of Kepler SC data of V2367 Cyg.}
	\label{tab:SDV_Kepler}
	\begin{tabular}{ c c c c c c c c c c c c c c c c  c } % four columns, alignment for each
		\hline
		ID    &  Q0.1 & Q6.1  & Q6.2  & Q6.3   & Q7.1  & Q7.2  & Q7.3  & Q8.1  &  Q8.2  & Q8.3  & Q9.1  &  Q9.2  & Q9.3  & Q10.1  & Q10.2  & Q10.3  \\
		\hline
		SD      & 115.8 & 118.6 & 118.7 &118.5 & 119.4 & 119.5 & 119.7 & 119.3 & 119.2  & 119.1 & 118.0 & 118.0  & 117.8 & 118.6  & 118.6  & 118.4 \\
		\hline
	\end{tabular}
\end{table*}
In this section, we analysed the photometric data delivered by two very  successful space missions: {\it Kepler}
\citep[e.g.\,][]{2010Sci...327..977B, 2010ApJ...713L..79K} and {\it TESS} \citep[e.g.\,][]{2015JATIS...1a4003R}.
{\it Kepler} observed V2367 Cyg in two modes: a long cadence (LC, $\sim$30 min exposure) and short cadence (SC, 1\,min exposure).
LC data are available for all {\it Kepler} quarters, i.e. Q0-Q17 from BJD\,245\,4953.539 to BJD\,245\,6424.001, whereas SC data 
are for Q0.1 from BJD\,245\,4953.529 to BJD\,245\,4963.255 and for Q6.1-Q10.3 from BJD\,245\,5372.460 to BJD\,245\,5833.276.
Eight years later the star was observed by {\it TESS} in its 2 min mode in sectors: S14-S15 from BJD\,245\,8683.356 to BJD 245\,8737.411,
S40-S41 from BJD\,245\,9390.655 to BJD\,245\,9446.581, and S54-S55 from BJD\,245\,9769.901 to BJD\,245\,9824.265.

Light curves from both missions can be downloaded from \href{https://archive.stsci.edu/}{MAST} portal. 
There are two types of flux data: simple aperture photometry (SAP) and Pre-search Data Conditioning SAP flux (PDCSAP).
Five-days cuts of {\it Kepler} and {\it TESS} light curves can be seen in Fig.\,\ref{fig:V2367Cyg_lc}, 
where the normalized and converted to ppt SAP data are shown.
\begin{figure*}
	\includegraphics[angle=270, width=14cm]{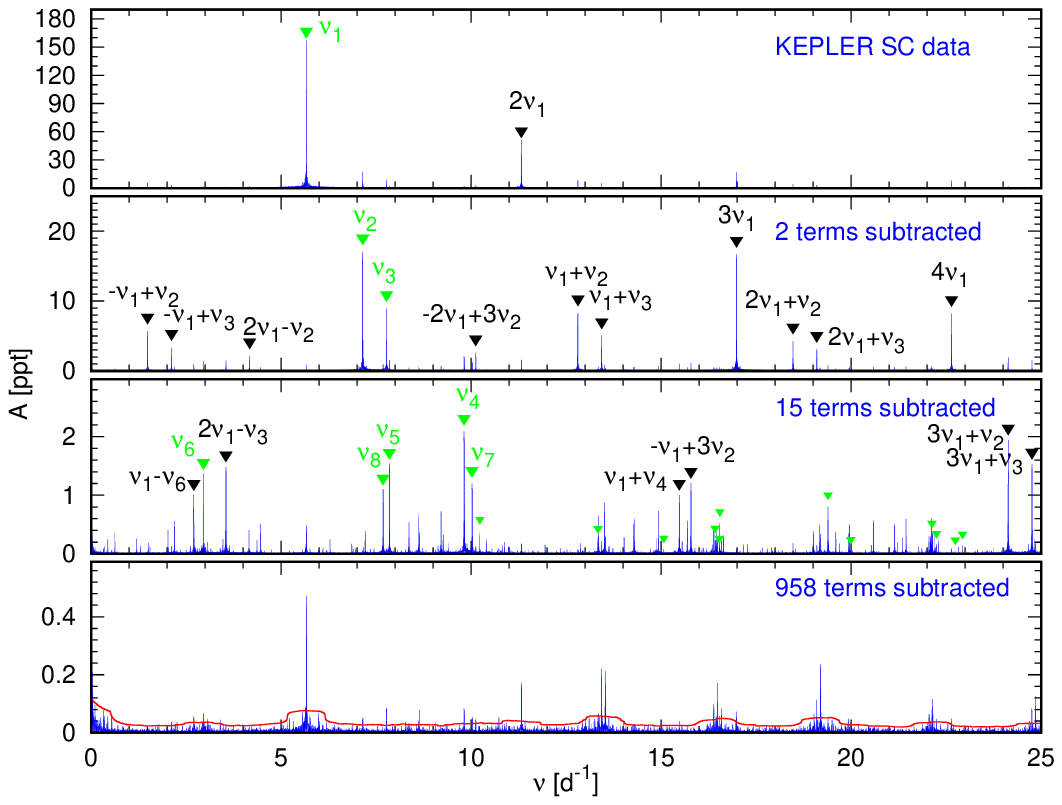}
	\caption{Amplitude periodograms for the {\it Kepler} SC Q6.1-Q10.3 data of V2367 Cyg. Panels from top to bottom show
		periodograms: for the original data set; after subtraction of 2 terms; after subtraction of 15 terms
		and after subtraction of 958 terms. Red solid line in the bottom panel indicates the $5S/N$ level. All frequency peaks 
		above 1\,ppt are marked by down-pointing triangles; green for independent signals and black for combinations or harmonics.
		Independent frequencies with amplitudes smaller than 1\,ppt are marked by smaller green triangles.}
	\label{fig:V2367Cyg_Kepler_SC_trf}
\end{figure*}

We started with the frequency analysis of the {\it Kepler} data. The Nyquist frequency for the SC and LC data is
720\,d$^{-1}$ and 24.5\,d$^{-1}$, respectively. Therefore, to circumvent aliasing issues, we first performed analysis on the SC data.
Each single part of this data, i.e. $Qn.m$, where $n$ is the quarter number and $m$ is one of the three parts in quarter, was normalized
by a linear function.

In the case of the {\it Kepler} data of V2367 Cyg, the SAP and PDCSAP fluxes are almost indistinguishable, thus we decided to analyse only the SAP fluxes.
It appeared that amplitudes in Q0.1 are considerably lower than in other {\it Kepler} quarters.
This is reflected in the standard deviation which is lower in Q0.1 than in other data parts ({\it cf}. Table\,\ref{tab:SDV_Kepler}).
Moreover, there is a 400-day gap between Q0.1 and the remaining data.
Therefore, the exclusion of Q0.1 seems to be justified and for the further analysis we used data from Q6.1 to Q10.3. 
This data set consists of 624\,096 observational points which span 460.8\,d. This range translates into the Rayleigh resolution 
of $\Delta \nu_{R,K}=0.0022\,\mathrm{d}^{-1}$.

In order to extract frequencies of the periodic light variability, we proceeded the standard pre-whitening procedure. 
The amplitude Lomb-Scargle periodograms \citep{1976Ap&SS..39..447L, 1982ApJ...263..835S} were calculated up to $130\,\mathrm{d}^{-1}$ 
with the fixed resolution of $10^{-5}\,\mathrm{d}^{-1}$. Above 100 $\,\mathrm{d}^{-1}$, no signal was detected. We used an algorithm based 
on the fast Fourier transforms for non-equally spaced data \citep[e.g.,][]{2012A&A...545A..50L}. As a significance criterion of a given frequency peak 
we chose the signal to noise ratio, $S/N=5$, which corresponds to the estimate made by \citet{2021AcA....71..113B} for TESS data.
%This threshold is higher than the standard value of 4 \citep{1993spct.conf..106B, 1997A&A...328..544K},
The noise $N$ was calculated as a mean amplitude in a one day window centred at the significant frequency before its extraction.
Four periodograms from the {\it Kepler} SC Q6.1-Q10.3 data are shown in Fig.\,\ref{fig:V2367Cyg_Kepler_SC_trf}.
For a better visibility only frequencies up to 25\,d$^{-1}$ were depicted.

In the case of {\it Kepler} data which have a high point-to-point precision, there is a risk of artificially introducing
spurious signals in the pre-whitening process. According to \cite{1978Ap&SS..56..285L}, in their most conservative case,
frequencies that are separated less than 2.5 times the Rayleigh resolution cannot be resolved properly and may be spurious.
Therefore, we excluded all lower-amplitude frequencies from the following range:
\begin{equation}
	\label{eq:prohibited}
	\nu \in \sum_{i=0}^n \left<\nu_i-2.5\Delta \nu_{R,K}, \nu_i+ \Delta \nu_{R,K}\right>,
\end{equation}
where $n$ is the number of all detected frequencies $\nu_i$ at a given pre-whitening step. 
In the bottom panel of Fig.\,\ref{fig:V2367Cyg_Kepler_SC_trf}, there are residual peaks clearly exceeding the detection threshold.
These are the frequencies that fall into the range defined in Eq.\,\ref{eq:prohibited}.
This procedure allowed us to detect 958 frequency peaks that met our criteria. However, during the non-linear fit, 13 frequencies 
drifted into the prohibited range given by Eq.\,\ref{eq:prohibited}. Therefore, they were also excluded from our ultimate frequency list. 
Our final frequencies, along with their corresponding amplitudes and phases, were determined
by the non-linear least-squares fitting function in the form:
\begin{equation}
	S(t)= \sum_{i=1}^N A_i \sin \left( 2\pi \left( \nu_i t + \phi_i \right)  \right) +c,
\end{equation}
where $N=945$ is the number of sinusoidal components, $A_i$, $\nu_i$, $\phi_i$ are the amplitude, frequency, and phase 
of the $i$-th component, respectively, while the $c$ is an offset. Finally, we applied corrections to the formal errors of frequencies 
according to \citet{1991MNRAS.253..198S}, the so-called post-mortem analysis. We got that the formal errors should be multiplied by about $2.7$.

In the last step, we looked for harmonics and combinations. A given frequency was considered a harmonic if it satisfied, 
up to the Rayleigh resolution, the equation:
\begin{equation}
	\nu_i=m\nu_j,
	\label{eq:harmonic}
\end{equation}
where $m$ is an integer from the range $\left<2,\,20\right>$. The frequency was considered a combination with two parents
if it met, up to the Rayleigh resolution, the equation
\begin{equation}
	\nu_i=m\nu_j+n\nu_k,
	\label{eq:combination2}
\end{equation}
where $m$ and $n$ are integers from the range $\left<-10,\,10\right>$, except for $\nu_1$, for which we assumed $m\in \left<-15,\,15\right>$.
Similarly, the  frequency was considered a combination with three parents if it satisfied the equation
\begin{equation}
	\nu_i=m\nu_j+n\nu_k + o \nu_l,
	\label{eq:combination3}
\end{equation}
where $m$, $n$, $o$ are integers from the range $\left<-2,\,2\right>$, except for $\nu_1$, for which we used $m\in \left<-5,\,5\right>$.
Additionally, when searching for combinations and harmonics, we excluded signals with very low frequencies, i.e. less than 0.1\,d$^{-1}$,
as they are most probably of instrumental origin.

After all steps of selection, we found that only a small part of the numerous signals detected in {\it Kepler} data are independent frequencies. 
They are listed in Table\,\ref{tab:Kepler_independent_freq} and the complete list of frequencies with $S/N>5$ 
are available in Appendix \ref{app1} in Table\,\ref{tab:Kepler_SC_ALL_freq}.
\setlength{\tabcolsep}{6pt}
\begin{table}
	\centering
	\caption{Independent frequencies detected in the {\it Kepler} SC data.
		The subsequent columns contain: our ID$_K$ number, frequency, amplitude, phase, signal to noise ratio
		and ID$_\mathrm B$ from Table\,1 of \citet{Balona2012}.}
	\label{tab:Kepler_independent_freq}
	\begin{tabular}{  c c c c c  c  } % four columns, alignment for each
		\hline
		ID$_K$  &   $\nu~[\mathrm{d}^{-1}]$   &     $A$ [ppt]   &   $\phi$ [0--1] & $S/N$ & ID$_\mathrm B$\\
		\hline
		1  &  5.661058(2) &    156.998(2)  &  0.8725(1) &  75.7 & $\nu_{1,\mathrm B}$ \\
		2  &  7.148954(7) &     17.051(2)  &  0.9992(1)  & 72.6 & $\nu_{2,\mathrm B}$ \\
		3  &  7.775571(9) &      8.922(2)  &  0.9548(1)  & 67.6 & $\nu_{3,\mathrm B}$ \\
		4  &  9.81587(1)  &    2.113(2)    &  0.5615(1)  & 43.8 & $\nu_{5,\mathrm B}$ \\
		5  &  7.85334(2)  &    1.528(2)    &  0.4139(2)  & 48.3 & $\nu_{6,\mathrm B}$ \\
		6  &  2.95671(2)  &    1.359(2)    &  0.3161(2)  & 42.9 & $\nu_{7,\mathrm B}$ \\
		7  & 10.02421(2)  &    1.223(2)    &  0.6596(3)  & 39.4 & $\nu_{8,\mathrm B}$ \\
		8  &  7.68347(2)  &    1.092(2)    &  0.3050(3)  & 52.8 & $\nu_{9,\mathrm B}$ \\
		9  & 19.39129(2)  &    0.806(2)    &  0.8494(4)  & 28.5 & $\nu_{10,\mathrm B}$ \\
		10  & 16.54373(3)  &    0.520(2)    &  0.1694(6)  & 19.6 & $\nu_{14,\mathrm B}$\\
		11  & 10.22938(3)  &    0.388(2)    &  0.8355(8)  & 27.6 & \\
		12  & 22.12973(4)  &    0.321(2)    &  0.231(1)   & 18.0 & \\
		13  & 16.42229(4)  &    0.239(2)    &  0.966(1)   & 16.8 & \\
		14  & 13.33679(4)  &    0.235(2)    &  0.870(1)   & 14.9 & \\
		15  & 39.09191(6)  &    0.168(2)    &  0.349(2)   & 23.9 & \\
		16  & 22.24766(6)  &    0.147(2)    &  0.743(1)   & 13.3 & \\
		17  & 22.93256(6)  &    0.138(2)    &  0.096(2)   & 20.9 & \\
		18  & 37.77743(8)  &    0.109(2)    &  0.086(3)   & 18.5 &\\
		19  & 15.0704(1)   &    0.071(2)    &  0.066(4)   & 11.0 & \\
		20  & 16.5302(1)   &    0.065(2)    &  0.460(5)   &  6.3 & \\
		21  & 19.9908(1)   &    0.047(2)    &  0.348(7)   &  7.6 & \\
		22  & 22.7419(2)   &    0.039(2)    &  0.935(8)   &  8.1 & \\
		23  & 50.4010(2)   &    0.039(2)    &  0.301(8)   & 13.4 & \\
		24  & 63.0535(6)   &    0.011(2)    &  0.54(2)    &  7.1 & \\
		25  & 77.4900(9)   &    0.007(2)    &  0.50(5)    &  6.8 & \\
		26  & 91.684(1)    &    0.006(2)    &  0.11(5)    &  7.6 & \\
		\hline
	\end{tabular}
\end{table}
A smaller part of the {\it Kepler} data, from Q0.1 and Q6.1-Q7.3, have already been analysed by \citet{Balona2012}.
%but the author used a combined dataset from Q0.1 and Q6.1-Q7.3, in contrast to our dataset from Q6.1-Q10.3.
We have confirmed all of their frequencies (see Table\,1 of \citet{Balona2012}). However, their $\nu_{\mathrm B,4}$,
$\nu_{\mathrm B,11}$, $\nu_{\mathrm B,12}$, $\nu_{\mathrm B,13}$, and $\nu_{\mathrm B,15}$ were identified by us as combinations. 
We provide Balona's frequency designations in the last column of Table\,\ref{tab:Kepler_independent_freq}.

High frequencies, which we define as those above 30\,d$^{-1}$ listed  in Table\,\ref{tab:Kepler_independent_freq} (i.e., $\nu_{15}$, $\nu_{18}$, $\nu_{23}$, $\nu_{24}$, $\nu_{25}$, and $\nu_{26}$), are of an unclear origin. All of them have very small amplitudes and, most probably, they are higher-order combinations,
i.e., with higher $m,\,n,\,o$ integers, than we assumed in our analysis. However, we know that as these integers increase, the probability of Equations\,\ref{eq:combination2} and \ref{eq:combination3}  being met by chance also increases.
The mentioned six frequencies can be expressed as: $\nu_{15}=9\nu_1 +  1\nu_3 -2\nu_4$, $\nu_{18}=1\nu_1 -  1\nu_2 +4\nu_4$,
$\nu_{23}=6\nu_1 -  2\nu_2 +4\nu_8$, $\nu_{24}=10\nu_1 +  2\nu_2 -\nu_5$, $\nu_{25}=9\nu_1 +  1\nu_2 + 1\nu_9$, $\nu_{26}=5\nu_1 +  5\nu_2 + 5\nu_4$.
Here, we note that $\nu_{15}$ and $\nu_{18}$ are also present in the  LC data, whereas $\nu_{23},~\nu_{24},~\nu_{53},~\nu_{26}$ 
as well as their aliases are absent in these data. Moreover, none of these high frequencies were detected in the {\it TESS} data.
Like \citet{Balona2012}, we also detected an independent low-frequency peak $\nu_6=2.95671$\,d$^{-1}$, which can only be associated 
with the high-order g mode. On the other hand, including higher-order combinations, we can express this low frequency
as $\nu_6 = -5\nu_1 +  3\nu_2 + 1\nu_4$. However, if both, this low frequency and above-mentioned high frequencies, correspond 
to pulsation modes, then we encounter a problem with their excitation. They would have to be driven by 
a mechanism other than the classical $\kappa$ mechanism or some modifications to the mean opacity profile would have to be made.

For the {\it TESS} data, we followed the same procedure as with the {\it Kepler} data.
However, this time, the SAP and PDCSAP fluxes showed a significant difference. Furthermore, the amplitudes based on PDCSAP flux in combined sectors
S14+S15 are significantly larger than those in combined sectors S40+S41 and S54+S55. This issue is greatly mitigated in the case of SAP data.
Therefore, we decided to analyse SAP data. The standard deviations calculated for the three pairs of the {\it TESS} sectors are listed in Table\,\ref{tab:SDV_TESS}. 
\setlength{\tabcolsep}{5pt}
\begin{table}
	\centering
	\caption{The standard deviation SD for the combined adjacent sectors of the {\it TESS} SAP and PDCSAP data, i.e., S14+S15, S40+S41, and S54+S55.}
	\label{tab:SDV_TESS}
	\begin{tabular}{ c c c c c c c c c c c c c c c c  c } % four columns, alignment for each
		\hline
		           &  S14+S15   & S40+S41   & S54+S55 \\
		\hline
		SD(SAP)     & 76.1 & 75.0 & 76.7 \\
		SD(PDCSAP)  & 90.8 & 85.7 & 86.2 \\
		\hline
	\end{tabular}
\end{table}
\begin{figure*}
	\includegraphics[angle=270, width=14cm]{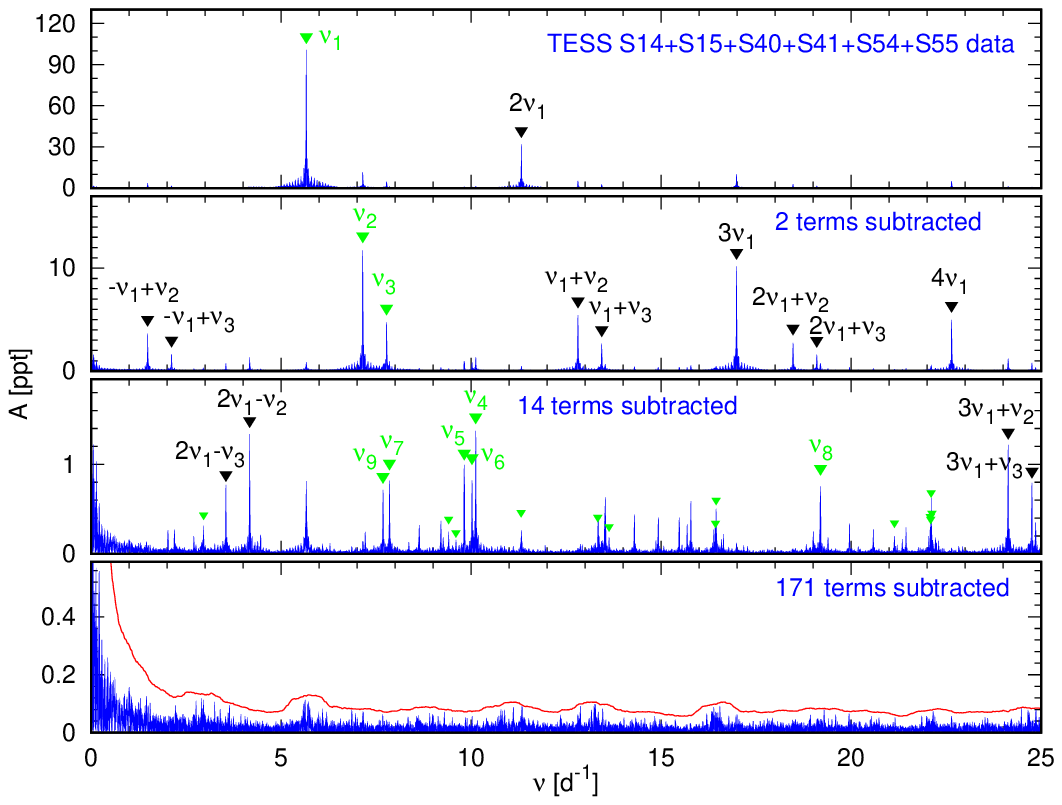}
	\caption{Amplitude periodogram for the combined {\it TESS} data of V2367 Cyg. Panels from top to bottom show
		periodograms: for the original data;  after subtraction of 2 terms; after subtraction of 14 terms and after subtraction of 171 terms.
		Red solid line in bottom panel indicates the $5S/N$ level. All frequency peaks above 0.63\,ppt ( the {\it TESS} amplitudes are smaller than the {\it Kepler} amplitudes by about this factor) 
		are marked by down-pointing triangles, green for independent signals, black for combinations or harmonics.
		Independent frequencies with amplitudes smaller than 0.63\,ppt are marked by small green triangles.}
	\label{fig:V2367Cyg_TESS_full_trf}
\end{figure*}
Moreover, since there are large time-gaps between data from sectors 14+15, 40+41 and 54+55, we analyzed the combined data as well as data from
the adjacent sectors. We started by analysing all combined sectors. This data set consists of 110\,891 observational points spanning 1140.9\,d 
what translates into the Rayleigh resolution $\Delta\nu_{R,T,all}=0.00088\,\mathrm{d}^{-1}$. The error corrections for this data were of about $1.6$.
Pre-whitening procedure resulted in 171 frequencies, most of which were combinations or harmonics. All independent frequencies are listed in Table\,\ref{tab:TESS_combined_independent_freq}. A sample of periodograms are shown in Fig.\,\ref{fig:V2367Cyg_TESS_full_trf} (only for $\nu<25\,\mathrm d^{-1}$). 

As in case of the {\it Kepler} data, we excluded frequencies lower than $0.1$\,d$^{-1}$. 
For their accurate determination, a long-term observations are necessary.
Therefore, we analysed the three data sets: the combined {\it Kepler} SC data , the  combined LC data and the  combined {\it TESS} data.
We cannot be certain whether frequencies from the LC observations  are real or aliases.  
Such low frequencies are often of instrumental origin. Since none of
them agrees  within the corresponding Rayleigh resolution in all three data sets, we consider them not intrinsic to the star.
All low frequencies are listed in Appendix\,\ref{app1} in Table\,\ref{tab:freq_low}.

Additionally, we excluded frequencies close to $\nu_1$, but exceeding the limit of 2.5$\Delta \nu_{R,T,all}$, and marked them as spurious. 
These frequencies may result from small differences in amplitudes in various sectors or imperfect subtraction of $\nu_1$, or both. 
%Although the amplitude differences in SAP photometry are small, they may still exist, as reflected in small differences in standard deviations (see Table\,\ref{tab:SDV_TESS}).
Such frequencies also appeared in the {\it Kepler} data, but they were identified as combinations (see Table\,\ref{tab:Kepler_SC_ALL_freq} in Appendix\,\ref{app1}). However, their origin can also be explained in a similar way.
All frequencies with $S/N>5$ extracted from the six combined {\it TESS} sectors are listed in Appendix\,\ref{app1} in Table\,\ref{tab:TESS_ALL_all_sec_fre}.
\setlength{\tabcolsep}{6pt}
\begin{table*}
	\centering
	\caption{Independent frequencies detected in {\it TESS} combined data.
		The subsequent columns contain: our ID$_T$ number, frequency, amplitude, phase, signal to noise ratio, ID$_K$
		and ID$_\mathrm B$. Tildes in the last and penultimate columns indicate that the difference in frequencies exceeds the Rayleigh resolution
		of the {\it Kepler} SC data but it is below 3 times this resolution.}
	\label{tab:TESS_combined_independent_freq}
	\begin{tabular}{  c c c c c c  c  } % four columns, alignment for each
		\hline
		ID$_{T}$  &   $\nu~[\mathrm{d}^{-1}]$   &     $A$ [ppt]   &   $\phi$ [0--1] & $S/N$ & ID$_{K}$ & ID$_{\mathrm B}$\\
		\hline
		1  &                 5.6610574(7)  &          100.24(1)  &         0.86115(2)  &        31.7 &  $\nu_{1,K}$                                 & $\nu_{1,\mathrm B}$ \\
		2  &                  7.148978(3)  &           11.76(1)  &          0.9321(2)  &        31.3 &  $\nu_{2,K}$                                 & $\nu_{2,\mathrm B}$   \\
		3  &                  7.775571(3)  &            4.71(1)  &          0.9577(4)  &        28.0 &  $\nu_{3,K}$                                 & $\nu_{3,\mathrm B}$ \\
		4  &                 10.122796(8)  &            1.39(1)  &           0.918(1)  &        17.3 &  $-2\nu_{1,K}+3\nu_{2,K}$                    & $\nu_{4,\mathrm B}$ \\
		5  &                   9.81861(1)  &            0.92(1)  &           0.881(2)  &        16.7 &  $\sim\nu_{4,K}$                             & $\sim\nu_{5,\mathrm B}$ \\
		6  &                  10.02425(1)  &            0.87(1)  &           0.513(2)  &        19.7 &  $\nu_{7,K}$                                 & $\nu_{8,\mathrm B}$ \\
		7  &                   7.85337(1)  &            0.81(1)  &           0.338(2)  &        16.8 &  $\nu_{5,K}$                                 & $\nu_{6,\mathrm B}$ \\
		8  &                  19.19308(1)  &            0.77(1)  &           0.665(2)  &        18.0 &  $9\nu_{4,K}-9\nu_{8,K}$                     &  \\
		9  &                   7.68344(1)  &            0.72(1)  &           0.331(2)  &        20.3 &  $\nu_{8,K}$                                 & $\nu_{9,\mathrm B}$\\
		10  &                  22.10602(1)  &            0.53(1)  &           0.288(4)  &        16.2 &  $\sim8\nu_{3,K}-4\nu_{7,K}$                 & $\sim\nu_{11,\mathrm B}$ \\
		11  &                  16.44771(2)  &            0.43(1)  &           0.067(4)  &        12.7 &  $5\nu_{1,K}+1\nu_{3,K}-2\nu_{4,K}$          & $\nu_{13,\mathrm B}$ \\
		12  &                   2.95974(3)  &            0.29(1)  &           0.752(6)  &         9.6 &  $\sim \nu_{6,K}$                            & $\sim\nu_{7,\mathrm B}$ \\
		13  &                  22.12713(3)  &            0.32(1)  &           0.162(6)  &         9.1 &  $\sim\nu_{12,K}$                            &  \\
		14  &                   9.40935(3)  &            0.25(1)  &           0.212(7)  &         9.0 &  $4\nu_{7,K}-3\nu_{11,K}$                    &  \\
		15  &                  22.08397(3)  &            0.28(1)  &           0.492(6)  &         8.9 &  $1\nu_{1,K}+1\nu_{13,K}$                    &  \\
		16  &                  13.33399(4)  &            0.27(1)  &           0.503(7)  &         7.0 & $\sim \nu_{14,K}$                            &  \\
		17  &                  21.13797(3)  &            0.21(1)  &           0.791(8)  &        11.0 & $2\nu_{1,K}+1\nu_{4,K}$                      &  \\
		18  &                  11.33598(3)  &            0.22(1)  &           0.272(9)  &         8.0 &  $\sim -2\nu_{14,K}+1\nu_{17,K}+1\nu_{19,K}$ & \\
		19  &                  13.63011(4)  &            0.16(1)  &            0.64(1)  &         6.9 & $8\nu_{6,K}-1\nu_{7,K}$                      &  \\
		20  &                  22.09581(4)  &            0.22(1)  &           0.098(9)  &         6.5 & $-1\nu_{4,K}+2\nu_{8,K}+1\nu_{10,K}$         &  \\
		21  &                  27.96166(6)  &            0.10(1)  &            0.35(2)  &         6.2 & $-2\nu_{9,K}+3\nu_{16,K}$                    & \\
		22  &                   9.60108(7)  &            0.10(1)  &            0.36(2)  &         6.0 & $1\nu_{2,K}-1\nu_{3,K}+1\nu_{11,K}$          &  \\
		23  &                  43.23259(7)  &            0.09(1)  &            0.10(2)  &         7.6  & $-8\nu_{1,K,K}+4\nu_{12,K}$                 & \\
		24  &                  32.11376(8)  &            0.08(1)  &            0.02(2)  &         5.6  & $\sim -1\nu_{2,K}+5\nu_{5,K}$               & \\
		25  &                  26.18343(9)  &            0.08(1)  &            0.24(2)  &         5.2  &                                             &  \\
		
		\hline
	\end{tabular}
\end{table*}

In the next step, we analysed TESS data from combined adjacent sectors, i.e., S14+S15, S40+S41 and S54+S55. Each of these data sets spans about 55 days 
and have the Rayleigh resolution of about 0.018\,d$^{-1}$. Here, we limited $m,~n$ in Eq.\,\ref{eq:combination2} to the range $<-10,\,10>$ 
for all detected frequencies. We found 61, 66 and 68 significant frequencies in S14+S15, S40+S41 and S54+S55, respectively. 
Independent frequencies are given in Appendix\,\ref{app1} in Table\,\ref{tab:TESS_adjacent_independent_freq}.
All significant frequencies (i.e. with $S/N>5$) are listed in Tables\,\ref{tab:TESS_ALL_S14_15_fre}--\ref{tab:TESS_ALL_S54_55_fre} 
in Appendix\,\ref{app1} along with a sample of periodograms calculated for the S14+S15 data shown in Fig.\,\ref{fig:V2367Cyg_TESS_S14S15_trf}.
We stress that due to the low Rayleigh resolution of the two-sector data, some frequencies identified in the full data set as independent are classified in Tables\,\ref{tab:TESS_ALL_S14_15_fre}--\ref{tab:TESS_ALL_S54_55_fre} as combinations. 
%They are mentioned in Appendix\,\ref{app1}. %Examples of such frequencies are marked by open green triangles in Fig.\,\ref{fig:V2367Cyg_TESS_S14S15_trf}.
%
\begin{figure}
	\includegraphics[angle=270, width=\columnwidth,]{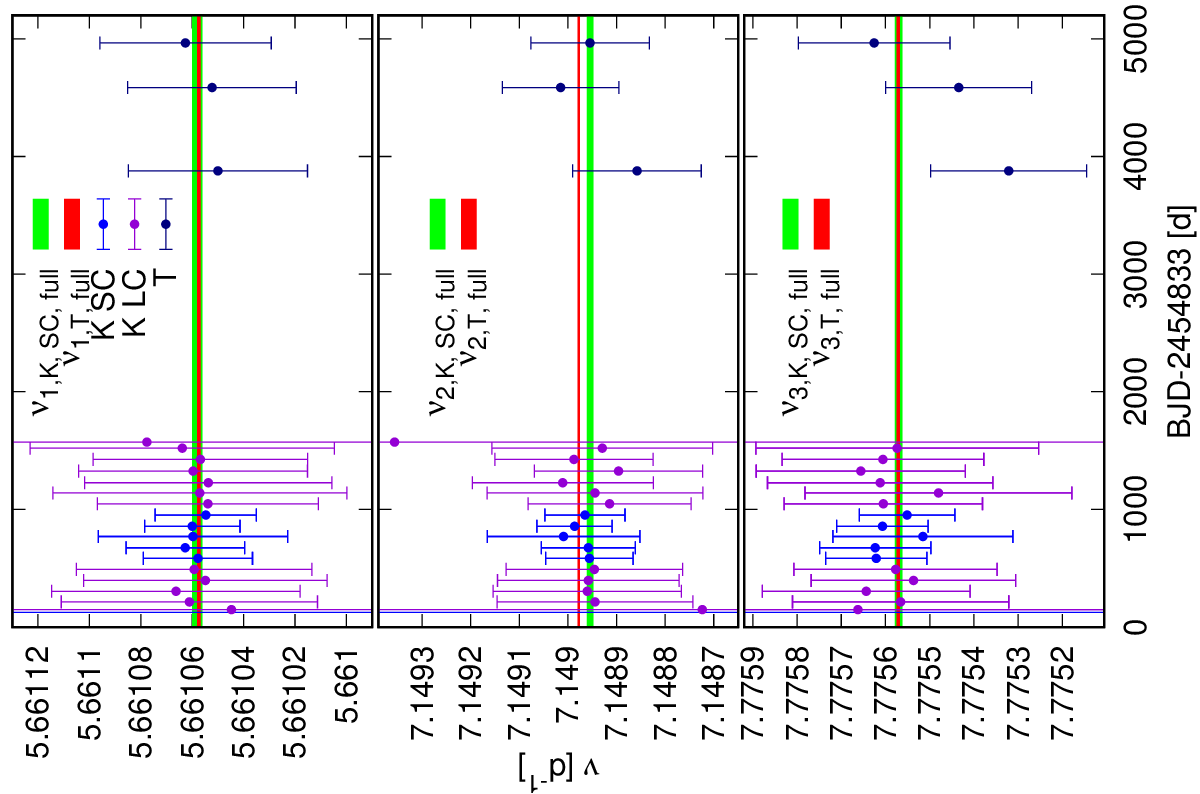}
	\caption{The values of $\nu_1$ (top panel), $\nu_2$ (middle panel), and $\nu_3$ (bottom panel) with  error bars determined in each {\it Kepler}
		 quarter and adjacent pairs of the  TESS sectors. For a better visibility, the error bars determined from Q0, Q1 and Q17 data are cut.
		 Horizontal green and red lines, whose width corresponds to the error, represent the values of frequencies determined from
		full {\it Kepler} and TESS data sets, respectively.}
	\label{fig:freq_evol_nu1}
\end{figure}

Subsequently, we looked for possible changes in the three highest-amplitude independent frequencies, $\nu_1,~\nu_2$ and $\nu_3$. 
To this end, we analyzed the {\it Kepler} data for each quarter separately. Whenever possible, we utilized SC data. In quarters where SC data were 
not available, we used LC data. However, due to a strong aliasing problem (see Fig.\,\ref{fig:LC_vs_SC} in Appendix\,\ref{app1}),
arising from the low Nyquist frequency for the LC data, we limited the pre-whitening process to several frequencies with the highest amplitudes.
The results are presented in Fig.\,\ref{fig:freq_evol_nu1}, where the values of three dominant frequencies are plotted as a function of time.
The  frequency values from adjacent {\it TESS} sectors were also added. As one can see,  all three frequencies, within errors,
are constant over the time considered.
The frequency values are listed in Table\,\ref{tab:freq_evol} in Appendix\,\ref{app1}.

\section{Identification of the mode degree $\ell$ from $UBVRI$ photometry}
\begin{figure*}
	\includegraphics[width=175mm,clip]{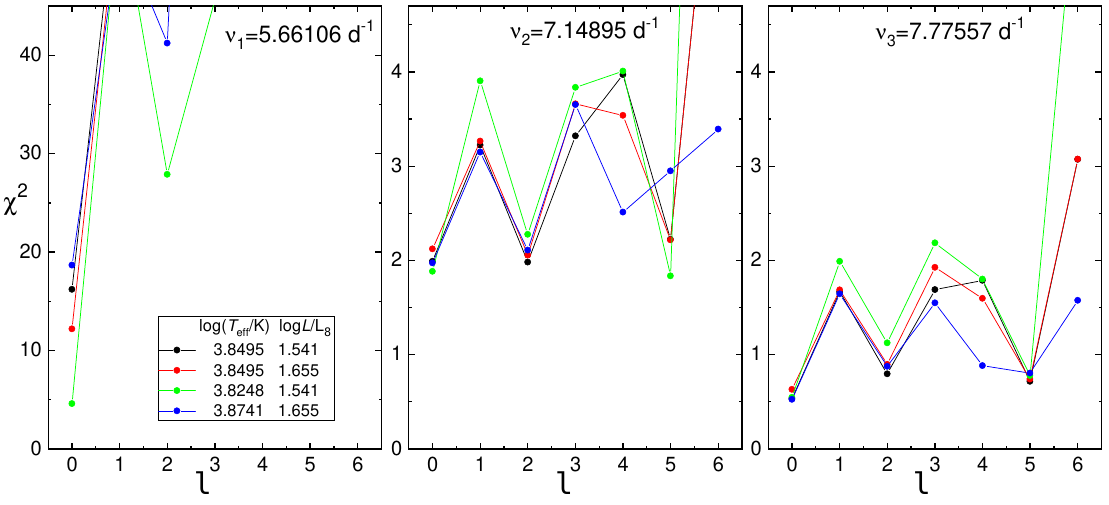}
	\caption{Identification of the mode degree $\ell$ for the three frequencies of V2367 Cyg with the highest amplitudes:
		$\nu_1=5.66106$\,d$^{-1}$ (left panel), $\nu_2=7.14898$\,d$^{-1}$ (middle panel) and
		$\nu_3=7.77558$\,d$^{-1}$ (right panel). NEMO model atmospheres were adopted with the metallicity [m/H]$=+0.2$ 
		and microturbulent velocity $\xi_t=4\,\kms$. Results fo the four pairs of $(T_{\rm eff}, L)$ are presented.}
	\label{ModeID}
\end{figure*}
Within the linear theory, the local radial displacement of the surface element as a result of a pulsational mode with 
the angular frequency $\omega$, harmonic degree $\ell$ and azimuthal order $m$, can be written as
$$\frac{\delta r}R= {\rm Re}\{ \varepsilon Y_\ell^m (\theta,\varphi)
{\rm e}^{-{\rm i}\omega t}\},  \eqno(6)$$
where $\varepsilon$ is the intrinsic mode amplitude, that cannot be determined within the linear approach, and 
$Y_\ell^m (\theta,\varphi)$ is a spherical harmonic defined as:
$$Y^m_\ell(\theta,\phi)=(-1)^{\frac{m+|m|}{2}}
\sqrt{ \frac{(2\ell+1)(\ell-|m|)!}{(\ell+|m|)!} }
P_{\ell}^{|m|}(\cos\theta){\rm e}^{{\rm i} m\phi}. \eqno(7)$$
With this normalization, $|\varepsilon|$ is the rms value of $\delta r/R$ over the stellar surface.
The corresponding changes of the bolometric flux, ${\cal F}_{\rm bol}$, and the local gravity, $g$, are given by
$$\frac{ \delta {\cal F}_{\rm bol} } { {\cal F}_{\rm bol} }= {\rm
	Re}\{ \varepsilon f Y_\ell^m(\theta,\varphi) {\rm e} ^{-{\rm i} \omega t} \}, \eqno(8a)$$
and
$$\frac{\delta g}{g} = - \left( 2 + \frac{\omega^2 R^3}{G M}
\right) \frac{\delta r}{R}. \eqno(8b)$$
The parameter $f$ describes the ratio of the relative flux variation to the
relative radial displacement of the surface. Both $\varepsilon$ and $f$ have to be regarded complex.
Since we can assume that both $\varepsilon$ and $f$ are constant in the
atmosphere, we can use the static plane-parallel approximation.
Then, the complex amplitude of the relative monochromatic flux
variation in a passband $\lambda$ can be expressed as \citep[e.g.,][]{JDD2005b}:
$${\cal A}_{\lambda} = {\cal D}_{\ell}^{\lambda} ({\tilde\varepsilon}
f) +{\cal E}_{\ell}^{\lambda} {\tilde\varepsilon}, \eqno(9)$$
where
$${\tilde\varepsilon}\equiv \varepsilon Y^m_{\ell}(i,0), \eqno(10a)$$
$${\cal D}_{\ell}^{\lambda} = b_{\ell}^{\lambda} \frac14
\frac{\partial \log ( {\cal F}_\lambda |b_{\ell}^{\lambda}| ) }
{\partial\log T_{\rm{eff}}},  \eqno(10b)$$
$${\cal E}_{\ell}^{\lambda}= b_{\ell}^{\lambda} \left[ (2+\ell
)(1-\ell ) -\left( \frac{\omega^2 R^3}{G M}
+ 2 \right) \frac{\partial \log ( {\cal F}_\lambda
	|b_{\ell}^{\lambda}| ) }{\partial\log g} \right], \eqno(10c)$$
and
$$b_{\ell}^{\lambda}=\int_0^1 h_\lambda(\mu) \mu P_{\ell}(\mu) d\mu. \eqno(10d)$$
The term ${\cal D}_{\ell}^\lambda$ describes the temperature effects and ${\cal E}_{\ell}^\lambda$ combines
the geometrical and pressure effect. $G,~M,~R$ have their usual meanings.
$h_\lambda(\mu)$ is the limb darkening law and $P_\lambda(\mu)$ is the Legendre polynomial.
The values of $b_{\ell}^{\lambda}$ and the partial derivatives of ${\cal F}_\lambda |b_{\ell}^{\lambda}|$ 
are calculated from model atmospheres.
The values of the amplitude and phase are given by $A_{\lambda}=|{\cal A}_\lambda|$, $\varphi_{\lambda}= {\rm arg}({\cal A}_\lambda)$, respectively.

The system of $N$(-passbands) complex equations (9) is solved for a given $\ell$ and $(T_{\rm eff},\log g)$ to determine $\tilde\varepsilon$ and $f$.
We considered the degree $\ell$ and associated complex values of $\tilde\varepsilon$ and $f$ as most probable if there is a clear
minimum in the difference between the calculated and observed photometric amplitudes and phases. 
The goodness of fit is defined as:
$$\chi^2=\frac1{2N-N_p} \sum_{i=1}^N  \frac{ \left|{\cal A}^{obs}_{\lambda_i} - {\cal A}^{cal}_{\lambda_i}\right|^2 }
{ |\sigma_{\lambda_i}|^2},\eqno(11)$$
where the superscripts $obs$ and $cal$  denote the observed and calculated complex amplitude, respectively.
$N_p$ is the number of parameters to be determined and $N_p=4$ because there are two complex parameters, $\tilde\varepsilon$ and $f$. 
The observational errors $\sigma_{\lambda}$ are computed as
$$|\sigma_\lambda|^2= \sigma^2 (A_{\lambda})  +  A_{\lambda}^2 \sigma^2(\varphi_\lambda), \eqno(12)$$
where $\sigma  (A_{\lambda})$ and $\sigma (\varphi)$ are the errors of the observed amplitude and phase 
in a passband $\lambda$, respectively.

\citet{Ulusoy2013} gathered multi-colour time series photometry of V2367 Cyg in 5 filters, $UBVRI$,
and determined the amplitudes and phases for the three main frequencies $\nu_1,~\nu_2,~\nu_3$.
Using these data, we applied the above described method to identify the mode degree $\ell$.
The results are shown in Fig.\,\ref{ModeID} for the four pairs of $(\log T_{\rm eff},\log L/{\rm L}_\odot)$.
We used Vienna model atmospheres \citep{Heiter2002} that include turbulent convection treatment from \citet{Canuto1996}.
For the limb darkening law, $h_\lambda(\mu)$, we computed coefficients assuming the non-linear formula of \citet{Claret2000}.
Here, we show results for the atmospheric metallicity [m/H]$=+0.2$ and microturbulent velocity $\xi_t=4\,\kms$.
For lower metallicity [m/H], the discriminant $\chi^2$ takes higher values but the dependence $\chi^2(\ell)$ is qualitatively the same.
As one can see the dominant frequency $\nu_1=5.66106$\,d$^{-1}$ is undoubtedly the radial mode. Taking into account the disc averaging effect, 
the most probable $\ell$ degrees for $\nu_2=7.14898$\,d$^{-1}$, $\nu_3=7.77557$\,d$^{-1}$ are 0 or 2.
Because of the values of $\nu_2$ and $\nu_3$, it is obvious that both of them cannot correspond to radial modes.
The amplitudes of $\nu_2$ and $\nu_3$ are smaller than the amplitude of $\nu_1$ by about 8.5 and 21.2 times, respectively.
 
\begin{figure}
	\includegraphics[width=\columnwidth,clip]{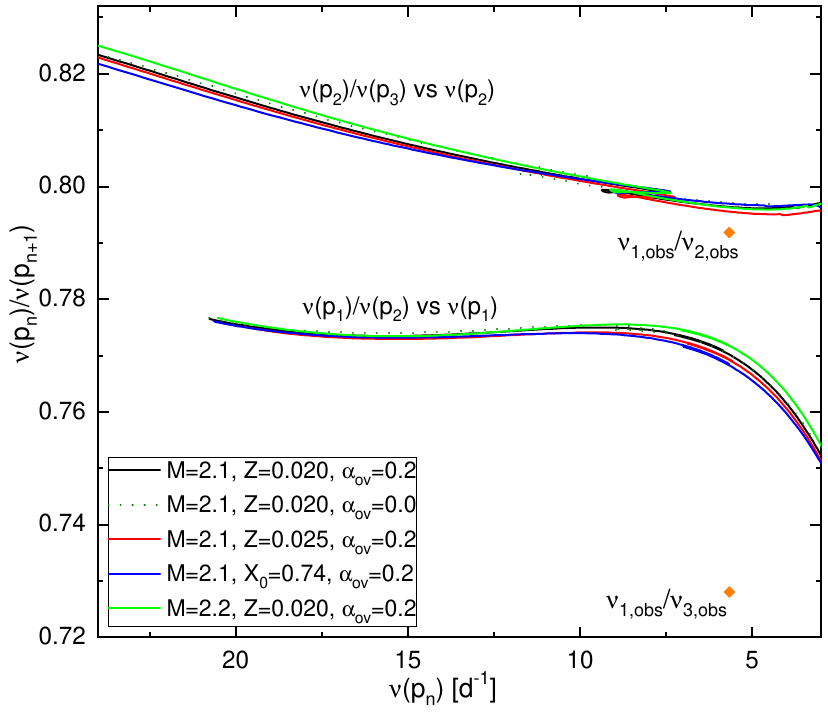}
	\caption{The Petersen diagram with a run of the frequency ratio for the two pairs of consecutive radial modes: fundamental and first overtone
		(lower lines), and the first and second overtone (upper lines). There is shown the effect of a mass $M$, initial hydrogen ($X_0$) 
		and metallicity $Z$ abundances as well as the overshooting parameter  $\alpha_{\rm ov}$. The mixing length parameter was $\alpha_{\rm MLT}=0.5$ and the initial rotation  $V_{\rm rot,0}=20~ \kms$.	The OPAL tables and AGSS09 chemical mixture were adopted. The observed ratios of frequencies of V2367 Cyg are marked as orange diamonds. }
	\label{Petersen1}
\end{figure}

Fig.\,\ref{Petersen1} shows the Petersen diagram with the marked values of the observed frequency ratios, $\nu_1/\nu_2=0.79187$ 
and $\nu_1/\nu_3=0.72807$, along with the run of theoretical ratios for the frequencies of radial fundamental and first overtone $\nu(p_1)/\nu(p_2)~vs~\nu(p_1)$ (lower lines) 
as well as the first and second overtones $\nu(p_2)/\nu(p_3)~vs~\nu(p_2)$ (upper lines).

All evolutionary computations were performed with the Warsaw-New Jersey code which takes into account the mean effect of the centrifugal force. 
It assumes the solid-body rotation and conservation of global angular momentum during evolution. Convection in stellar envelope is treated in the framework
of standard mixing-length theory (MLT). The OPAL2005 equation of state was used \citep{Rogers1996,Rogers2002},  the OPAL opacity tables 
and AGSS09 chemical mixture. Overshooting from a convective core is treated according to the formula of \citet{DziemPamy2008}.
In our pulsational computations, we rely on the nonadiabatic code of W.\,Dziembowski \citep{Dziembowski1977a,Pamyatnykh1999}, which 
includes the effects of rotation on pulsational frequencies within the perturbation approach taking into account non-spherically symmetric 
distortion due to the centrifugal force and the second- and third-order effects of the Coriolis force.
The convective flux freezing approximation is used, which is justified if convection is not very efficient in the envelope.

In Fig.\,\ref{Petersen1}, there is shown the effect of mass ($M=2.1$ vs 2.2 M$_\odot$),
metallicity ($Z=0.02$ vs 0.025), initial hydrogen abundance ($X_0=0.70$ vs 0.74) and overshooting parameter ($\alpha_{\rm ov}=0.2$ vs 0.0).
The mixing length parameter was $\alpha_{\rm MLT}=0.5$ and the initial rotation $V_{\rm rot,0}=20~ \kms$.	

As one can see, the hypothesis that $\nu_3$ is a radial mode can be safely rejected because of much too low value of $\nu_1/\nu_3$.
Changing any parameter (within a reasonable range) will not allow this value to be achieved.
On the other hand a ratio of the dominant and second frequency $\nu_1/\nu_2$ is much closer to the radial frequency ratios, in particular, 
for the $p_2\& p_3$ hypothesis. However, even in that case, we would have to assume high values of metallicity ($Z> 0.03$)
and very large overshooting $\alpha_{\rm ov}\gtrsim0.4$. Moreover, such models have $M\approx 1.7$\,M$_\odot$,
$\log T_{\rm eff}\approx3.67$  and $\log L/{\rm L}_\odot\approx 0.9$, which are far outside  the observed error box  in the HR diagram.
To explain the observed ratio $\nu_1/\nu_2$ by the fundamental and first overtone (the $p_1\& p_2$ hypothesis)
we had to assume a very low metallicity ($Z\lesssim0.002$). In this case,  the models have $M\approx 2 .1- 3.1$\,M$_\odot$,
$\log T_{\rm eff}\approx4.0$  and $\log L/{\rm L}_\odot\approx 2.1 - 2.6$, again far outside the observed error box.
%as mentioned also by \citet{Balona2012}. 

\section{The effects of rotation}
As we mentioned in Sect.\,2, the are two different determinations of the projected rotational velocity of V2367 Cyg:
$V_{\rm rot}\sin i\approx 100~\kms$ \citep{Balona2012,Ulusoy2013} and $V_{\rm rot}\sin i\approx 16~\kms$ \citep{Niemczura2017}.
In any case, this is the lower limit of rotation. Therefore, in our modelling, we will consider the whole range up to the  half of critical rotation,
which amounts to about $V_{\rm rot}^{\rm crit} = 340~ \kms$, if $M\approx 2.2$\,M$_\odot$ and $R\approx 3.7$\,R$_\odot$. 
The corresponding critical value of rotational frequency is about $\nu_{\rm rot}^{\rm crit}\approx 1.7$\,d$^{-1}$. 
If the correct value of the projected rotational velocity is $100~\kms$ then the minimum value of the rotation frequency is approximately
$\nu_{\rm rot}^{\rm min}\approx 0.53$\,d$^{-1}$, if it is $16~\kms$ then $\nu_{\rm rot}^{\rm min}\approx 0.08$\,d$^{-1}$.
So, even at half the critical rotation velocity, the ratio  $\nu_{\rm rot}/\nu_{\rm puls}$ reaches the values of about 0.15 and 0.12
for the observed frequencies $\nu_1$ and $\nu_2$, respectively. Therefore, using a perturbation approach is justified.

Let us recall the third order expression for a rotationally splitting frequency \citep[e.g.,][]{Gough1990,Soufi1998,Pamyatnykh2003}
$$\nu_{n\ell m}=\nu_0+m(1-C_{n\ell})\nu_{\rm rot}+\frac{\nu_{\rm rot}^2}{\nu_0}(D_0+m^2D_1)+m\frac{\nu_{\rm rot}^3}{\nu_0^2}T, \eqno(13)$$
where $n$ is the radial order, the frequency $\nu_0$ includes already the effect of rotation on the equilibrium model,
the Ledoux constant $C_{n\ell}$ depends on the mode and model parameters, and the term $m\nu_{\rm rot}$ transforms the co-rotating coordinate system
to the inertial coordinate system of the observer. The coefficients $D_0, ~D_1,~T$ describe % the second- and third-order effects, i.e.,
non-spherically symmetric distortion due to the centrifugal force as well as the second- and third- order effects of the Coriolis force.
In the case of radial mode, this formula reduces to:
$$\nu=\nu_0 + D_0 \frac{\nu_{\rm rot}^2}{\nu_0}. \eqno(14)$$
As was shown long time ago by \citet{Simon1969}, the coefficient $D_0$ amounts to about $\frac43$ independently of the radial order $n$.
Thus, if $V_{\rm rot}\approx100~\kms$ then $\nu_{\rm rot}^{\rm min}\approx 0.5$\,d$^{-1}$ and a correction for the frequencies $\nu_1$ and $\nu_2$ is about 0.06 and 0.05, respectively.

\begin{figure}
	\includegraphics[width=\columnwidth,clip]{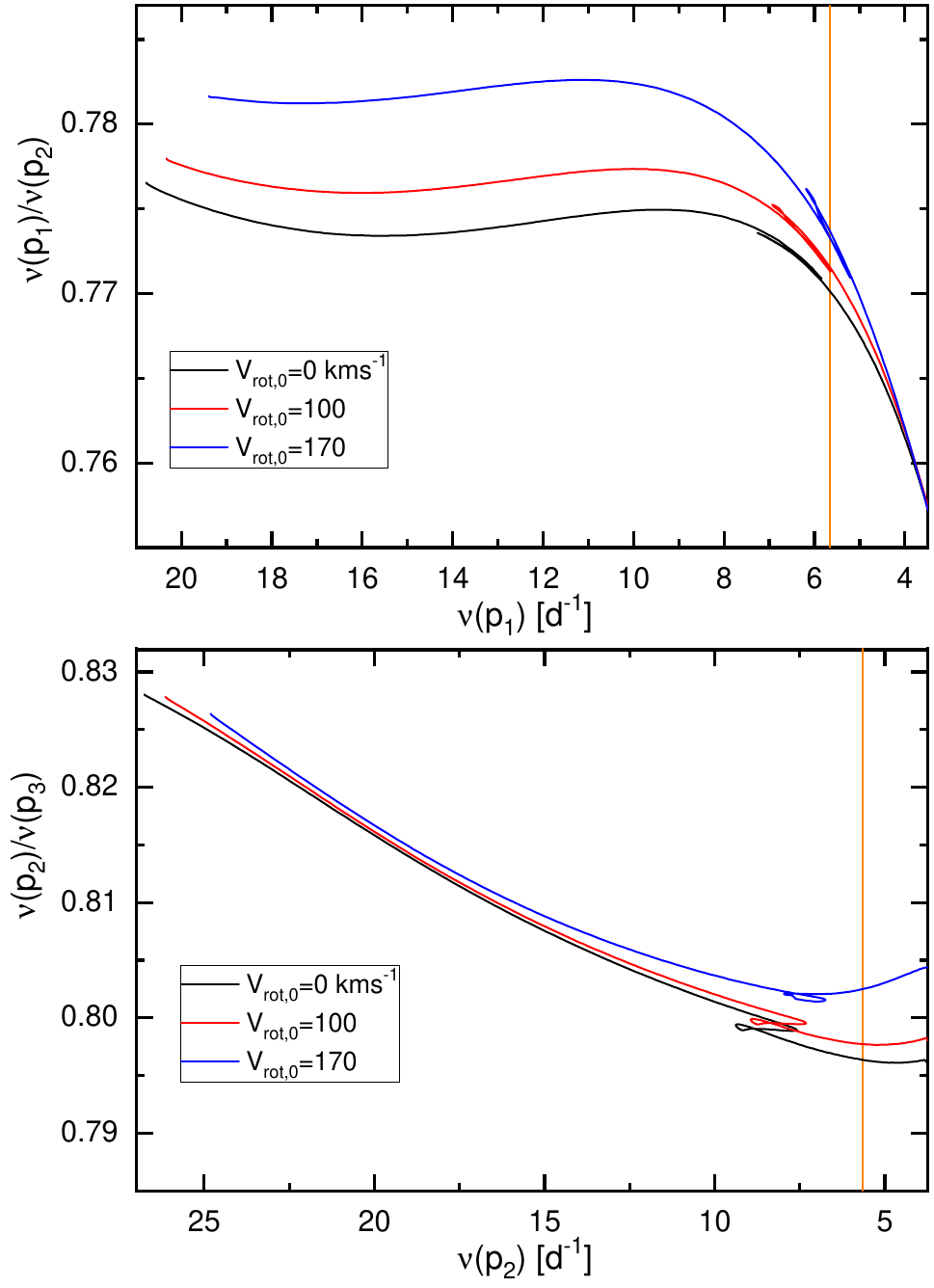}
	\caption{The Petersen diagrams for the two pairs of radial modes: fundamental and first overtone (top panel), first and second
		overtone (bottom panel). The model parameters are: $M=2.1$\,M$_\odot$, $X_0=0.70$, $Z=0.02$, $\alpha_{\rm MLT}=0.5$	and $\alpha_{\rm ov}=0.2$.
		The OPAL tables and AGSS09 chemical mixture were adopted. There is shown the effect of rotation (up to the second order) on pulsational frequencies. Models were computed for the three values of the initial rotational velocity $V_{\rm rot,0} =0,~100,~170~ \kms$.}
	\label{Petersen_Vrot}
\end{figure}

In Fig.\,\ref{Petersen_Vrot}, we compare the run of frequency ratios of radial modes on the Petersen diagrams for three values 
of the initial rotation: $V_{\rm rot,0}= 0,~100,~170~\kms$, where $V_{\rm rot,0}=170~\kms$ corresponds to about half 
the critical rotation. The model parameters are given in the caption and vertical line marks the dominant frequency of V2367 Cyg.
The top panel of Fig.\,\ref{Petersen_Vrot} shows the frequency ratio of the radial fundamental and first overtone modes, $\nu(p_1)/\nu(p_2)$,
as a function of the fundamental mode frequency $\nu(p_1)$. As one can see, the effect of rotation is huge, in particular,
in the main-sequence (MS) phase of evolution. The higher the rotational velocity, the higher the frequency ratio.
During the post-main sequence (post-MS) phase the differences in $\nu(p_1)/\nu(p_2)$ 
decrease and very quickly all three lines converge. The largest difference in the frequency ratio  $\nu(p_1)/\nu(p_2)$ between the rotational velocity 
0 and 170\,$\kms$ reaches a value of approximately 0.01. This is well above the observational error and much above the accuracy with which we want 
to match the theoretical frequencies to the observed ones. Typical fitting accuracy is at least to the fourth decimal place in the frequency ratio.

In the bottom panel of Fig.\,\ref{Petersen_Vrot}, we depicted the frequency ratio of the first and second overtone modes, $\nu(p_2)/\nu(p_3)$,
as a function of the first overtone frequency $\nu(p_2)$. Again, the effect of rotation is huge but the situation is opposite to the previous one.
During most of the MS phase, all three lines  $\nu(p_2)/\nu(p_3)(V_{\rm rot})$ nearly converge, whereas the differences between various $V_{\rm rot}$
start to increase at the end of MS and becomes larger and larger in the post-MS phase. The largest difference in $\nu(p_2)/\nu(p_3)$
 between 0 and 170\,$\kms$ is about 0.008. Again, this is well above the observational error and above our requirements in seismic modelling.

The same effects of rotation on the Petersen diagram, $\nu(p_1)/\nu(p_2)~vs~\nu(p_1)$, have already been studied by \citet{Suarez2006} 
for the rotational velocity up to $V_{\rm rot}=50~\kms$. Even at such low rotation, the authors obtained differences in $\nu(p_1)/\nu(p_2)$ 
on the third decimal place.

One more important effect of rotation, that is not included in Eq.\,13, is the rotational coupling of pulsation modes  \citep{Chandrasekhar1962,Soufi1998}.
This effect occurs for the close-frequency modes with the degrees $\ell$ differing by 2 and with the same azimuthal order $m$, 
i.e., $\ell_j=\ell_i+2$ and $m_i=m_j$. Two (or more) modes have close frequencies if their difference is of the order of rotational frequency or less. 
%  the displacement eigenfunction of an individual mode is given as a sum
%$$\vec{\xi}=\sum_{k}a_{k} \vec{\xi}_{0k}, \eqno(10)$$
%
%where $\vec{\xi}_{0k}$ are the eigenfunctions of pure modes.

\begin{figure}
	\includegraphics[width=\columnwidth,clip]{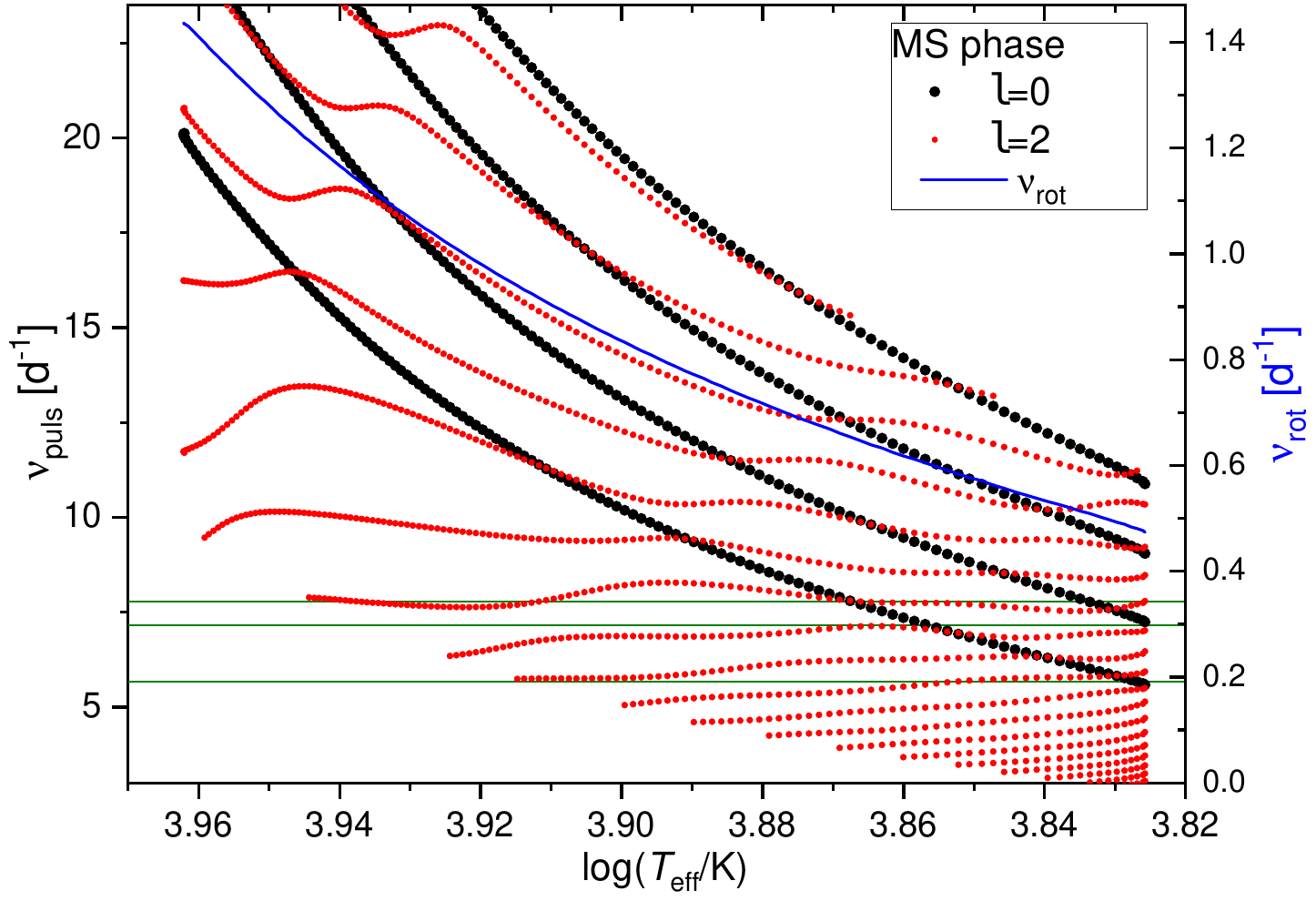}
	\includegraphics[width=\columnwidth,clip]{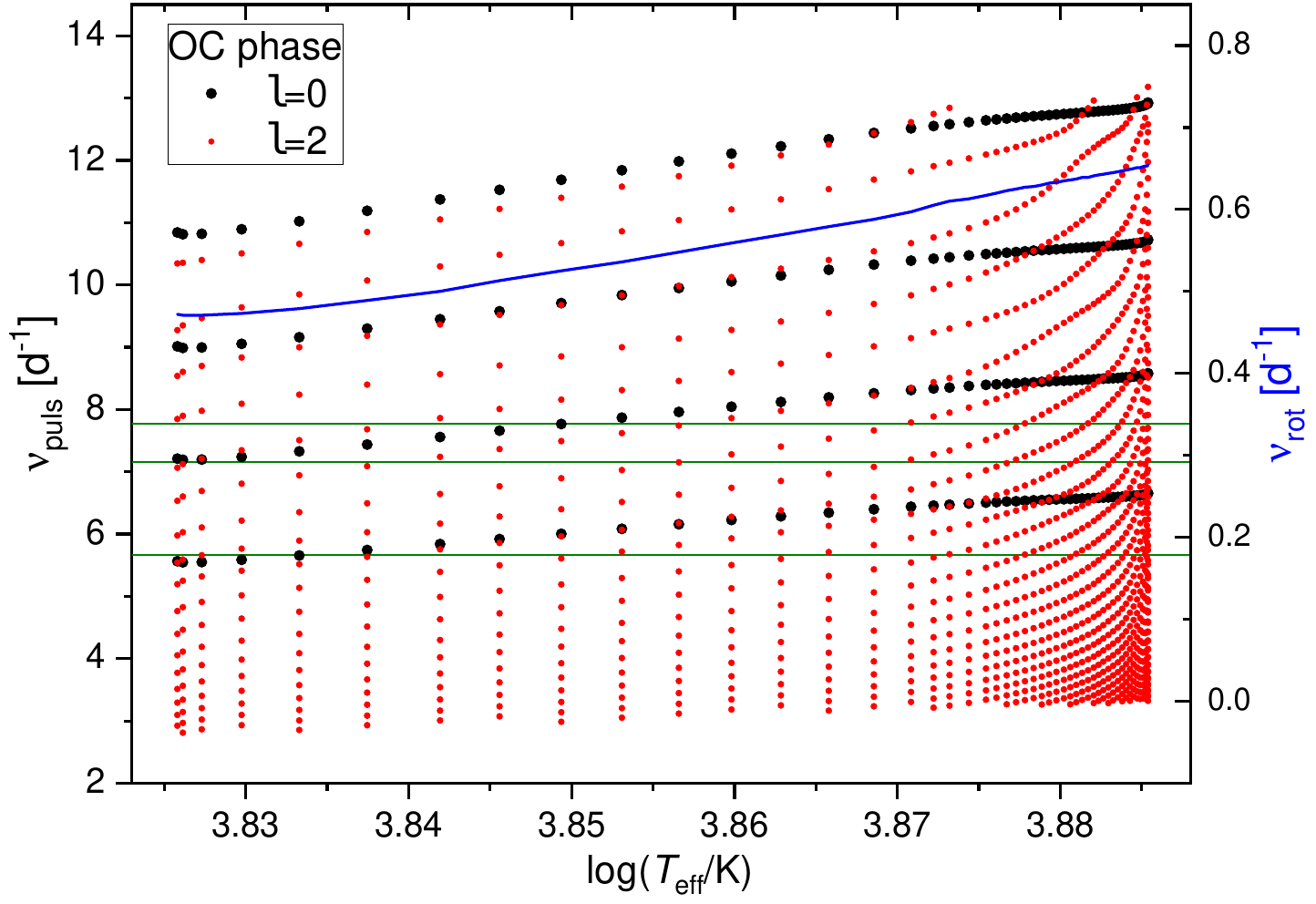}
	\includegraphics[width=\columnwidth,clip]{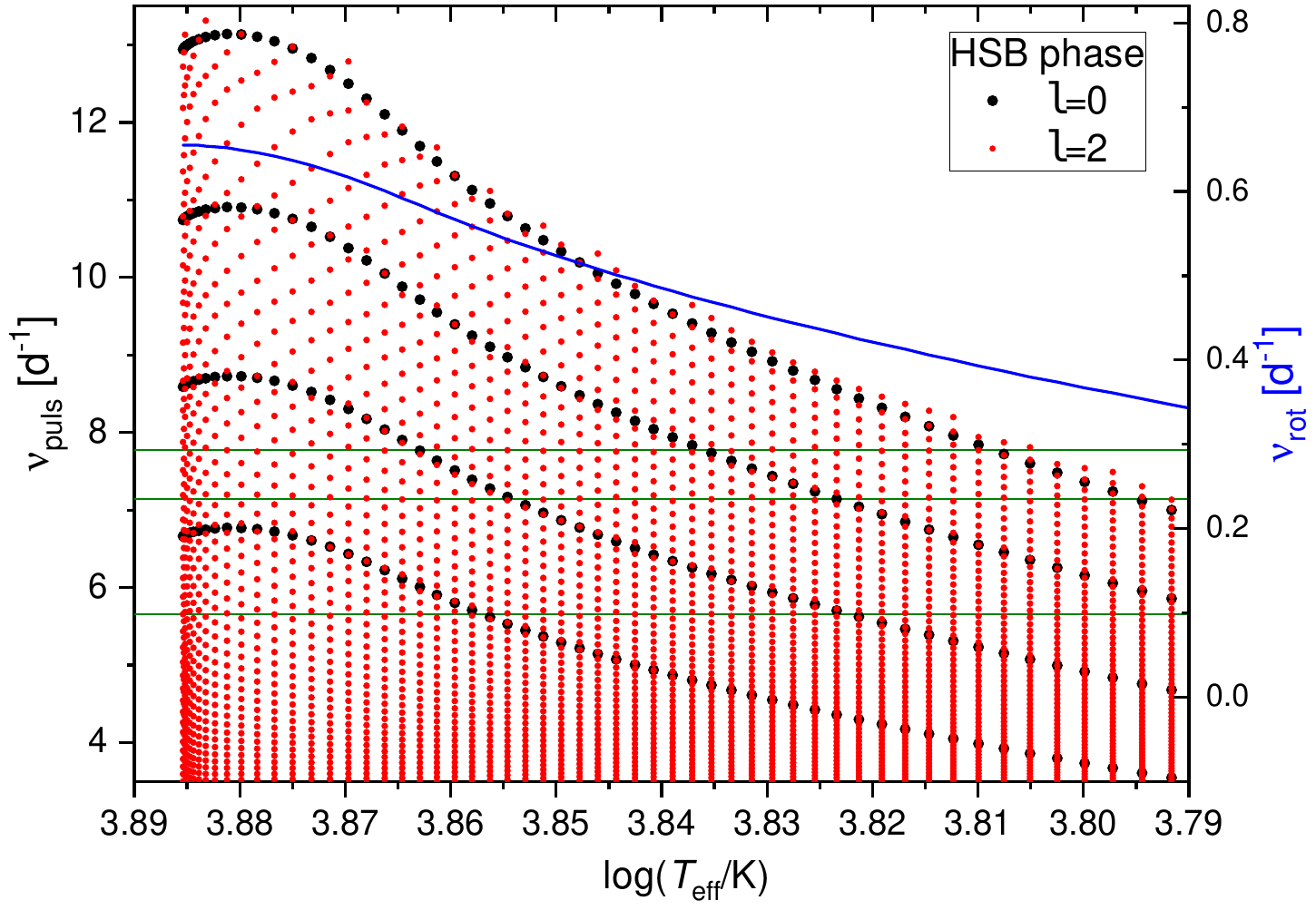}
	\caption{The evolution of pulsational frequencies of modes with $\ell=0$ and $\ell=2$ for the model with the same parameters as in Fig.\,\ref{Petersen_Vrot} except for the initial rotation which is here $V_{\rm rot,0} =120~ \kms$. Each phases of evolution is presented separately:
	MS (top panel), OC (middle panel) and HSB (bottom panel).  The value of rotational frequency is given on the right-hand Y axis.
	The horizontal lines mark the values of the three dominant frequencies of V2367 Cyg.}
	\label{freq_evol}
\end{figure}

In Fig.\,\ref{freq_evol}, we show the evolution of frequencies of the pulsational modes with $\ell=0$ and $\ell=2$, for the model with parameters: $M=2.1$\,M$_\odot$, $X_0=0.70$, $Z=0.02$, $\alpha_{\rm MLT}=0.5,~\alpha_{\rm ov}=0.2$ and the initial rotation of  $V_{\rm rot,0} =120~ \kms$. Three phases of evolutions are presented separately: MS (top panel), overall contraction (OC) (middle panel)
and hydrogen shell burning (HSB) (bottom panel).  We indicated also the value of the rotational frequency on the right-hand Y axis.
The three dominant frequencies of V2367\,Cyg ($\nu_1=5.66106$\,d$^{-1}$,  $\nu_2=7.14898$\,d$^{-1}$, $\nu_3=7.77557$\,d$^{-1}$) 
are marked with horizontal lines. As one can see, a close encounter of radial and quadrupole modes occurs very often, in particular
during the HSB phase, because of very dense spectrum of nonradial modes. 

\begin{figure}
	\includegraphics[width=\columnwidth,clip]{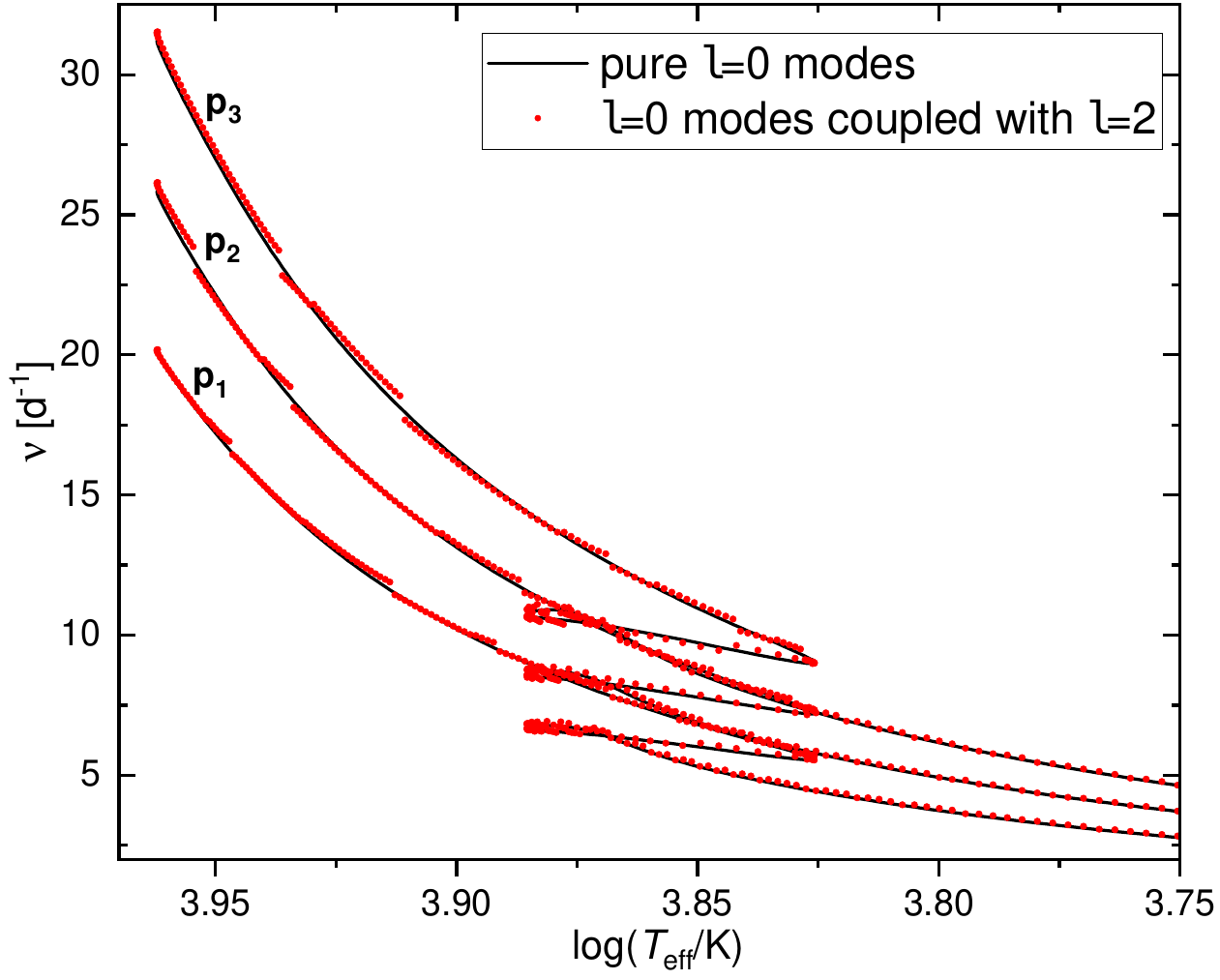}
	\caption{ The evolution of pulsational frequencies of the pure (solid lines) and coupled (dots) radial modes.
		The following modes are shown:  fundamental ($p_1$), first overtone  ($p_2$)
		and secoond overtone ($p_3$).  The same model as in Fig.\,\ref{freq_evol} was considered. }
	\label{freq_evol_coup}
\end{figure}

\begin{figure}
	\includegraphics[width=\columnwidth,clip]{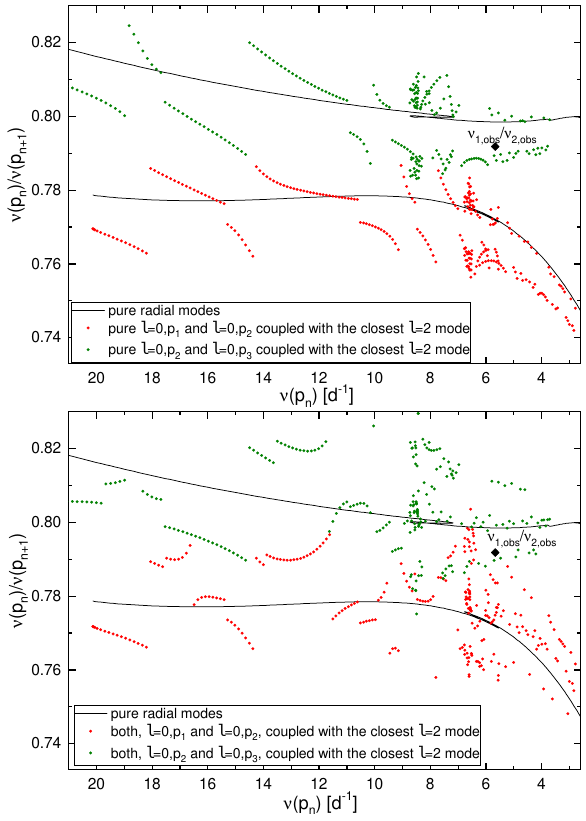}
	\caption{The Petersen diagrams for the two pairs of radial modes: fundamental and first overtone (lower solid lines), first and second
		overtone (upper solid lines). The model parameters are the same as in the case of Fig.\,\ref{freq_evol} with $V_{\rm rot,0} =120~ \kms$.
	%, i.e., $M=2.1$\,M$_\odot$, $X_0=0.70$, $Z=0.02$,  $\alpha_{\rm MLT}=0.5$, $\alpha_{\rm ov}=0.2$ and $V_{\rm rot,0} =120~ \kms$.
		The top panel shows the case if one radial mode (a higher overtone) is coupled by rotation with the nearest $\ell=2$ mode, whereas 
		in the lower panel both radial modes are coupled.}
	\label{Petersen_coup}
\end{figure}

Using the extended version of W.\,Dziembowski code \citep[e.g.,][]{JDD2002}, which relies on the third order formalism of \citet{Soufi1998}, 
we computed eigenfrequencies of radial modes coupled with the nearest quadruple mode.
In Fig.\,\ref{freq_evol_coup}, we show the effect of rotational mode coupling for the first three radial modes. 
The frequency evolution of pure radial modes and those coupled with the nearest $\ell=2$ mode was compared.
The same model as before is considered with the initial rotation $V_{\rm rot,0} =120~ \kms$. 
In the case of the MS phase, it is clearly visible that the frequencies of coupled radial modes follow 
the avoided crossing phenomenon of the $\ell=2$ modes.
In the HSB phase, the frequencies of coupled radial modes are systematically higher. This comparison suggests rather a small effect
of the $\ell=0\&2$ rotational coupling on the frequencies of radial modes. However, these small differences produce a huge effect on the Petersen 
diagrams presented in Fig.\,\ref{Petersen_coup}.
The top panel shows the effect of rotational mode coupling if only one radial mode, i.e., the higher overtone, is rotationally coupled with
the nearest quadrupole mode. Solid lines represent the frequency ratios of pure radial modes; the lower 
line corresponds to $\nu(p_1)/\nu(p_2)$ vs $\nu(p_1)$ and the upper line corresponds to $\nu(p_2)/\nu(p_3)$ vs $\nu(p_2)$.
Red dots corresponds to the frequency ratio $\nu(p_1)/\nu(p_2)$  if the first overtone is coupled and should be 
compared with the lower solid line. Green dots corresponds to the frequency ratio $\nu(p_2)/\nu(p_3)$  if  the second overtone
is coupled and should be compared with the upper solid line.
In the bottom panel of Fig.\,\ref{Petersen_coup}, we show a similar comparison but if both radial modes are rotational coupled.
As one can see, the second frequency  $\nu_2$ can be easily associated with a radial mode if the effect of rotational mode coupling 
is taken into account.


\section{Seismic modelling of V2367 Cyg}
	%with the Monte Carlo-based Bayesian analysis }

Using the Baysian analysis based on Monte Carlo  simulations, we made an extensive seismic modelling of V2367 Cyg assuming 
that $\nu_1$ and $\nu_2$ are both radial modes and %taking into account the effect of rotation according to the formula given in Eg.\,(8) 
including the rotational mode coupling  between the radial and quadrupole modes. 
Because of the very unique identification of $\ell=0$ for the dominant mode $\nu_1$ ({\it cf}. the left panel of Fig.\,\ref{ModeID}) 
and its high amplitude  (about 10 times larger than the amplitude of $\nu_2$), we made the assumption that only the frequency $\nu_2$ 
is coupled with the nearest $\ell=2$ mode. The same approach %, but without a justification, 
was adopted by \citet{Balona2012}, who made the first attempt to explain $\nu_2$ as radial mode through the rotational mode coupling.
They provided a sample model which had the radial fundamental mode with the frequency $\nu_1$ and the first overtone 
with the frequency $7.138$\,d$^{-1}$, which is $0.01$ lower  than the observed  value of $\nu_2=7.14898$\,d$^{-1}$ .  
%When the pulsational frequencies of the radial and quadrupole modes are close, i.e., their difference is smaller than the rotational frequency, 
%a phenomenon of rotational mode coupling occurs. As a result these two modes repel each other in the frequency values.

It is worth to mention that in the oscillation spectrum of V2367\,Cyg, we detected two close-frequency triplets that are nearly equidistant.
According to the analysis of the Kepler data these are $(\nu_3,~\nu_5,~\nu_8)=(7.77557,~7.85334,~7.68347)$\,d$^{-1}$ and
$(\nu_4,~\nu_7,~\nu_{11})= (9.81587,~10.02421,~10.22938)$\,d$^{-1}$. 
%We did not get the frequency $10.22938$\,d$^{-1}$ from the TESS light curve. Instead, we found the frequency $10.122796$d$^{-1}$. 
Comparing with pulsational models, these two sequences could be consecutive overtones 
of $\ell=2$ and $\ell=1$ modes, respectively.
%according to the theoretical frequencies of the HSB models. 
However, without unique identification of the angular numbers $(\ell,~m)$ of these modes, we cannot include them in seismic modelling. 
Such modes would provide a very strong selection of parameters and evolutionary stage.

\subsection{Frequency uncertainty from rotational mode coupling}
When two modes are coupled by rotation then the distance between their frequencies increases;  in other words, the modes repel each other.
This distance becomes larger the closer the frequencies of the two modes are before coupling and the higher the rotational velocity.
As a consequence, the frequency of a radial mode can significantly differ from the value given by Eq.\,(14), hereafter $\nu_{\rm Eq14}$.  
Here, we assumed that for the closest approach of $\ell=0$ and $\ell=2$ modes, 
the average difference $\overline{\Delta\nu}(\ell=0)=\nu_{\rm coup} - \nu_{\rm Eq14}$ 
can be approximated by a linear function of the rotational velocity, i.e, $\overline{\Delta\nu}=a\cdot V_{\rm rot}$.

\begin{figure}
	\includegraphics[width=\columnwidth,clip]{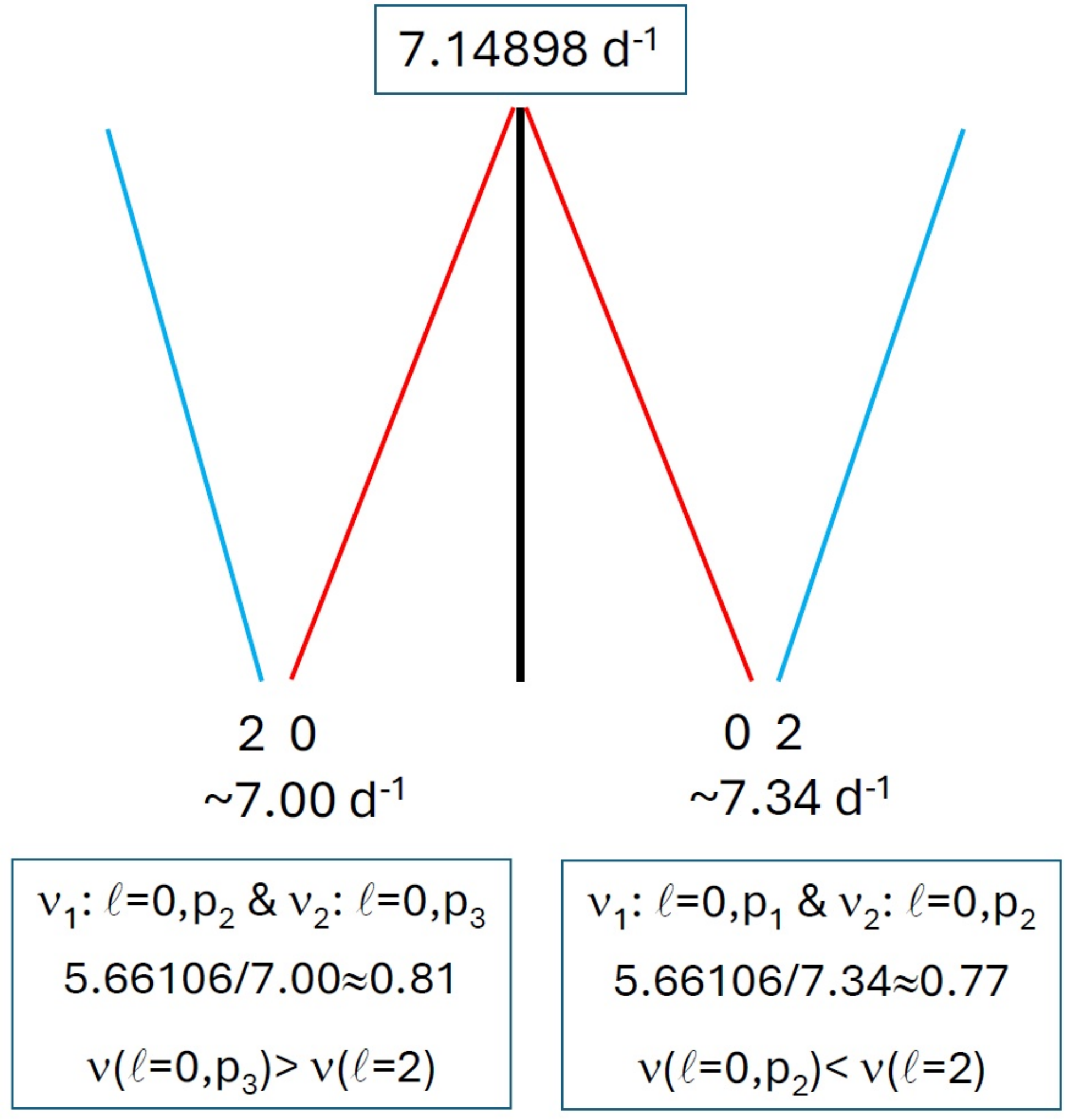}
	\caption{A schematic sketch showing the repulsion of modes $\ell=0$ and $\ell=2$ due to rotational coupling. %illustration
		The observed frequency $\nu_2$ is given in the upper box. To the right of a vertical black line,  a hypothesis of the radial 
		fundamental mode and first overtone ($p_1~\&~p_2$) is presented. To the left - a hypothesis of the first and second overtone.}
	\label{sketch_coup}
\end{figure}

In case of the frequencies $\nu_1$ and $\nu_2$, we have to consider two scenarios: 1) they correspond to the radial fundamental and first overtone mode, respectively, (the $p_1~\&~p_2$ hypothesis),  2) they correspond to the radial first and second overtone mode, respectively (the $p_2~\&~p_3$ hypothesis). A schematic sketch  of these two scenarios is presented in Fig.\,\ref{sketch_coup}.
Depending on the scenario a coefficient $a$ can be positive (first case) or negative (second case)  to reach the observed frequency $\nu_2$.
We limited our calculations up to the rotational velocity equal to half the critical value, i.e, to $V_{\rm rot,0}\approx 170~\kms$.

In case of the $p_1~\&~p_2$ hypothesis, the theoretical frequency of the first overtone has to be greater than the observed value
 $\nu_2=7.14808$\,d$^{-1}$ to give the frequency ratio $\nu(p_1)/\nu(p_2)$ of about 0.77-0.78 as in the case of non-coupling. 
Moreover, $\nu(p_2)$ has to be less than the nearest $\ell=2$ mode to be repelled in the right direction, i.e., towards $\nu_2$.
In this case, we got the allowed range  $\nu(p_2)\in (7.14898,~\nu_{\rm max})$~d$^{-1}$,   where 
$\nu_{\rm max}=+0.0011236\cdot V_{\rm rot}+7.14898$.

If the $p_2~\&~p_3$ hypothesis is considered, then the theoretical frequency of the second overtone has to be less than the observed value 
$\nu_2$ to give the frequency ratio $\nu(p_2)/\nu(p_3)$ of about 0.80-0.81 as in non-coupling case.
In turn, $\nu(p_3)$ has to be greater than the nearest $\ell=2$ mode to be repelled towards the observed value $\nu_2$.
Then, the allowed range for the coupled second overtone was $\nu(p_3)\in(\nu_{\rm min}, ~7.14898)$\,d$^{-1}$, where 
$\nu_{\rm min}=-0.0008765\cdot V_{\rm rot}+7.14898$.

\subsection{Seismic models for the two hypotheses}
In our seismic modelling, we fitted the dominant frequency $\nu_1$, its empirical value of the non-adiabatic complex parameter $f=(f_R,~f_I)$ 
and the frequency $\nu_2$ allowing its uncertainty due to rotational coupling within the above estimated ranges.
Including the parameter $f$ into seismic modelling  is crucial to get constraints on the efficiency of convection in the outer layers,
described by the mixing length parameter $\alpha_{\rm MLT}$ \citep{JDD2003}.
An additional constraint was the consistency of effective  temperatures and luminosities of the models with the observed values.

Thus, we had six observables ($\nu_1, ~\nu_2, ~f_R(\nu_1), f_I(\nu_1),~T_{\rm eff}, ~L$)  and one could expect that at least five parameters 
could be derived from our seismic modelling based on MC simulations. However, because of a large "uncertainty"  in the second frequency 
due to allowing for the rotational mode coupling with the $\ell=2$ mode, we fixed the initial hydrogen abundance at $X_0=0.70$ 
and the metallicity at three values  $Z=0.014, 0.020, 0.030$ from the observed range $Z\in [0.0134,~0.0356]$. 
The parameters to be determined were $(M,~V_{\rm rot}, ~\alpha_{\rm ov},~\alpha_{\rm MLT})$.
Guided by our recent results from seismic modelling of double-mode radial HADS stars \citep{JDD2023} that
only with the OPAL opacity tables could we match all observables,
we performed all computations  with these opacity data. \\
%\subsection{Hypothesis of the fundamental and first overtone modes}
%\subsection{The $p_1\&p_2$ hypothesis}

{\it The $p_2\&p_3$ hypothesis}\\
In the first step, we considered that $\nu_1$ is the first overtone  and $\nu_2$ is the second overtone.  We searched for models 
which reproduce the observables  ($\nu_1,~f(\nu_1),~T_{\rm eff}, ~L$) within the 1-$\sigma$ error and for the frequency $\nu_2$,
we accepted all values from the range $(\nu_{\rm min}, ~7.14898)$\,d$^{-1}$, as described in the previous subsection.
We found seismic models of V2367\,Cyg which reproduce our observables in the HSB and OC phase, for $Z=0.020$ and $Z=0.030$
in each phase.
The seismic models computed with $Z=0.014$ have too low effective temperature and luminosity.   
In the case of MS models, the first overtone mode never reaches the observed value of the dominant frequency $\nu_1=5.66106$\,d$^{-1}$.
Both radial modes are excited in all seismic models.
In Table\,5, we give the median values for the derived parameters  $(M,~V_{\rm rot},~\alpha_{\rm ov},~\alpha_{\rm MLT})$
and for the three considered metallicites $Z$.  For a comparison, we gave also the initial values of rotation $V_{\rm rot,0}$.
The seismic models for $Z=0.014$ were added in Table \,5 to demonstrate how their ($T_{\rm eff},~L$) are far from the observed 
values, i.e., from $\log (T_{\rm eff}/{\rm K})=3.8495 \pm 0.0246$. and  $\log L/{\rm L}_\odot = 1.5407 \pm 0.1138$. 
In the last two columns, we put the empirical values of the intrinsic amplitude of the dominant mode $|\varepsilon|(\nu_1)$, 
as defined by Eq.\,6, and the difference $\Delta\nu_2$ between theoretical and observed values of $\nu_2$.
\begin{table*}
	\centering
	\caption{The median of parameters of seismic models of V2367 Cyg  from MC simulations assuming the $p_2\&p_3$ hypothesis. The models were computed with the OPAL opacity tables, initial hydrogen abundance $X_0=0.70$ and the three values of metallicity $Z=0.014, 0.020, 0.030$. Two phases of evolution are possible: HSB and OC. The uncertainties were calculated from quantiles  0.84 and 0.16. The models with $Z=0.014$ are outside the error box on the HR diagram and their parameters are given for a comparison.}
	\begin{tabular}{cccccccccc }
		\hline
		& & && HSB phase && & &  \\
       \hline
 $Z$ (fixed) & $\log (T_{\rm eff}/{\rm K})$ & $\log L/$L$_\odot$ & $M$ & $\alpha_{\rm ov}$ & $\alpha_{\rm MLT}$ & $V_{\rm rot,0}$ & $V_{\rm rot}$ & $|\varepsilon|(\nu_1)$ & $\Delta\nu_2$ \\
	                   &                                          &                               &[M$_\odot$] &            &                     &  [km$\cdot$s$^{-1}$] & [km$\cdot$s$^{-1}$] &  & [d$^{-1}$]\\
		\hline

$0.014$  &$3.7815^{+0.0030}_{-0.0054}$     & $1.352^{+0.009}_{-0.032}$   & $1.770^{+0.051}_{-0.024}$     &  $0.16^{+0.12}_{-0.13}$ &  $0.08^{+0.06}_{-0.04}$     & $140^{+24}_{-17}$     &  $109^{+21}_{-11}$ &  $0.0120^{+0.0006}_{-0.0003}$     & $-0.092^{+0.008}_{-0.019}$   \\
&&&&&&&&&\\

$0.020$ & $3.8275^{+0.0019}_{-0.0018}$     & $1.588^{+0.007}_{-0.008}$     & $2.151^{+0.039}_{-0.035}$     &  $0.15^{+0.13}_{-0.1}$ &  $0.08^{+0.05}_{-0.05}$     & $135^{+42}_{-45}$     &  $107^{+38}_{-37}$ &  $0.0126^{+0.0003}_{-0.0002}$     & $-0.063^{+0.014}_{-0.024}$   \\
&&&&&&&&&\\

$0.030$ & $3.8324^{+0.0066}_{-0.0077}$     & $1.636^{+0.032}_{-0.037}$     &  $2.370^{+0.048}_{-0.051}$ &  $0.15^{+0.12}_{-0.10}$     &  $0.20^{+0.11}_{-0.13}$ &  $114^{+48}_{-48}$     & $91^{+42}_{-39}$     & $0.0122^{+0.0003}_{-0.0005}$     & $-0.040^{+0.008}_{-0.016}$    \\

	  	\hline	  	
	  	
		& &  &&OC phase && & & &  \\
\hline
$Z$ (fixed) & $\log (T_{\rm eff}/{\rm K})$ & $\log L/$L$_\odot$ & $M$ & $\alpha_{\rm ov}$ & $\alpha_{\rm MLT}$ & $V_{\rm rot,0}$ & $V_{\rm rot}$ & $|\varepsilon|(\nu_1)$ & $\Delta\nu_2$ \\
                      &                                             &                               &[M$_\odot$] &              &       &    [km$\cdot$s$^{-1}$]   &  [km$\cdot$s$^{-1}$] &  & [d$^{-1}$] \\
\hline

$0.014$  &  $3.7824^{+0.0021}_{-0.0079}$ &  $1.357^{+0.009}_{-0.045}$     & $1.807^{+0.030}_{-0.085}$     &  $0.41^{+0.07}_{-0.02}$ &  $0.11^{+0.02}_{-0.05}$     & $151^{+26}_{-14}$     &  $119^{+13}_{-15}$ &  $0.0119^{+0.0008}_{-0.0001}$     & $-0.099^{+0.009}_{-0.013}$   \\
&&&&&&&&&\\

$0.020$  &  $3.8274^{+0.0018}_{-0.0017}$     & $1.590^{+0.009}_{-0.010}$     & $2.168^{+0.057}_{-0.068}$     &  $0.34^{+0.06}_{-0.05}$ &  $0.07^{+0.05}_{-0.05}$     & $141^{+42}_{-52}$     &  $105^{+35}_{-39}$ &  $0.0126^{+0.0002}_{-0.0002}$     & $-0.063^{+0.015}_{-0.021}$    \\
&&&&&&&&&\\

$0.030$  &  $3.8326^{+0.0064}_{-0.0078}$ &  $1.640^{+0.031}_{-0.037}$     & $2.398^{+0.082}_{-0.098}$     &  $0.28^{+0.07}_{-0.05}$ &  $0.20^{+0.12}_{-0.13}$     & $120^{+52.9}_{-49.3}$     &  $91^{+40}_{-38}$ &  $0.0122^{+0.0003}_{-0.0005}$     & $-0.042^{+0.011}_{-0.015}$   \\

    \hline
	\end{tabular}
\end{table*}
The median is more informative statistic for non-Gaussian distributions or distributions with outliers, which is the case  for some parameters. 
The histogram for $(M,~V_{\rm rot},~\alpha_{\rm ov},~\alpha_{\rm MLT})$ of the seismic models computed for $Z=0.02$ 
and 0.03 are shown in the Appendix\,B. 
Uncertainties of parameters in Table\,5 were calculated as the 0.84 quantile minus the median and the median minus 
the 0.16 quantile. These quantiles  correspond to one standard deviation from the mean value in the case of a normal distribution.

As one can see, the values of $(\log T_{\rm eff}, \log L/$L$_\odot,~M)$ are, within the errors, very close in both evolutionary
phases. As a consequence for the models with the same $Z$, we also got the close empirical values of the mixing length parameter $\alpha_{\rm MLT}$, i.e., $\alpha_{\rm MLT}\approx 0.07$ for $Z=0.020$ and $\alpha_{\rm MLT}\approx 0.20$ for $Z=0.030$. 
It means that convection in the envelope of V2367\,Cyg is hardly efficient. 
In turn, significant differences are in the overshooting parameter $\alpha_{\rm ov}$, which amounts to $\alpha_{\rm ov}\approx 0.15$
for the HSB seismic models and to $\alpha_{\rm ov}\approx 0.30$ for the OC models, if $Z=0.020$ and $Z=0.030$ are considered. 
The rotational velocity is similar in both phases of evolution with $V_{\rm rot}\approx 106~\kms$ for the models with the metallicity $Z=0.020$ 
and $V_{\rm rot}\approx 91~\kms$ for models with $Z=0.030$. However, we would like to point out that the uncertainties
of $V_{\rm rot}$ are of about 40\,$\kms$. 
\begin{figure}
	\includegraphics[width=\columnwidth,clip]{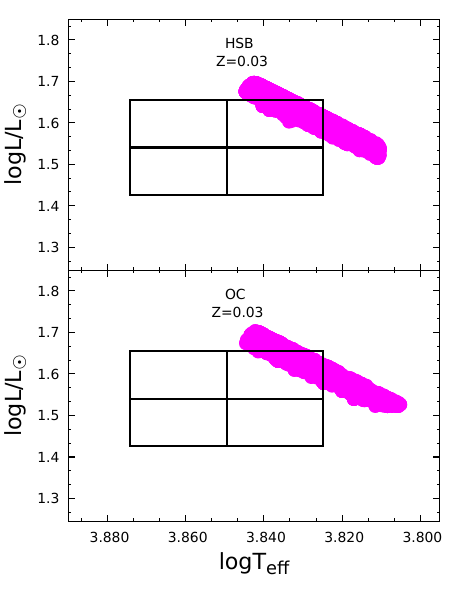}
	\caption{The HR diagrams with the position of the HSB (top panel) and OC (bottom panel) seismic models of V2367 Cyg computed 
		with the MC-based Bayesian analysis assuming the $p_2\&p_3$ hypothesis. The OPAL table, initial hydrogen abundance $X_0=0.70$ 
		and $Z=0.030$ were adopted.}
	\label{HR_p2_Z030}
\end{figure}
The positions of seismic models for $Z=0.030$ in the HR diagram along with the error box are shown in Fig.\,\ref{HR_p2_Z030}.
The upper panel contains the seismic models in the HSB phase and the lower panel -- in the OC phase.\\

\begin{figure}
	\includegraphics[width=\columnwidth,clip]{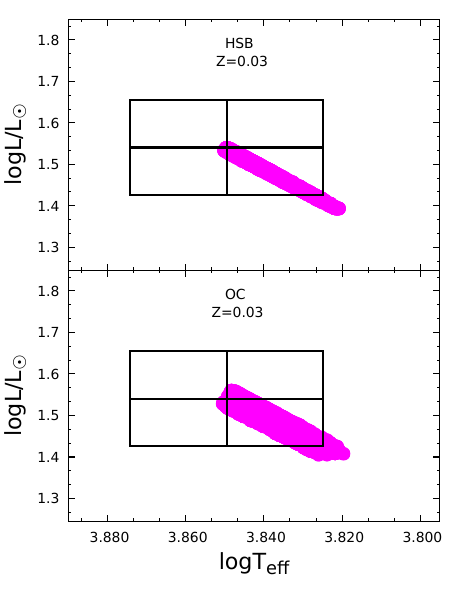}
	\caption{The HR diagrams with the position of the HSB (top panel) and OC (bottom panel) seismic models of V2367 Cyg computed 
		with the MC-based Bayesian analysis assuming the $p_1\&p_2$ hypothesis. The OPAL table, initial hydrogen abundance $X_0=0.70$ 
		and $Z=0.030$ were adopted.}
	\label{HR_p1_Z030}
\end{figure}

{\it The $p_1\&p_2$ hypothesis}\\
Next, we made seismic modelling assuming  that $\nu_1$ is the radial fundamental mode and $\nu_2$ is the radial first overtone.
All requirements to be fulfilled are the same as above except that for the frequency $\nu_2$, we accepted all values from the range
$(7.14898, \nu_{\rm max})$\,d$^{-1}$. In this case, the number of solutions is much smaller. 
First of all, only the seismic models with $Z=0.030$  reproduce the empirical values of $f$ (both, the real and imaginary part)
within the 1-$\sigma$ error. Such seismic models were found in the OC and HSB evolutionary stages. 
Their parameters are given in  Table\,6 and their position on the HR diagram is shown in Fig.\,\ref{HR_p1_Z030}.
As before, the considered radial modes are unstable. 
\begin{table*}
	\centering
	\caption{The median of parameters of seismic models of V2367 Cyg from MC simulations assuming the $p_1\&p_2$ hypothesis. Two phases of evolution are possible : HSB and OC.
		Only models computed with $Z=0.030$ meet  all the requirements described in the text.}
	\begin{tabular}{ccccccccccc }
		\hline
%		& &  $Z=0.03$ (fixed)&& & & 
		\multicolumn{9}{c}{$Z=0.03$ (fixed)} & \\
		\hline
 phase of & $\log (T_{\rm eff}/{\rm K})$ & $\log L/$L$_\odot$ &  $M$ &  $\alpha_{\rm ov}$ & $\alpha_{\rm MLT}$ & $V_{\rm rot,0}$ & $V_{\rm rot}$ & $|\varepsilon|(\nu_1)$ & $\Delta\nu_2$\\
evolution &                                       	               &                               &  [M$_\odot$]  &                     &   &  [km$\cdot$s$^{-1}$]    &  [km$\cdot$s$^{-1}$] &  &  [d$^{-1}$] \\
		\hline
HSB   &   $3.8420^{+0.0048}_{-0.0061}$     & $1.498^{+0.024}_{-0.031}$     &  $2.237^{+0.030}_{-0.038}$ &  $0.08^{+0.06}_{-0.05}$     & $0.19^{+0.12}_{-0.13}$     &  $182^{+12}_{-11}$ &  $162^{+12}_{-11}$     & $0.0124^{+0.0006}_{-0.0002}$     & $0.161^{+0.006}_{-0.006}$ \\ 
  &&&&&&&&&\\			
OC &  $3.8426^{+0.0041}_{-0.0059}$     & $1.513^{+0.023}_{-0.033}$     &  $2.31^{+0.087}_{-0.120}$ &  $0.09^{+0.06}_{-0.05}$     & $0.19^{+0.12}_{-0.12}$     &  $185^{+11}_{-10}$ &  $152^{+13}_{-11}$     & $0.0124^{+0.0005}_{-0.0003}$     & $0.152^{+0.016}_{-0.011}$  \\  			  			
		\hline
	\end{tabular}
\end{table*}

In case of the $p_1\&p_2$ hypothesis, the seismic models have much higher rotational velocity, with a median of about 160\,$\kms$ 
in the HSB phase and about 150\,$\kms$ in the OC phase. The overshooting parameter $\alpha_{\rm ov}$ is lower than in case of the $p_2\&p_3$ hypothesis, i.e., $\alpha_{\rm ov}\approx 0.1$. The mixing length parameter $\alpha_{\rm MLT}$ has a similar value of about 0.2.

Independent of the adopted hypothesis, the intrinsic mode amplitudes $|\varepsilon|$ of the dominant frequency $\nu_1$ 
is about 0.012. %In this paper, $|\varepsilon|$ is defined as the rms value of $\delta r/R$ over the surface.

\section{Summary}
Asteroseismology  of moderate and fast rotators is a challenging task. Rotation affects equilibrium stellar models,
their pulsation as well as mode identification based on the photometric amplitudes and phases. We cannot also expect 
equidistant patters resulting from the rotational splitting. 
Besides the main effects of rotation, i.e., centrifugal and Coriolis forces, we have also to take into account a possibility of mode coupling
even if the rotation is not too fast.

In this paper, we presented a comprehensive analysis of the HADS star V2367\,Cyg using all available observational data.
We analysed the whole space photometry of V2367\,Cyg, both from {\it Kepler} and {\it TESS} mission.
We confirmed all frequencies found by \citet{Balona2012},  including the three dominant ones,
and found much more. %No regularities were found in the oscillation spectrum, e.g., suggesting rotational splitting.

Then, we performed identification of the mode degree $\ell$ from the photometric amplitudes and phases using data of \citet{Ulusoy2013} for the three main
frequencies.We got an unambiguous identification for the dominant frequency $\nu_1$, i.e., the radial mode.
Determination of $\ell$ was less successful for $\nu_2$ and $\nu_3$ but, because of the visibility effect, the most probable degrees were $\ell=0$ or 
$\ell=2$. From a comparison of the theoretical values of frequency ratios, we concluded that only $\nu_2$ can be a radial mode.
but only if the higher order effects of rotation, including mode coupling, are taken into account. It has already  been
 suggested by \citet{Balona2012} who initiated seismic study of V2367\,Cyg but in a limited space of parameters. 
 Here we performed the first extensive seismic modelling that takes into account mode coupling by rotation.
 
Assuming that $\nu_1$ and $\nu_2$ are both radial modes and including the non-adiabatic parameter $f$ of the dominant frequency,
we made extended seismic modelling using the Bayesian analysis based on MC simulations.  About one million  models  were computed in total.
Two hypotheses had to be considered: 1) $\nu_1$ and $\nu_2$ correspond to the radial first and second overtone, respectively ( the $p_2\&p_3$ hypothesis)
and 2) $\nu_1$ and $\nu_2$ correspond to the radial fundamental and first overtone, respectively ( the $p_1\&p_2$ hypothesis).
All models which met our requirements are in the post-MS phase of evolution, either the hydrogen-shell burning (HSB) or overall contraction (OC),
regardless of the assumed hypothesis. Besides, all have a moderate or quite fast rotation to account for the frequency ratio of the two radial modes.
%The radial fundamental mode in the MS models, never reaches th observed values of $\nu_1$.
%If, both $\nu_1$ and $\nu_2$ are radial modes that V2367\,Cyg must be a fast rotator to account for their frequency ratio,
%regardless of whether we accept the $p_2\&p_3$ hypothesis or the $p_1\&p_2$ hypothesis.
The rotational velocity is much higher for the seismic models obtained with the $p_1\& p_2$ hypothesis, with the median value of about $150 - 160~\kms$.
If the $p_2\& p_3$ hypothesis is adopted then the median of $V_{\rm rot}$ amounts to $90 - 110~\kms$.%, depending on the evolutionary phase and metallicity.

Independent of the adopted hypothesis, from our seismic analysis we obtained small values of the mixing length parameter in the total range of
$\alpha_{\rm MLT}\in(0.01,~0.33)$. Thus, convective transport of energy in the outer layers of V2367\,Cyg is hardly efficient
or even completely inefficient.

In the case of the $p_2\& p_3$ hypothesis, the value of the overshooting parameter depends on the evolutionary phase.
We obtained $\alpha_{\rm ov}\in (0.02,~0.29)$ for the HSB models  and $\alpha_{\rm ov}\in (0.23,~0.40)$ for the OC models.
Adopting the $p_1\& p_2$ hypothesis, we got similar values of $\alpha_{\rm ov}$ for both evolutionary stages (HSB and OC), 
which amount to about $\alpha_{\rm ov}\in (0.03,~0.15)$.

Our seismic analysis of V2367\,Cyg clearly indicates that the star is in a post-MS phase of evolution, either in HSB or in OC. However, it is difficult to say which stage is more preferred and which hypothesis for two radial modes, $p_2\&p_3$ or $p_1\&p_2$, is more likely.
% In case of the $p_2\&p_3$ hypothesis, we obtained a much larger number of  models that meet our requirements. But on this basis, of course, the $p_1\&p_2$ hypothesis cannot be rejected.
Perhaps, new time-series multi-colour photometry and spectroscopy would allow for the unambiguous mode identification 
of a larger number of frequencies. Then, the inclusion of additional modes in seismic modelling would provide additional selection.
%and probably also allow resolving between the HSB and OC phases.
For example, one could try to incorporate the frequency $\nu_3$ as a quadrupole mode. %provided the azimuthal order $m$ is known. 
However, to make this possible, it would first be necessary to identify its azimuthal order $m$.

\section*{Acknowledgements}
This work has made use of data from the European Space Agency
(ESA) mission Gaia (https://www.cosmos.esa.int/gaia), processed
by the Gaia Data Processing and Analysis Consortium (DPAC; https:
//www.cosmos.esa.int/web/gaia/dpac/consortium). Funding for the
DPAC has been provided by national institutions, in particular the
institutions participating in the Gaia Multilateral Agreement.

This paper includes data collected by the TESS mission. Funding
for the TESS mission is provided by the NASA Explorer Program.

This paper includes data collected by the Kepler mission and obtained from the MAST
data archive at the Space Telescope Science Institute (STScI). Funding 
for the Kepler mission is provided by the NASA Science Mission
Directorate. STScI is operated by the Association of Universities
for Research in Astronomy, Inc., under NASA contract NAS 5-26555. 

Calculations were carried out using resources provided by Wroclaw Centre for Networking and
Supercomputing (http://wcss.pl), grant no. 265.

\section*{Data Availability}
The Kepler and TESS data are available from the NASA MAST portal  https://archive.stsci.edu/.
Theoretical computations will be shared on reasonable request to the corresponding author.

\bibliographystyle{mnras}
\bibliography{JDD_biblio2} % if your bibtex file is called example.bib

\appendix

\section{Frequencies from the Fourier analysis of the Kepler and TESS light curves}
%Here we list all detected significant frequencies extracted from the Kepler and TESS photometry of V2367 Cygni, i.e., those with $S/N>5$.
%\section{Full lists of significant freqencies}
\label{app1}

\clearpage
\onecolumn

\begin{longtable}{  c c c c c  c  }
	\caption{All frequencies detected in {\it Kepler} SC data with $S/N>5$. %They are listed in the order they appeared during pre-whitening. 
		Columns contain: number, our ID$_K$, frequency, amplitude, phase and  signal to noise ratio.}\\
	\hline
	\# & ID$_K$  &   $\nu~[\mathrm{d}^{-1}]$   &     $A$ [ppt]   &   $\phi$ [$0-1$] & $S/N$ \\
	\hline
	\endfirsthead
	\caption[]{continued}
	\label{tab:Kepler_SC_ALL_freq}\\
	%	\begin{tabular}{  c c c c c  c  } % four columns, alignment for each
		\hline
		\# & ID$_K$  &   $\nu~[\mathrm{d}^{-1}]$   &     $A$ [ppt]   &   $\phi$ [$0-1$] & $S/N$ \\
		\hline
		\endhead
		1  &          $\nu_{1}$  &        5.661058(2)  &         156.998(2)  &        0.872513(3)  &        75.7   \\
		2  &  $2\nu_{1}$  &       11.322118(3)  &          50.863(2)  &        0.650089(8)  &        76.0   \\
		3  &          $\nu_{2}$  &        7.148954(7)  &          17.050(2)  &         0.99924(2)  &        72.6   \\
		4  &  $3\nu_{1}$  &       16.983178(5)  &          16.695(2)  &         0.44384(2)  &        74.6   \\
		5  &          $\nu_{3}$  &        7.775571(9)  &           8.922(2)  &         0.95486(5)  &        67.6   \\
		6  &  $1\nu_{1}+1\nu_{2}$  &       12.810015(8)  &           8.300(2)  &         0.71374(5)  &        71.4   \\
		7  &  $4\nu_{1}$  &       22.644241(7)  &           8.227(2)  &         0.19484(5)  &        73.0   \\
		8  &  $-1\nu_{1}+1\nu_{2}$  &        1.487893(9)  &           5.715(2)  &          0.4046(1)  &        70.2   \\
\end{longtable}

\begin{longtable}{  c c c c c  c  }
	\caption{All frequencies detected in six combined {\it TESS} sectors with $S/N>5$. Columns contain: number, our ID$_T$, frequency, amplitude, phase, signal to noise ratio.}\\
	\hline
	\# & ID$_T$  &   $\nu~[\mathrm{d}^{-1}]$   &     $A$ [ppt]   &   $\phi$ [$0-1$] & $S/N$ \\
	\hline
	\endfirsthead
	\caption[]{continued}
	\label{tab:TESS_ALL_all_sec_fre}\\
	%	\begin{tabular}{  c c c c c  c  } % four columns, alignment for each
		\hline
		\# & ID$_T$  &   $\nu~[\mathrm{d}^{-1}]$   &     $A$ [ppt]   &   $\phi$ [$0-1$] & $S/N$ \\
		\hline
		\endhead
		1  &          $\nu_{1}$  &       5.6610574(7)  &          100.24(1)  &         0.86115(2)  &        31.7   \\
		2  &  $2\nu_{1}$  &       11.322115(1)  &           31.99(1)  &         0.64295(6)  &        31.6   \\
		3  &          $\nu_{2}$  &        7.148978(3)  &           11.76(1)  &          0.9321(2)  &        31.3   \\
		4  &  $3\nu_{1}$  &       16.983173(2)  &           10.14(1)  &          0.4398(2)  &        31.5   \\
		5  &  $1\nu_{1}+1\nu_{2}$  &       12.810036(4)  &            5.42(1)  &          0.6431(3)  &        30.3   \\
		6  &  $4\nu_{1}$  &       22.644231(4)  &            4.99(1)  &          0.1819(4)  &        30.5   \\
		7  &          $\nu_{3}$  &        7.775571(3)  &            4.71(1)  &          0.9577(4)  &        28.0   \\
		8  &  $-1\nu_{1}+1\nu_{2}$  &        1.487924(4)  &            3.63(1)  &          0.3245(5)  &        27.7   \\
\end{longtable}

\begin{table}
	\centering
	\caption{Low frequencies detected in three data sets, i.e., {\it Kepler} LC, {\it Kepler} SC, and {\it TESS}.}
	\label{tab:freq_low}
	\begin{tabular}{  c c c } % four columns, alignment for each
		\hline
		Data set  &   ID   & $\nu~[\mathrm{d}^{-1}]$ \\
		\hline
		{\it Kepler} LC & $\nu_{a,K,LC}$ &  0.10449(4) \\
		{\it Kepler} LC & $\nu_{b,K,LC}$ &  0.00967(3) \\
		{\it Kepler} LC & $\nu_{c,K,LC}$ &  0.09778(5) \\
		{\it Kepler} LC & $\nu_{d,K,LC}$ &  0.02055(5) \\
		{\it Kepler} LC & $\nu_{e,K,LC}$ &  0.03103(5) \\
		{\it Kepler} SC & $\nu_{a,K,SC}$ &  0.00255(3) \\
		{\it Kepler} SC & $\nu_{b,K,SC}$ &  0.02188(4)  \\
		{\it Kepler} SC & $\nu_{c,K,SC}$ &  0.01577(4)  \\
		{\it Kepler} SC & $\nu_{d,K,SC}$ &  0.01019(4)  \\
		{\it Kepler} SC & $\nu_{e,K,SC}$ &  0.03466(5)  \\
		{\it Kepler} SC & $\nu_{f,K,SC}$ &  0.05427(7)  \\
		TESS  &  $\nu_{a,T}$  & 0.067193(8) \\
		TESS  &  $\nu_{b,T}$  & 0.055213(9) \\
		TESS  &  $\nu_{c,T}$  & 0.13889(1)  \\
		\hline
	\end{tabular}
\end{table}
\begin{figure*}
	\includegraphics[angle=270, width=14 cm]{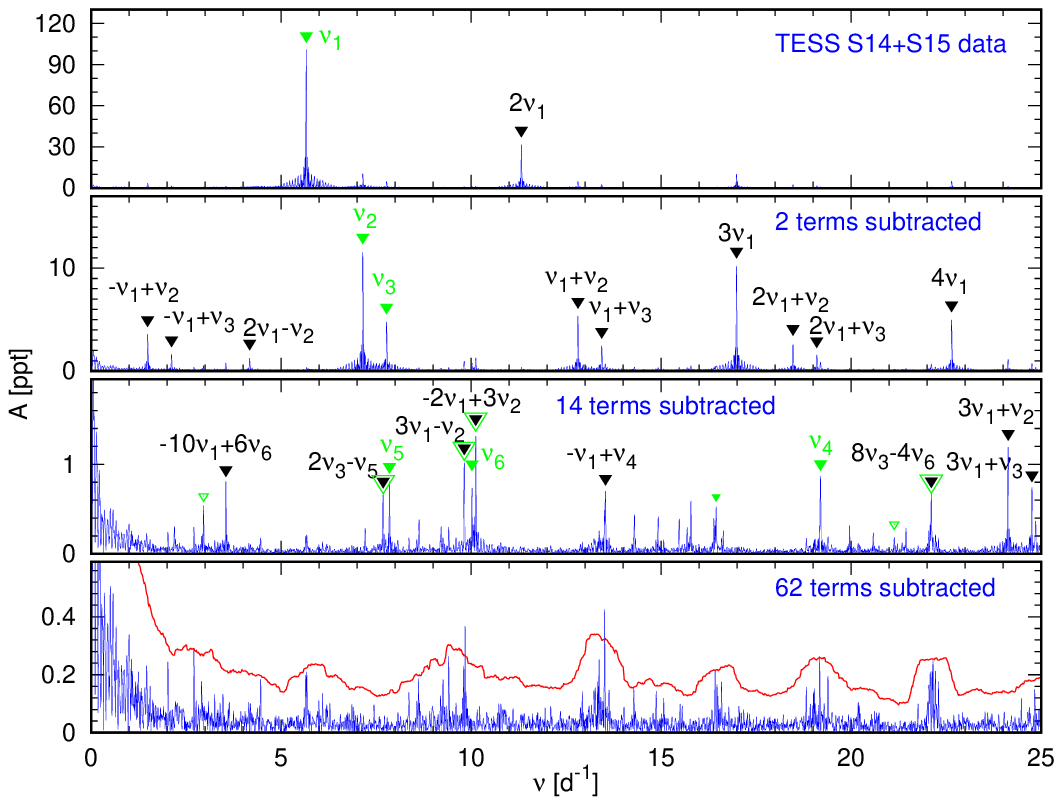}
 	\caption{Amplitude periodogram for the combined S14+S15 {\it TESS} data of V2367 Cyg. Panels from top to bottom show
 		periodograms: for the original S14+S15 data;  after subtraction of 2 terms; after subtraction of 14 terms and
 		after subtraction of 171 terms. Red solid line in bottom panel indicates the $5S/N$ level. All frequency peaks above 0.63\,ppt are marked by down-pointing triangles, green for independent signals, black for combinations or harmonics.
 		Independent frequencies with amplitudes smaller than 0.63\,ppt are marked by small green triangles.	Additionally, big green open triangles mark frequencies that were identified as independent in the full {\it TESS} dataset but combinations in S14+S15 dataset, whereas small green open triangles mean the same but for frequencies with amplitudes below 0.63 ppt.}
 	\label{fig:V2367Cyg_TESS_S14S15_trf}
\end{figure*}

\setlength{\tabcolsep}{6pt}
\begin{table*}
	\centering
	\caption{Independent frequencies detected in the {\it TESS} data from combined adjacent sectors, i.e., S14+S15, S40+S41 and S54+S55.
		Columns contain: our ID$_T$, frequency, amplitude, phase, signal to noise ratio, ID$_K$ and ID$_\mathrm B$.}
	\label{tab:TESS_adjacent_independent_freq}
	\begin{tabular}{  c c c c c c  c  } % four columns, alignment for each
		\hline
		ID$_{T}$  &   $\nu~[\mathrm{d}^{-1}]$   &     $A$ [ppt]   &   $\phi$ [0--1] & $S/N$ & ID$_{K}$ & ID$_{\mathrm B}$\\
		\hline
		\multicolumn{7}{c}{S14+S15}\\
		\hline
		1  &                   5.66105(3)  &          100.97(3)  &         0.88792(4)  &        17.2   &  $\nu_{1,K}$  & $\nu_{1,\mathrm B}$ \\
		2  &                    7.1489(1)  &           11.63(3)  &          0.3888(4)  &        16.4   &  $\nu_{2,K}$  & $\nu_{2,\mathrm B}$\\
		3  &                    7.7753(2)  &            4.82(3)  &          0.9053(9)  &        15.1   &  $\nu_{3,K}$  & $\nu_{3,\mathrm B}$\\
		4  &                   19.1940(6)  &            0.87(3)  &           0.054(5)  &        10.7   &  $9\nu_{4,K}-9\nu_{8,K}$ &\\
		5  &                   10.0244(7)  &            0.80(3)  &           0.024(5)  &        10.3   &  $\nu_{7,K}$   & $\nu_{8,\mathrm B}$\\
		6  &                    16.447(1)  &            0.50(3)  &           0.009(9)  &         7.2   &  $5\nu_{1,K}+1\nu_{3,K}-2\nu_{4,K}$  & $\nu_{13,\mathrm B}$\\
		\hline
		\multicolumn{7}{c}{S40+S41}\\
		\hline
		1  &                   5.66105(3)  &           99.19(2)  &         0.88294(3)  &        17.0 &  $\nu_{1,K}$  & $\nu_{1,\mathrm B}$ \\
		2  &                    7.1490(1)  &           11.55(2)  &          0.7651(3)  &        17.3 &  $\nu_{2,K}$  & $\nu_{2,\mathrm B}$ \\
		3  &                    7.7754(2)  &            4.60(2)  &          0.5743(7)  &        14.9 &  $\nu_{3,K}$  & $\nu_{3,\mathrm B}$   \\
		4  &                   10.0243(5)  &            0.82(2)  &           0.308(4)  &        10.7 &  $\nu_{7,K}$   & $\nu_{8,\mathrm B}$\\
		5  &                    7.8533(5)  &            0.80(2)  &           0.581(4)  &         9.8 &  $\nu_{5,K}$   & $\nu_{6,\mathrm B}$\\
		6  &                   16.4455(8)  &            0.45(2)  &           0.185(7)  &         6.8 &  $5\nu_{1,K}+1\nu_{3,K}-2\nu_{4,K}$  & $\nu_{13,\mathrm B}$\\
		\hline
		\multicolumn{7}{c}{S54+S55}\\
		\hline
		1  &                   5.66106(3)  &          101.55(2)  &         0.83449(3)  &        17.1 &  $\nu_{1,K}$  & $\nu_{1,\mathrm B}$ \\
		2  &                    7.1490(1)  &           11.94(2)  &          0.0485(3)  &        17.0 &  $\nu_{2,K}$  & $\nu_{2,\mathrm B}$ \\
		3  &                    7.7756(2)  &            4.62(2)  &          0.6908(7)  &        15.9 &  $\nu_{3,K}$  & $\nu_{3,\mathrm B}$ \\
		4  &                   10.0239(5)  &            0.94(2)  &           0.183(3)  &        11.8 &  $\nu_{7,K}$   & $\nu_{8,\mathrm B}$\\
		5  &                    7.8534(5)  &            0.78(2)  &           0.008(4)  &         9.0 &  $\nu_{5,K}$   & $\nu_{6,\mathrm B}$\\
		6  &                   16.4468(7)  &            0.54(2)  &           0.473(6)  &         7.2 & $5\nu_{1,K}+1\nu_{3,K}-2\nu_{4,K}$  & $\nu_{13,\mathrm B}$\\
		\hline
	\end{tabular}
\end{table*}

\begin{longtable}{  c c c c c  c  }
	\caption{All frequencies detected in combined {\it TESS} sectors 14 and 15. Columns contain: number, our ID$_T$, frequency, amplitude, phase, signal to noise ratio.}\\
	\hline
	\# & ID$_{T,S14-15}$  &   $\nu~[\mathrm{d}^{-1}]$   &     $A$ [ppt]   &   $\phi$ [$0-1$] & $S/N$ \\
	\hline
	\endfirsthead
	\caption[]{continued}
	\label{tab:TESS_ALL_S14_15_fre}\\
	%	\begin{tabular}{  c c c c c  c  } % four columns, alignment for each
		\hline
		\# & ID$_{T,S14-15}$  &   $\nu~[\mathrm{d}^{-1}]$   &     $A$ [ppt]   &   $\phi$ [$0-1$] & $S/N$ \\
		\hline
		\endhead
		1  &          $\nu_{1}$  &         5.66105(3)  &          100.97(3)  &         0.88792(4)  &        17.2   \\
		2  &  $2\nu_{1}$  &        11.32210(6)  &           32.10(3)  &          0.6850(1)  &        17.1   \\
		3  &          $\nu_{2}$  &          7.1489(1)  &           11.63(3)  &          0.3888(4)  &        16.4   \\
		4  &  $3\nu_{1}$  &         16.9831(1)  &           10.25(3)  &          0.6458(4)  &        16.9   \\
		5  &  $1\nu_{1}+1\nu_{2}$  &         12.8099(2)  &            5.37(3)  &          0.9766(8)  &        16.0   \\
		6  &  $4\nu_{1}$  &         22.6442(2)  &            5.02(3)  &          0.2615(9)  &        16.6   \\
		7  &          $\nu_{3}$  &          7.7753(2)  &            4.82(3)  &          0.9053(9)  &        15.1   \\
		8  &  $-1\nu_{1}+1\nu_{2}$  &          1.4876(2)  &            3.59(3)  &           0.593(1)  &        14.0   \\
		9  &  $2\nu_{1}+1\nu_{2}$  &         18.4710(3)  &            2.74(3)  &           0.664(2)  &        15.3   \\
		10  &  $1\nu_{1}+1\nu_{3}$  &         13.4363(3)  &            2.45(3)  &           0.975(2)  &        12.9   \\
		11  &  $5\nu_{1}$  &         28.3053(3)  &            2.38(3)  &           0.964(2)  &        16.1   \\
		12  &  $-1\nu_{1}+1\nu_{3}$  &          2.1146(4)  &            1.61(3)  &           0.848(3)  &        12.4   \\
		13  &  $2\nu_{1}+1\nu_{3}$  &         19.0972(4)  &            1.54(3)  &           0.203(3)  &        10.8   \\
		14  &  $2\nu_{1}-1\nu_{2}$  &          4.1725(5)  &            1.32(3)  &           0.520(3)  &        14.2   \\
		15  &  $-2\nu_{1}+3\nu_{2}$  &         10.1224(4)  &            1.37(3)  &           0.375(3)  &         8.8   \\
		16  &  $3\nu_{1}+1\nu_{2}$  &         24.1323(5)  &            1.20(3)  &           0.678(4)  &        14.1   \\
		17  &  $6\nu_{1}$  &         33.9666(5)  &            1.15(3)  &           0.775(4)  &        14.8   \\
		18  &  $3\nu_{1}-1\nu_{2}$  &          9.8208(5)  &            1.04(3)  &           0.599(4)  &         9.2   \\
		19  &          $\nu_{4}$  &         19.1940(6)  &            0.87(3)  &           0.054(5)  &        10.7   \\
		20  &  $-2\nu_{1}+1\nu_{4}$  &          7.8535(7)  &            0.82(3)  &           0.668(5)  &         9.4   \\
		21  &  $-10\nu_{1}+6\nu_{5}$  &          3.5457(7)  &            0.80(3)  &           0.053(5)  &        11.4   \\
		22  &          $\nu_{5}$  &         10.0244(7)  &            0.80(3)  &           0.024(5)  &        10.3   \\
		23  &  $4\nu_{1}+1\nu_{2}$  &         29.7928(7)  &            0.75(3)  &           0.542(6)  &        13.4   \\
		24  &  $3\nu_{1}+1\nu_{3}$  &         24.7583(7)  &            0.74(3)  &           0.770(6)  &        10.0   \\
		25  &  $-1\nu_{1}+1\nu_{4}$  &         13.5340(7)  &            0.70(3)  &           0.404(6)  &         8.2   \\
		26  &  $8\nu_{3}-4\nu_{5}$  &         22.1082(8)  &            0.68(3)  &           0.904(6)  &         9.8   \\
		27  &  $2\nu_{1}+2\nu_{3}-1\nu_{4}$     &          7.6836(8)  &            0.67(3)  &           0.611(6)  &        11.0   \\
		28  &  $-1\nu_{1}+3\nu_{2}$  &         15.7842(9)  &            0.60(3)  &           0.336(7)  &         8.3   \\
		29  &  $7\nu_{1}$  &         39.6273(9)  &            0.60(3)  &           0.059(7)  &        12.8   \\
		30  &          $\nu_{6}$  &          16.447(1)  &            0.50(3)  &           0.009(9)  &         7.2   \\
		31  &  $-2\nu_{1}+2\nu_{2}$  &           2.959(1)  &            0.50(3)  &           0.395(9)  &         7.7   \\
		32  &  $5\nu_{1}+1\nu_{2}$  &          35.454(1)  &            0.45(3)  &            0.51(1)  &        11.7   \\
		33  &  $2\nu_{2}$  &          14.298(1)  &            0.43(3)  &            0.35(1)  &         9.4   \\
		34  &  $1\nu_{2}+1\nu_{3}$  &          14.924(1)  &            0.42(3)  &            0.73(1)  &         9.0   \\
		35  &  $4\nu_{1}+1\nu_{3}$  &          30.420(1)  &            0.41(3)  &            0.94(1)  &         9.5   \\
		36  &  $2\nu_{1}+1\nu_{6}$  &          27.768(1)  &            0.41(3)  &            0.62(1)  &         8.7   \\
		37  &  $-1\nu_{1}+2\nu_{2}$  &           8.633(1)  &            0.39(3)  &            0.44(1)  &         8.1   \\
		38  &  $4\nu_{1}-1\nu_{2}$  &          15.478(1)  &            0.41(3)  &            0.80(1)  &         7.6   \\
		39  &  $1\nu_{1}+1\nu_{4}$  &          24.856(1)  &            0.37(3)  &            0.23(1)  &         7.8   \\
		40  &  $2\nu_{3}-1\nu_{4}+2\nu_{5}$     &          16.389(1)  &            0.36(3)  &            0.89(1)  &         6.1   \\
		41  &  $1\nu_{1}+1\nu_{5}$  &          15.686(1)  &            0.33(3)  &            0.02(1)  &         8.5   \\
		42  &  $-9\nu_{1}+6\nu_{5}$  &           9.210(2)  &            0.31(3)  &            0.58(1)  &         5.6   \\
		43  &  $1\nu_{1}+2\nu_{2}$  &          19.961(2)  &            0.31(3)  &            0.20(1)  &         7.4   \\
		44  &  $-3\nu_{1}+1\nu_{4}$  &           2.193(2)  &            0.30(3)  &            0.45(1)  &         5.2   \\
		45  &  $-6\nu_{2}+5\nu_{5}$  &           7.214(2)  &            0.28(3)  &            0.48(2)  &         7.0   \\
		46  &  $3\nu_{2}$  &          21.445(2)  &            0.28(3)  &            0.81(2)  &         6.5   \\
		47  &  $8\nu_{1}$  &          45.289(2)  &            0.27(3)  &            0.37(2)  &         9.5   \\
		48  &  $4\nu_{4}-6\nu_{5}$  &          16.637(2)  &            0.25(3)  &            0.95(2)  &         5.1   \\
		49  &  $4\nu_{1}+2\nu_{5}-1\nu_{6}$     &          26.247(2)  &            0.24(3)  &            0.46(2)  &         6.4   \\
		50  &  $-1\nu_{3}-1\nu_{4}+2\nu_{5}$     &          20.584(2)  &            0.23(3)  &            0.99(2)  &         6.4   \\
		51  &  $6\nu_{1}+1\nu_{2}$  &          41.116(2)  &            0.23(3)  &            0.54(2)  &         7.8   \\
		52  &  $3\nu_{1}+1\nu_{6}$  &          33.429(2)  &            0.21(3)  &            0.28(2)  &         6.0   \\
		53  &  $5\nu_{1}+1\nu_{3}$  &          36.081(2)  &            0.21(3)  &            0.52(2)  &         7.5   \\
		54  &  $-10\nu_{1}+10\nu_{3}$  &          21.137(3)  &            0.19(3)  &            0.14(2)  &         5.5   \\
		55  &  $2\nu_{1}+1\nu_{4}$  &          30.517(3)  &            0.19(3)  &            0.09(2)  &         6.3   \\
		56  &  $-10\nu_{1}+5\nu_{6}$  &          25.622(3)  &            0.18(3)  &            0.38(2)  &         5.3   \\
		57  &  $4\nu_{1}+1\nu_{6}$  &          39.091(3)  &            0.16(3)  &            0.17(3)  &         5.9   \\
		58  &  $9\nu_{1}$  &          50.950(3)  &            0.15(3)  &            0.82(3)  &         5.9   \\
		59  &  $2\nu_{1}+1\nu_{5}$  &          21.343(3)  &            0.15(3)  &            0.10(3)  &         5.4   \\
		60  &  $6\nu_{1}+1\nu_{3}$  &          41.741(3)  &            0.14(3)  &            0.64(3)  &         6.0   \\
		61  &  $7\nu_{1}+1\nu_{2}$  &          46.776(4)  &            0.13(3)  &            0.21(3)  &         5.4   \\
		62  &  $1\nu_{1}+2\nu_{3}$  &          21.216(4)  &            0.11(3)  &               0(9)  &         5.0   \\
		\hline
		%	\end{tabular}
\end{longtable}

\begin{longtable}{  c c c c c  c  }
	\caption{All frequencies detected in combined {\it TESS} sectors 40 and 41. Columns contain: number, our ID$_T$, frequency, amplitude, phase, signal to noise ratio.}\\
	\hline
	\# & ID$_{T,S40-41}$  &   $\nu~[\mathrm{d}^{-1}]$   &     $A$ [ppt]   &   $\phi$ [$0-1$] & $S/N$ \\
	\hline
	\endfirsthead
	\caption[]{continued}
	\label{tab:TESS_ALL_S40_41_fre}\\
	%	\begin{tabular}{  c c c c c  c  } % four columns, alignment for each
		\hline
		\# & ID$_{T,S40-41}$  &   $\nu~[\mathrm{d}^{-1}]$   &     $A$ [ppt]   &   $\phi$ [$0-1$] & $S/N$ \\
		\hline
		\endhead
		1  &          $\nu_{1}$  &         5.66105(3)  &           99.19(2)  &         0.88294(3)  &        17.0   \\
		2  &  $2\nu_{1}$  &        11.32209(5)  &           31.59(2)  &          0.7505(1)  &        17.0   \\
		3  &          $\nu_{2}$  &          7.1490(1)  &           11.55(2)  &          0.7651(3)  &        17.3   \\
		4  &  $3\nu_{1}$  &         16.9831(1)  &            9.99(2)  &          0.8321(3)  &        17.0   \\
		5  &  $1\nu_{1}+1\nu_{2}$  &         12.8100(2)  &            5.38(2)  &          0.8358(6)  &        16.7   \\
		6  &  $4\nu_{1}$  &         22.6441(2)  &            4.88(2)  &          0.5498(7)  &        16.4   \\
		7  &          $\nu_{3}$  &          7.7754(2)  &            4.60(2)  &          0.5743(7)  &        14.9   \\
		8  &  $-1\nu_{1}+1\nu_{2}$  &          1.4881(2)  &            3.59(2)  &          0.6663(9)  &        16.1   \\
		9  &  $2\nu_{1}+1\nu_{2}$  &         18.4711(2)  &            2.73(2)  &           0.292(1)  &        16.4   \\
		10  &  $1\nu_{1}+1\nu_{3}$  &         13.4366(2)  &            2.68(2)  &           0.053(1)  &        13.3   \\
		11  &  $5\nu_{1}$  &         28.3052(2)  &            2.31(2)  &           0.409(1)  &        15.9   \\
		12  &  $2\nu_{1}+1\nu_{3}$  &         19.0977(3)  &            1.60(2)  &           0.317(2)  &        11.9   \\
		13  &          instrumental  &          0.0662(4)  &            1.46(2)  &           0.539(2)  &         5.1   \\
		14  &  $-1\nu_{1}+1\nu_{3}$  &          2.1139(3)  &            1.56(2)  &           0.930(2)  &        13.9   \\
		15  &  $-2\nu_{1}+3\nu_{2}$  &         10.1225(4)  &            1.37(2)  &           0.174(2)  &         9.4   \\
		16  &  $2\nu_{1}-1\nu_{2}$  &          4.1727(3)  &            1.33(2)  &           0.696(2)  &        15.1   \\
		17  &  $3\nu_{1}+1\nu_{2}$  &         24.1321(4)  &            1.19(2)  &           0.466(3)  &        15.1   \\
		18  &  $6\nu_{1}$  &         33.9662(4)  &            1.14(2)  &           0.596(3)  &        15.3   \\
		19  &  $3\nu_{1}-1\nu_{2}$  &          9.8206(5)  &            0.95(2)  &           0.156(3)  &         8.9   \\
		20  &          $\nu_{4}$  &         10.0243(5)  &            0.82(2)  &           0.308(4)  &        10.7   \\
		21  &  $3\nu_{1}+1\nu_{3}$  &         24.7587(5)  &            0.80(2)  &           0.428(4)  &        10.8   \\
		22  &          $\nu_{5}$  &          7.8533(5)  &            0.80(2)  &           0.581(4)  &         9.8   \\
		23  &  $4\nu_{1}+1\nu_{2}$  &         29.7932(5)  &            0.76(2)  &           0.199(4)  &        14.2   \\
		24  &  $-10\nu_{1}+6\nu_{4}$  &          3.5470(5)  &            0.75(2)  &           0.083(4)  &        12.3   \\
		25  &  $2\nu_{1}+1\nu_{5}$  &         19.1926(6)  &            0.73(2)  &           0.832(4)  &        10.4   \\
		26  &  $1\nu_{1}+1\nu_{5}$  &         13.5302(6)  &            0.68(2)  &           0.279(5)  &         8.5   \\
		27  &  $2\nu_{3}-1\nu_{5}$  &          7.6833(6)  &            0.70(2)  &           0.894(5)  &        11.5   \\
		28  &  $7\nu_{1}$  &         39.6275(7)  &            0.58(2)  &           0.042(6)  &        13.7   \\
		29  &  $-1\nu_{1}+3\nu_{2}$  &         15.7840(7)  &            0.56(2)  &           0.149(6)  &         8.4   \\
		30  &  $8\nu_{3}-4\nu_{4}$  &         22.1051(7)  &            0.55(2)  &           0.230(6)  &         8.2   \\
		31  &  $1\nu_{1}+2\nu_{3}-1\nu_{5}$     &         13.3424(7)  &            0.52(2)  &           0.740(6)  &         7.4   \\
		32  &  $4\nu_{1}+1\nu_{3}$  &         30.4202(8)  &            0.45(2)  &           0.234(7)  &        10.6   \\
		33  &          $\nu_{6}$  &         16.4455(8)  &            0.45(2)  &           0.185(7)  &         6.8   \\
		34  &  $2\nu_{2}$  &         14.2967(9)  &            0.43(2)  &           0.598(8)  &        11.0   \\
		35  &  $5\nu_{1}+1\nu_{2}$  &         35.4544(9)  &            0.42(2)  &           0.197(8)  &        12.2   \\
		36  &  $-9\nu_{1}+6\nu_{4}$  &          9.2040(9)  &            0.41(2)  &           0.113(8)  &         7.6   \\
		37  &  $2\nu_{1}+1\nu_{6}$  &          27.765(1)  &            0.38(2)  &           0.393(9)  &         7.8   \\
		38  &  $4\nu_{1}-1\nu_{2}$  &          15.479(1)  &            0.40(2)  &           0.766(8)  &         7.5   \\
		39  &  $1\nu_{2}+1\nu_{3}$  &          14.924(1)  &            0.39(2)  &           0.266(8)  &         9.4   \\
		40  &  $1\nu_{1}+2\nu_{2}$  &          19.958(1)  &            0.36(2)  &           0.055(9)  &        11.0   \\
		41  &  $1\nu_{1}+1\nu_{4}$  &          15.686(1)  &            0.34(2)  &           0.231(9)  &         9.2   \\
		42  &  $3\nu_{1}+1\nu_{5}$  &          24.854(1)  &            0.32(2)  &            0.17(1)  &         7.4   \\
		43  &  $-4\nu_{3}+5\nu_{4}$  &          19.003(1)  &            0.32(2)  &            0.97(1)  &         6.2   \\
		44  &  $-1\nu_{1}+2\nu_{2}$  &           8.636(1)  &            0.31(2)  &            0.56(1)  &         8.1   \\
		45  &  $8\nu_{1}$  &          45.288(1)  &            0.29(2)  &            0.86(1)  &        10.1   \\
		46  &  $6\nu_{1}+1\nu_{2}$  &          41.115(1)  &            0.28(2)  &            0.94(1)  &        10.8   \\
		47  &  $3\nu_{2}$  &          21.445(1)  &            0.28(2)  &            0.86(1)  &         6.6   \\
		48  &  $-8\nu_{4}+5\nu_{6}$  &           2.023(1)  &            0.28(2)  &            0.87(1)  &         5.7   \\
		49  &  $3\nu_{1}+2\nu_{4}-1\nu_{6}$     &          20.586(1)  &            0.28(2)  &            0.07(1)  &         8.5   \\
		50  &  $-1\nu_{1}+1\nu_{5}$  &           2.192(1)  &            0.28(2)  &            0.45(1)  &         6.3   \\
		51  &  $4\nu_{1}+2\nu_{4}-1\nu_{6}$     &          26.246(1)  &            0.25(2)  &            0.11(1)  &         8.1   \\
		52  &  $-6\nu_{2}+5\nu_{4}$  &           7.212(1)  &            0.25(2)  &            0.35(1)  &         6.8   \\
		53  &  $5\nu_{1}+1\nu_{3}$  &          36.081(1)  &            0.24(2)  &            0.95(1)  &         8.8   \\
		54  &  $-10\nu_{1}+5\nu_{6}$  &          25.619(2)  &            0.22(2)  &            0.15(1)  &         8.1   \\
		55  &  $3\nu_{1}+1\nu_{6}$  &          33.426(2)  &            0.22(2)  &            0.20(1)  &         6.8   \\
		56  &  $-3\nu_{1}+3\nu_{2}$  &           4.462(2)  &            0.20(2)  &            0.05(2)  &         6.6   \\
		57  &  $-10\nu_{1}+10\nu_{3}$  &          21.138(2)  &            0.20(2)  &            0.77(2)  &         6.0   \\
		58  &  $2\nu_{1}+1\nu_{4}$  &          21.346(2)  &            0.19(2)  &            0.68(2)  &         6.5   \\
		59  &  $5\nu_{1}+2\nu_{4}-1\nu_{6}$     &          31.908(2)  &            0.16(2)  &            0.79(2)  &         5.4   \\
		60  &  $4\nu_{1}+1\nu_{5}$  &          30.513(2)  &            0.14(2)  &            0.16(2)  &         5.5   \\
		61  &  $9\nu_{1}$  &          50.950(2)  &            0.14(2)  &            0.11(2)  &         6.3   \\
		62  &  $4\nu_{1}+1\nu_{6}$  &          39.092(3)  &            0.12(2)  &            0.38(3)  &         5.2   \\
		63  &  $6\nu_{1}+1\nu_{3}$  &          41.743(3)  &            0.12(2)  &            0.95(3)  &         5.6   \\
		64  &  $7\nu_{1}+1\nu_{2}$  &          46.776(3)  &            0.12(2)  &            0.05(3)  &         6.2   \\
		65  &  $-9\nu_{1}+5\nu_{6}$  &          31.281(3)  &            0.12(2)  &            0.40(3)  &         5.4   \\
		66  &  $5\nu_{1}+1\nu_{2}+1\nu_{3}$     &          43.229(3)  &            0.10(2)  &               1(9)  &         5.3   \\
		\hline
		%	\end{tabular}
\end{longtable}

\begin{longtable}{  c c c c c  c  }
	\caption{All frequencies detected in combined TESS sectors 54 and 55. Columns contain: number, our ID$_T$, frequency, amplitude, phase, signal to noise ratio.}\\
	\hline
	\# & ID$_{T,S54-55}$  &   $\nu~[\mathrm{d}^{-1}]$   &     $A$ [ppt]   &   $\phi$ [$0-1$] & $S/N$ \\
	\hline
	\endfirsthead
	\caption[]{continued}
	\label{tab:TESS_ALL_S54_55_fre}\\
	%	\begin{tabular}{  c c c c c  c  } % four columns, alignment for each
		\hline
		\# & ID$_{T,S54-55}$  &   $\nu~[\mathrm{d}^{-1}]$   &     $A$ [ppt]   &   $\phi$ [$0-1$] & $S/N$ \\
		\hline
		\endhead
		1  &          $\nu_{1}$  &         5.66106(3)  &          101.55(2)  &         0.83449(3)  &        17.1   \\
		2  &  $2\nu_{1}$  &        11.32209(6)  &           32.36(2)  &          0.7408(1)  &        17.0   \\
		3  &          $\nu_{2}$  &          7.1490(1)  &           11.94(2)  &          0.0485(3)  &        17.0   \\
		4  &  $3\nu_{1}$  &         16.9832(1)  &           10.28(2)  &          0.4881(3)  &        16.9   \\
		5  &  $1\nu_{1}+1\nu_{2}$  &         12.8100(2)  &            5.51(2)  &          0.7926(6)  &        16.5   \\
		6  &  $4\nu_{1}$  &         22.6442(2)  &            5.05(2)  &          0.1771(6)  &        16.5   \\
		7  &          $\nu_{3}$  &          7.7756(2)  &            4.62(2)  &          0.6908(7)  &        15.9   \\
		8  &  $-1\nu_{1}+1\nu_{2}$  &          1.4881(2)  &            3.70(2)  &          0.5804(8)  &        15.7   \\
		9  &  $2\nu_{1}+1\nu_{2}$  &         18.4709(2)  &            2.82(2)  &           0.423(1)  &        16.4   \\
		10  &  $1\nu_{1}+1\nu_{3}$  &         13.4362(2)  &            2.78(2)  &           0.064(1)  &        13.7   \\
		11  &  $5\nu_{1}$  &         28.3053(2)  &            2.40(2)  &           0.841(1)  &        16.2   \\
		12  &          instrumental  &          0.0658(3)  &            1.73(2)  &           0.076(2)  &         5.6   \\
		13  &  $-1\nu_{1}+1\nu_{3}$  &          2.1144(3)  &            1.66(2)  &           0.771(2)  &        14.3   \\
		14  &  $2\nu_{1}+1\nu_{3}$  &         19.0974(3)  &            1.61(2)  &           0.760(2)  &        12.1   \\
		15  &  $-2\nu_{1}+3\nu_{2}$  &         10.1225(3)  &            1.44(2)  &           0.600(2)  &        10.2   \\
		16  &  $2\nu_{1}-1\nu_{2}$  &          4.1726(3)  &            1.36(2)  &           0.308(2)  &        15.0   \\
		17  &  $3\nu_{1}+1\nu_{2}$  &         24.1318(4)  &            1.25(2)  &           0.013(2)  &        15.6   \\
		18  &  $6\nu_{1}$  &         33.9662(4)  &            1.16(2)  &           0.470(3)  &        15.1   \\
		19  &  $3\nu_{1}-1\nu_{2}$  &          9.8195(4)  &            1.04(2)  &           0.276(3)  &        10.0   \\
		20  &          $\nu_{4}$  &         10.0239(5)  &            0.94(2)  &           0.183(3)  &        11.8   \\
		21  &          $\nu_{5}$  &          7.8534(5)  &            0.78(2)  &           0.008(4)  &         9.0   \\
		22  &  $3\nu_{1}+1\nu_{3}$  &         24.7586(5)  &            0.80(2)  &           0.025(4)  &        11.1   \\
		23  &  $4\nu_{1}+1\nu_{2}$  &         29.7931(5)  &            0.78(2)  &           0.624(4)  &        14.5   \\
		24  &  $2\nu_{3}-1\nu_{5}$  &          7.6831(5)  &            0.77(2)  &           0.098(4)  &        11.8   \\
		25  &  $-10\nu_{1}+6\nu_{4}$  &          3.5464(6)  &            0.71(2)  &           0.735(4)  &        11.7   \\
		26  &  $2\nu_{1}+1\nu_{5}$  &         19.1928(6)  &            0.71(2)  &           0.026(4)  &         9.2   \\
		27  &  $8\nu_{3}-4\nu_{4}$  &         22.1062(6)  &            0.68(2)  &           0.577(5)  &         9.4   \\
		28  &  $1\nu_{1}+2\nu_{3}-1\nu_{5}$     &         13.3428(6)  &            0.64(2)  &           0.309(5)  &         7.2   \\
		29  &  $-1\nu_{1}+3\nu_{2}$  &         15.7832(6)  &            0.62(2)  &           0.771(5)  &         8.8   \\
		30  &  $7\nu_{1}$  &         39.6269(6)  &            0.60(2)  &           0.132(5)  &        13.8   \\
		31  &  $1\nu_{1}+1\nu_{5}$  &         13.5287(7)  &            0.56(2)  &           0.284(6)  &         7.8   \\
		32  &          $\nu_{6}$  &         16.4468(7)  &            0.54(2)  &           0.473(6)  &         7.2   \\
		33  &  $5\nu_{1}+1\nu_{2}$  &         35.4543(8)  &            0.48(2)  &           0.484(6)  &        12.5   \\
		34  &  $2\nu_{2}$  &         14.2970(8)  &            0.45(2)  &           0.161(7)  &        11.4   \\
		35  &  $4\nu_{1}+1\nu_{3}$  &         30.4193(8)  &            0.45(2)  &           0.701(7)  &        10.2   \\
		36  &  $4\nu_{1}-1\nu_{2}$  &         15.4779(8)  &            0.42(2)  &           0.160(7)  &         9.0   \\
		37  &  $-9\nu_{1}+6\nu_{4}$  &          9.2047(8)  &            0.43(2)  &           0.085(7)  &         7.7   \\
		38  &  $2\nu_{1}+1\nu_{6}$  &         27.7659(8)  &            0.44(2)  &           0.676(7)  &         8.9   \\
		39  &  $1\nu_{2}+1\nu_{3}$  &         14.9253(9)  &            0.40(2)  &           0.923(8)  &        10.6   \\
		40  &  $1\nu_{1}+1\nu_{4}$  &         15.6848(9)  &            0.37(2)  &           0.802(8)  &        11.2   \\
		41  &  $1\nu_{1}+2\nu_{2}$  &          19.957(1)  &            0.34(2)  &           0.765(9)  &         9.3   \\
		42  &  $-4\nu_{3}+5\nu_{4}$  &          19.004(1)  &            0.34(2)  &           0.788(9)  &         6.1   \\
		43  &  $3\nu_{1}+1\nu_{5}$  &          24.855(1)  &            0.33(2)  &           0.753(9)  &         7.6   \\
		44  &  $3\nu_{1}+2\nu_{4}-1\nu_{6}$     &          20.585(1)  &            0.32(2)  &            0.70(1)  &         8.9   \\
		45  &  $3\nu_{2}$  &          21.444(1)  &            0.32(2)  &            0.42(1)  &         7.1   \\
		46  &  $8\nu_{1}$  &          45.289(1)  &            0.29(2)  &            0.10(1)  &        10.0   \\
		47  &  $6\nu_{1}+1\nu_{2}$  &          41.116(1)  &            0.27(2)  &            0.14(1)  &         9.4   \\
		48  &  $2\nu_{2}+1\nu_{3}-2\nu_{4}$     &           2.023(1)  &            0.28(2)  &            0.77(1)  &         5.1   \\
		49  &  $-1\nu_{1}+2\nu_{2}$  &           8.638(1)  &            0.28(2)  &            0.52(1)  &         6.8   \\
		50  &  $4\nu_{1}+2\nu_{4}-1\nu_{6}$     &          26.245(1)  &            0.25(2)  &            0.41(1)  &         8.3   \\
		51  &  $5\nu_{1}+1\nu_{3}$  &          36.081(1)  &            0.25(2)  &            0.48(1)  &         8.9   \\
		52  &  $-6\nu_{2}+5\nu_{4}$  &           7.214(1)  &            0.25(2)  &            0.96(1)  &         7.3   \\
		53  &  $3\nu_{1}+1\nu_{6}$  &          33.427(1)  &            0.24(2)  &            0.90(1)  &         7.3   \\
		54  &  $-10\nu_{1}+10\nu_{3}$  &          21.138(2)  &            0.21(2)  &            0.19(1)  &         6.9   \\
		55  &  $2\nu_{1}+1\nu_{4}$  &          21.346(2)  &            0.19(2)  &            0.01(2)  &         7.1   \\
		56  &  $-3\nu_{1}+3\nu_{2}$  &           4.462(2)  &            0.19(2)  &            0.38(2)  &         5.5   \\
		57  &  $-10\nu_{1}+5\nu_{6}$  &          25.618(2)  &            0.19(2)  &            0.34(2)  &         7.6   \\
		58  &  $3\nu_{1}+2\nu_{3}-1\nu_{5}$     &          24.664(2)  &            0.19(2)  &            0.24(2)  &         5.3   \\
		59  &  $5\nu_{1}+2\nu_{4}-1\nu_{6}$     &          31.906(2)  &            0.17(2)  &            0.92(2)  &         6.8   \\
		60  &  $6\nu_{1}+1\nu_{3}$  &          41.742(2)  &            0.16(2)  &            0.32(2)  &         6.9   \\
		61  &  $9\nu_{1}$  &          50.948(2)  &            0.16(2)  &            0.48(2)  &         7.5   \\
		62  &  $-9\nu_{1}+5\nu_{6}$  &          31.280(2)  &            0.16(2)  &            0.62(2)  &         6.4   \\
		63  &  $4\nu_{1}+1\nu_{6}$  &          39.088(2)  &            0.15(2)  &            0.22(2)  &         6.2   \\
		64  &  $4\nu_{1}+1\nu_{5}$  &          30.514(2)  &            0.15(2)  &            0.21(2)  &         5.3   \\
		65  &  $-8\nu_{1}+6\nu_{4}$  &          14.866(3)  &            0.13(2)  &            0.64(2)  &         5.2   \\
		66  &  $7\nu_{1}+1\nu_{2}$  &          46.777(3)  &            0.13(2)  &            0.89(2)  &         6.8   \\
		67  &  $-8\nu_{1}+5\nu_{6}$  &          36.939(3)  &            0.13(2)  &            0.36(2)  &         6.4   \\
		68  &  $5\nu_{1}+1\nu_{6}$  &          44.752(3)  &            0.12(2)  &               2(8)  &         5.1   \\
		\hline
		%	\end{tabular}
\end{longtable}

\begin{figure}
	\includegraphics[angle=0, width=\columnwidth]{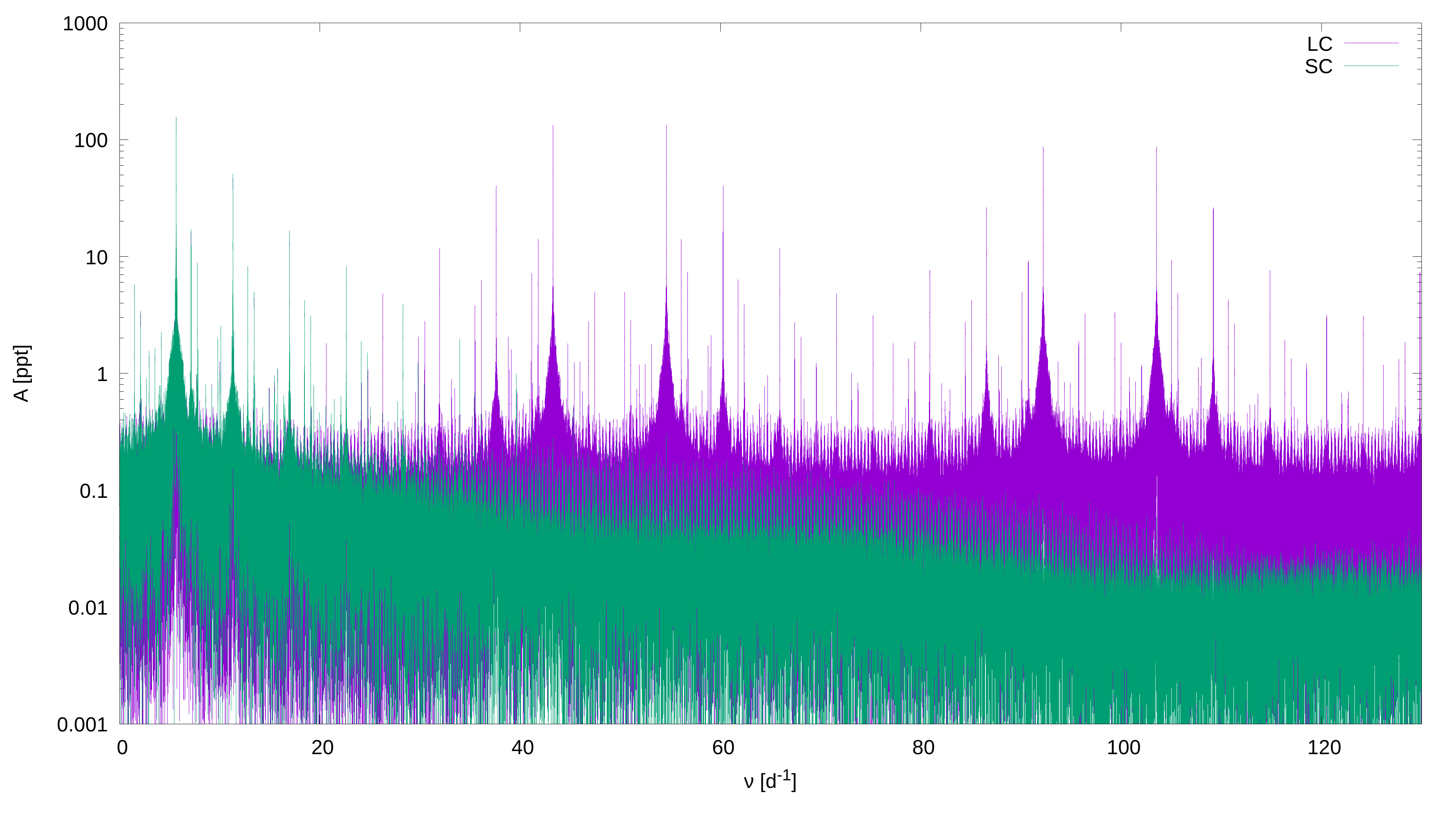}
	\caption{Comparison of periodograms calculated for LC (Q0-Q17) and SC (Q6-Q10) {\it Kepler} data.}
	\label{fig:LC_vs_SC}
\end{figure}

\setlength{\tabcolsep}{6pt}
\begin{table*}
	\centering
	\caption{The values of three independent frequencies with highest-amplitudes determined separately in each {\it Kepler} quarter and pairs of adjacent {\it TESS} sectors.
		Columns contain: the name of data set, median value of time in each data set, and frequencies of $\nu_1$, $\nu_2$, and $\nu_3$.}
	\label{tab:freq_evol}
	\begin{tabular}{  c c c c c } % four columns, alignment for each
		\hline
		Data set  &   BJD-2454833 [d]   &$\nu_1~[\mathrm{d}^{-1}]$  &   $\nu_2~[\mathrm{d}^{-1}]$  &   $\nu_3~[\mathrm{d}^{-1}]$ \\
		\hline
		{\it Kepler} Q0SC    &   125.39      &   5.661(1)   &    7.150(4)  &   7.775(5)  \\
		{\it Kepler} Q1LC    &   148.23      &  5.6610(2)   &    7.1487(8) &  7.7757(9)  \\
		{\it Kepler} Q2LC    &   213.98      &  5.66106(5)  &   7.1489(2)  & 7.7756(2)  \\
		{\it Kepler} Q3LC    &   304.75      &  5.66107(5)  &   7.1490(2)  & 7.7756(2) \\
		{\it Kepler} Q4LC    &   395.78      &  5.66105(5)  &   7.1490(2)  & 7.7755(2) \\
		{\it Kepler} Q5LC    &   491.14      &  5.66106(5)  &   7.1489(2)  & 7.7756(2) \\
		{\it Kepler} Q6SC    &   584.31      &  5.66106(2)  &   7.14896(9) & 7.7756(1) \\
		{\it Kepler} Q7SC    &   674.53      &  5.66106(2)  &   7.1490(1)  & 7.7756(1) \\
		{\it Kepler} Q8SC    &   770.20      &  5.66106(4)  &   7.1490(2)  & 7.7755(2) \\
		{\it Kepler} Q9SC    &   857.23      &  5.66106(2)  &   7.14899(8) & 7.7756(1) \\
		{\it Kepler} Q10SC   &   953.41      &  5.66105(2)  &   7.14897(8) & 7.7756(1) \\
		{\it Kepler} Q11LC   &  1048.57      &  5.66105(4)  &   7.1489(2)  & 7.7756(2) \\
		{\it Kepler} Q12LC   &  1139.95      &  5.66106(6)  &   7.1489(2)  & 7.7755(3) \\
		{\it Kepler} Q13LC   &  1227.63      &  5.66105(5)  &   7.1490(2)  & 7.7756(3) \\
		{\it Kepler} Q14LC   &  1326.43      &  5.66106(4)  &   7.1489(2)  & 7.7757(2) \\
		{\it Kepler} Q15LC   &  1425.29      &  5.66106(4)  &   7.1490(2)  & 7.7756(2) \\
		{\it Kepler} Q16LC   &  1520.22      &  5.66106(6)  &   7.1489(2)  & 7.7756(3) \\
		{\it Kepler} Q17LC   &  1572.55      &  5.6611(3)   &   7.149(1)   & 7.776(1) \\
		TESS S14S15    &  3876.74      &  5.66105(3)  &   7.1489(1)  & 7.7753(2) \\
		TESS S40S41    &  4584.96      &  5.66105(3)  &   7.1490(1)  &  7.7754(2) \\
		TESS S54S55    &  4964.78      &  5.66106(3)  &   7.1490(1)  & 7.7756(2)  \\
		\hline
	\end{tabular}
\end{table*}

\section{Asteroseismic modelling with Monte-Carlo-based Bayesian Analysis}
%We perform seismic modelling of V2367\,Cyg with the MC-based Bayesian analysis assuming two hypotheses:
%1) the observed frequencies $\nu_1$ and $\nu_2$ are the first and second overtone radial mode, respectively, 
%2) the observed frequencies $\nu_1$ and $\nu_2$ are the fundamental  and first overtone radial mode, respectively, 
%We took into account the effects of rotation on pulsational frequencies including the effect of rotational mode coupling.
%We fitted the dominant frequency $\nu_1$ and its non-adiabatic complex parameter $f$ within the errors, the frequency $\nu_2$
%with an uncertainty resulting from coupling with the nearest $\ell=2$ mode, as well as the observed  range of $(T_{\rm eff},~L)$

%A set of parameters  ($M,~V_{\rm rot,0},~\alpha_{\rm ov}$ and $\alpha_{\rm MLT}$) was randomly selected
%to calculated evolutionary and pulsational models. The number of  simulations  was about 100\,000.
%The initial hydrogen abundance was  $X_0=0.70$ and  three metallicities were considered $Z=0.014,~0.020, ~0.030$.
%The OPAL tables were adopted in all calculations.

\begin{figure*}
	\centering
	\includegraphics[clip,width=0.495\linewidth,height=60mm]{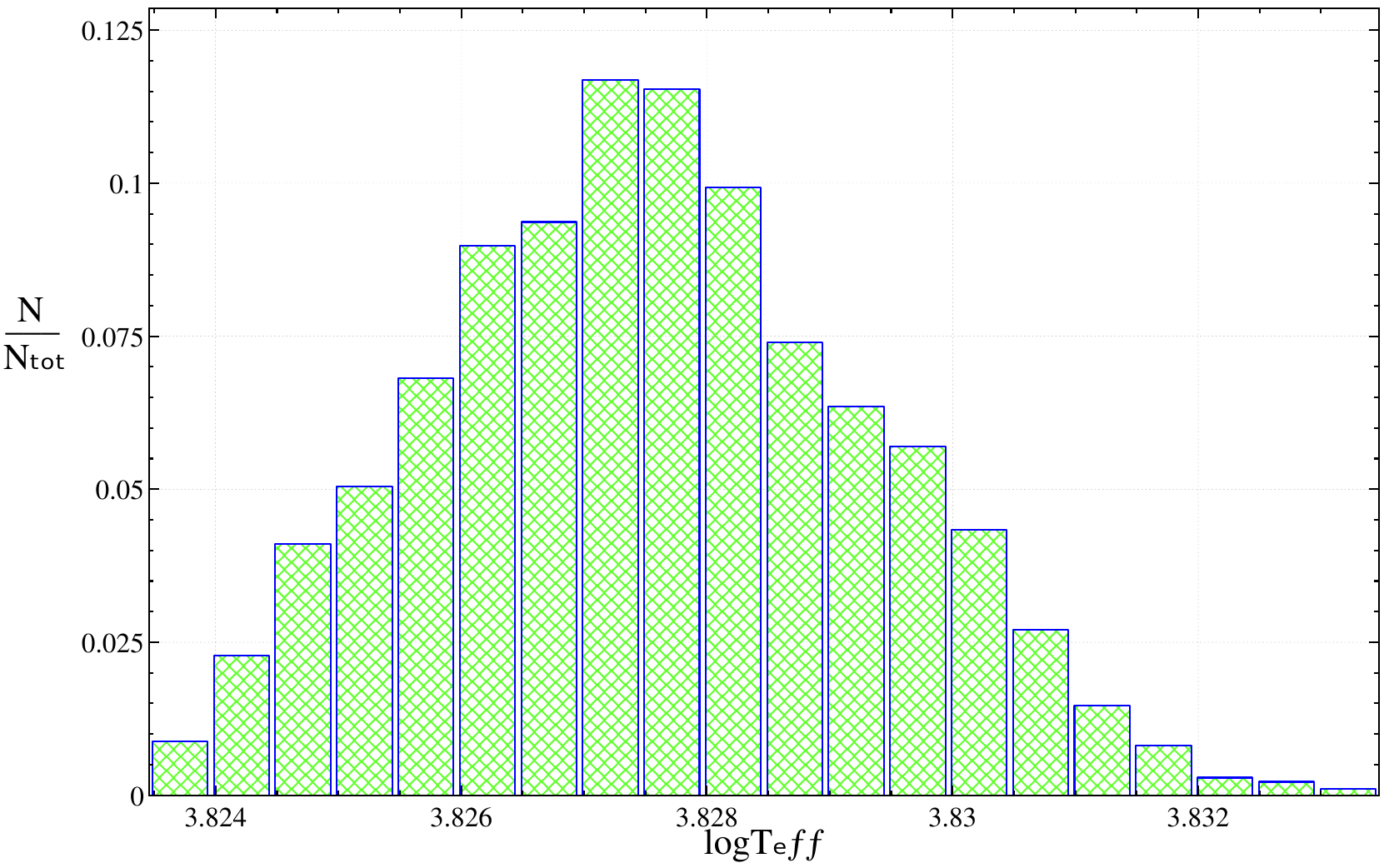}
	\includegraphics[clip,width=0.495\linewidth,height=60mm]{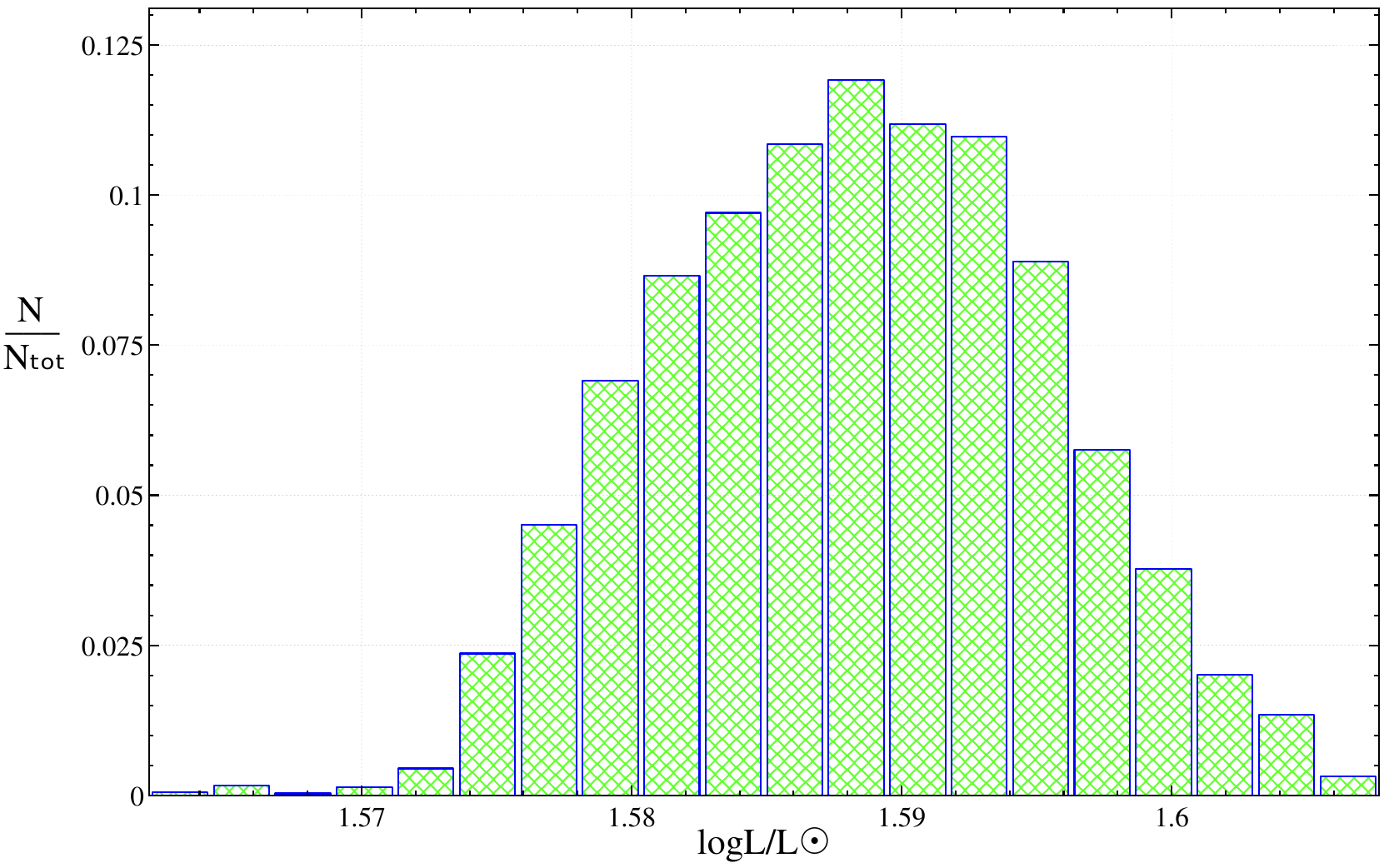}
	\includegraphics[clip,width=0.495\linewidth,height=60mm]{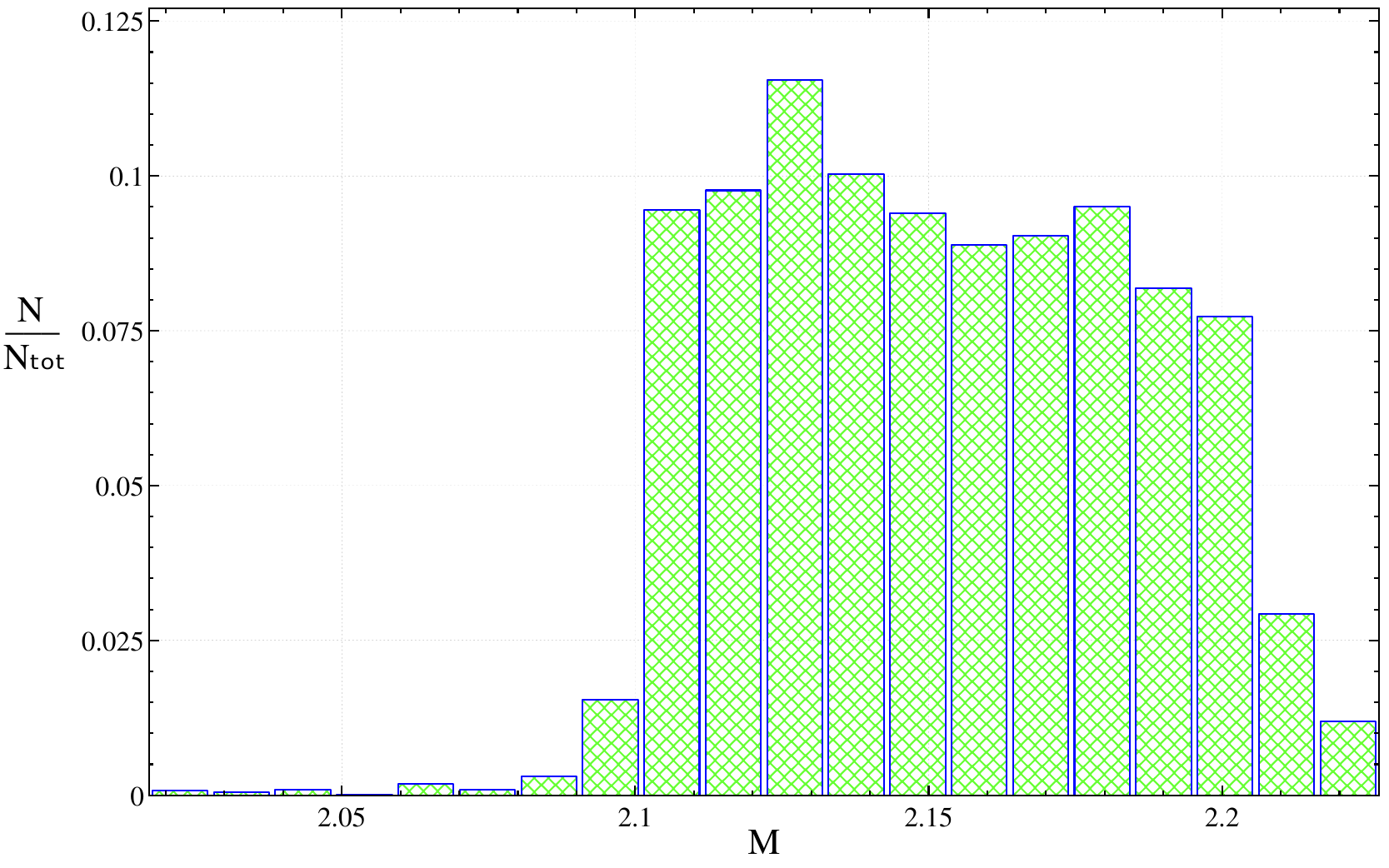}
	\includegraphics[clip,width=0.495\linewidth,height=60mm]{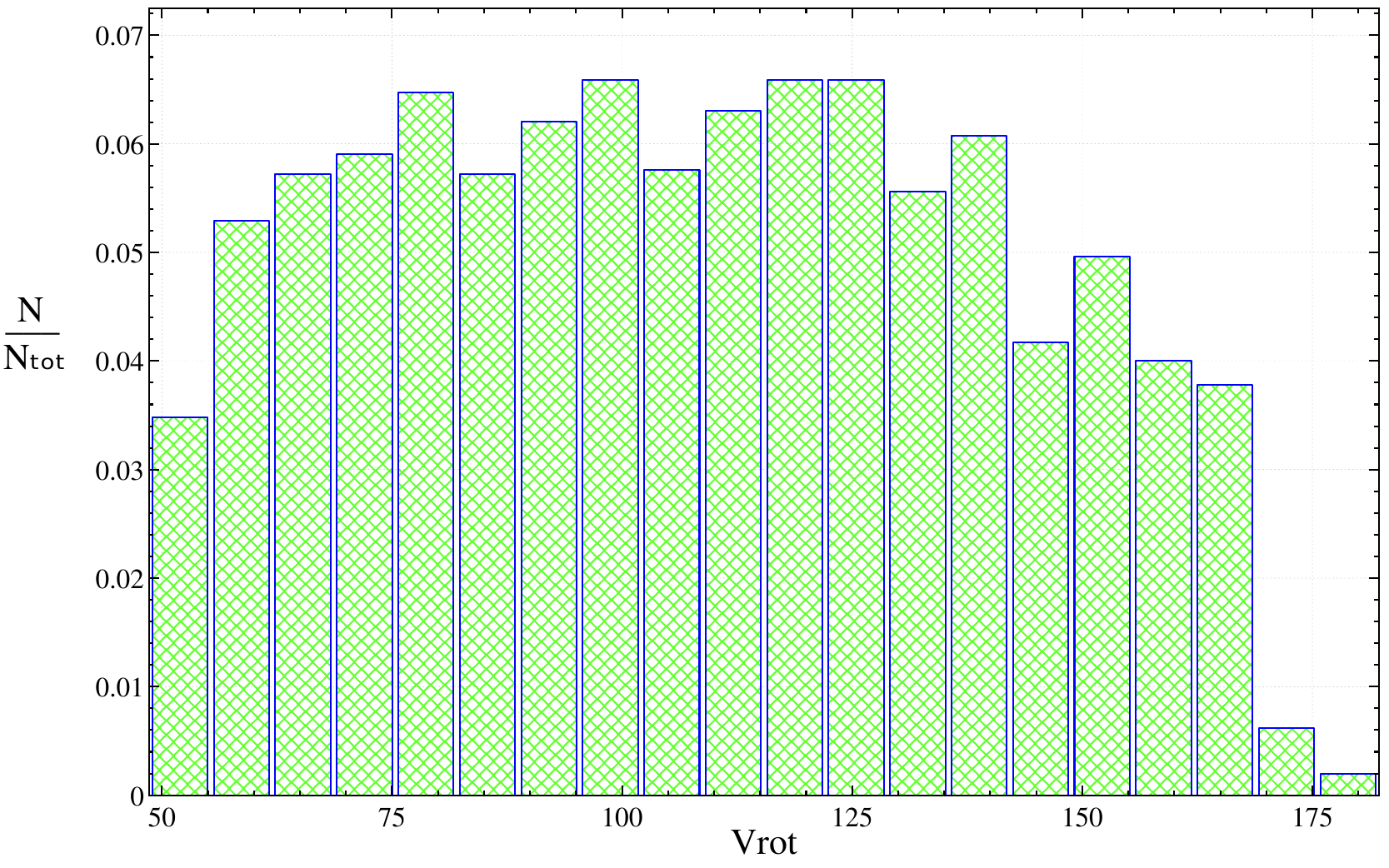}
	\includegraphics[clip,width=0.495\linewidth,height=60mm]{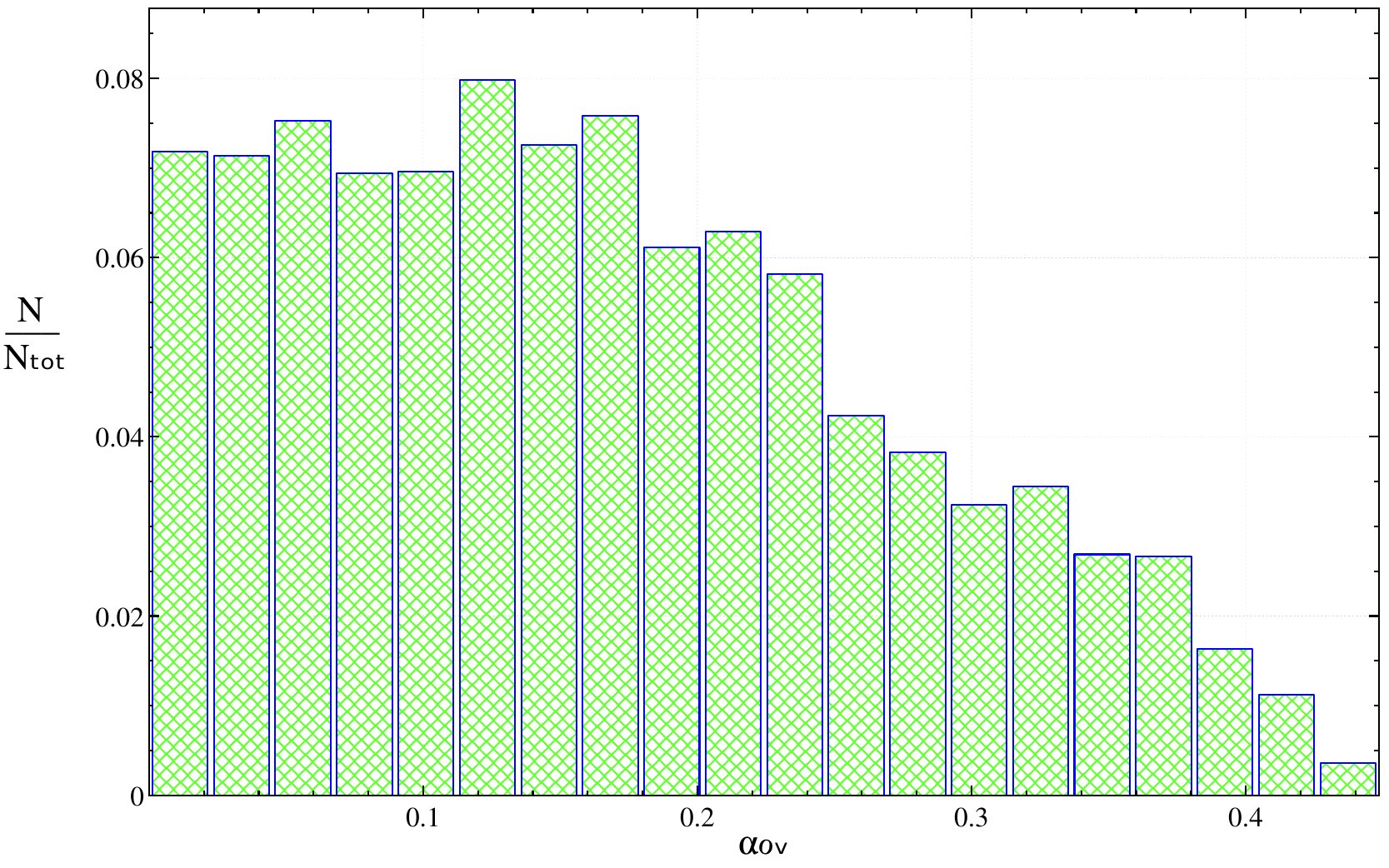}
	\includegraphics[clip,width=0.495\linewidth,height=60mm]{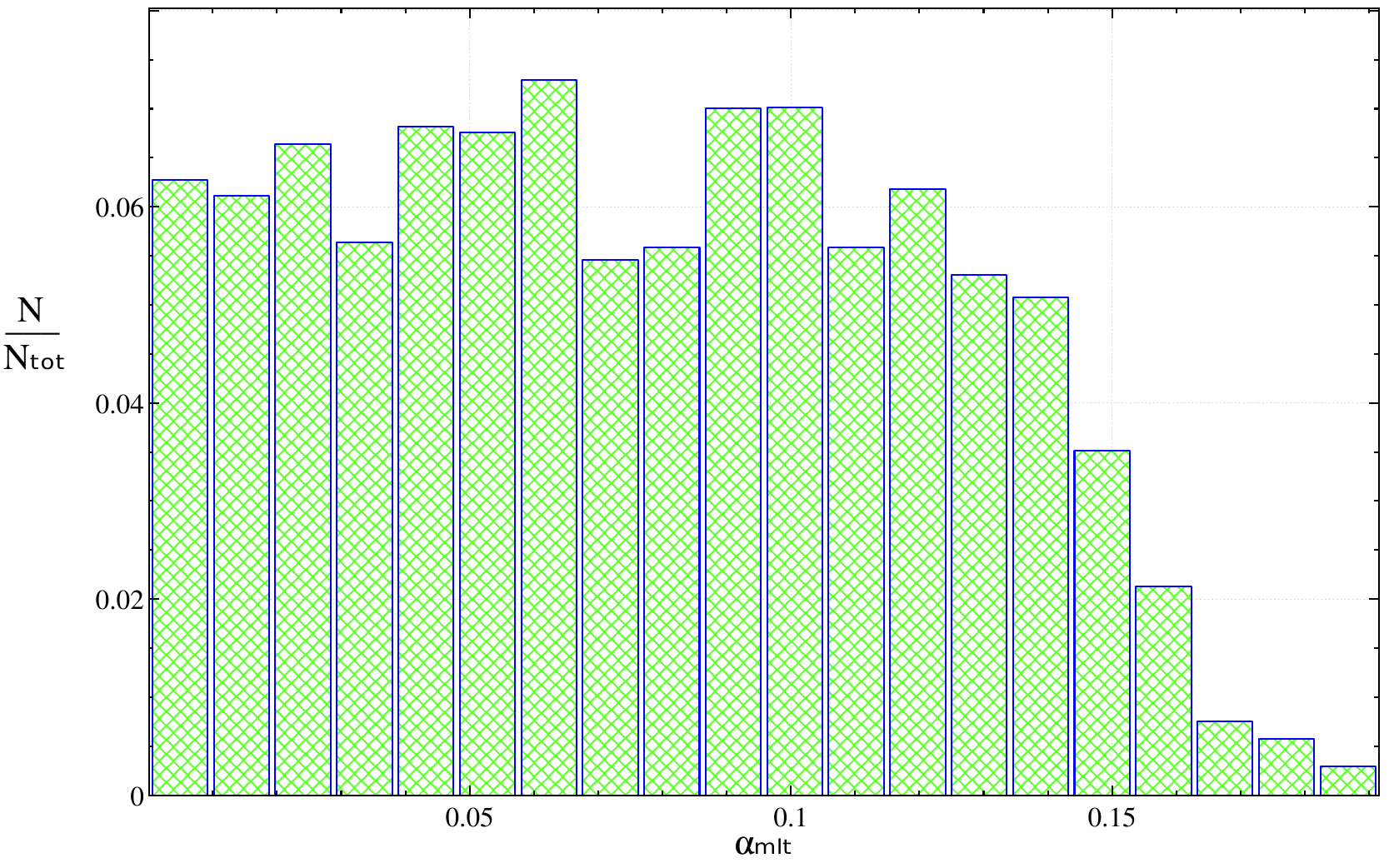}
	\caption{The normalized histograms for the parameters of the HSB seismic models of V2367\,Cyg, calculated assuming 
		that the observed frequencies $\nu_1$ and $\nu_2$ are the first and second radial overtone, respectively 
		(the $p_2\& p_3$ hypothesis).  The OPAL opacities were adopted and  $X_0=0.70,~Z=0.020$ were fixed.}
	\label{seismic_HSB_Z020}
\end{figure*}

\begin{figure*}
	\centering
	\includegraphics[clip,width=0.495\linewidth,height=60mm]{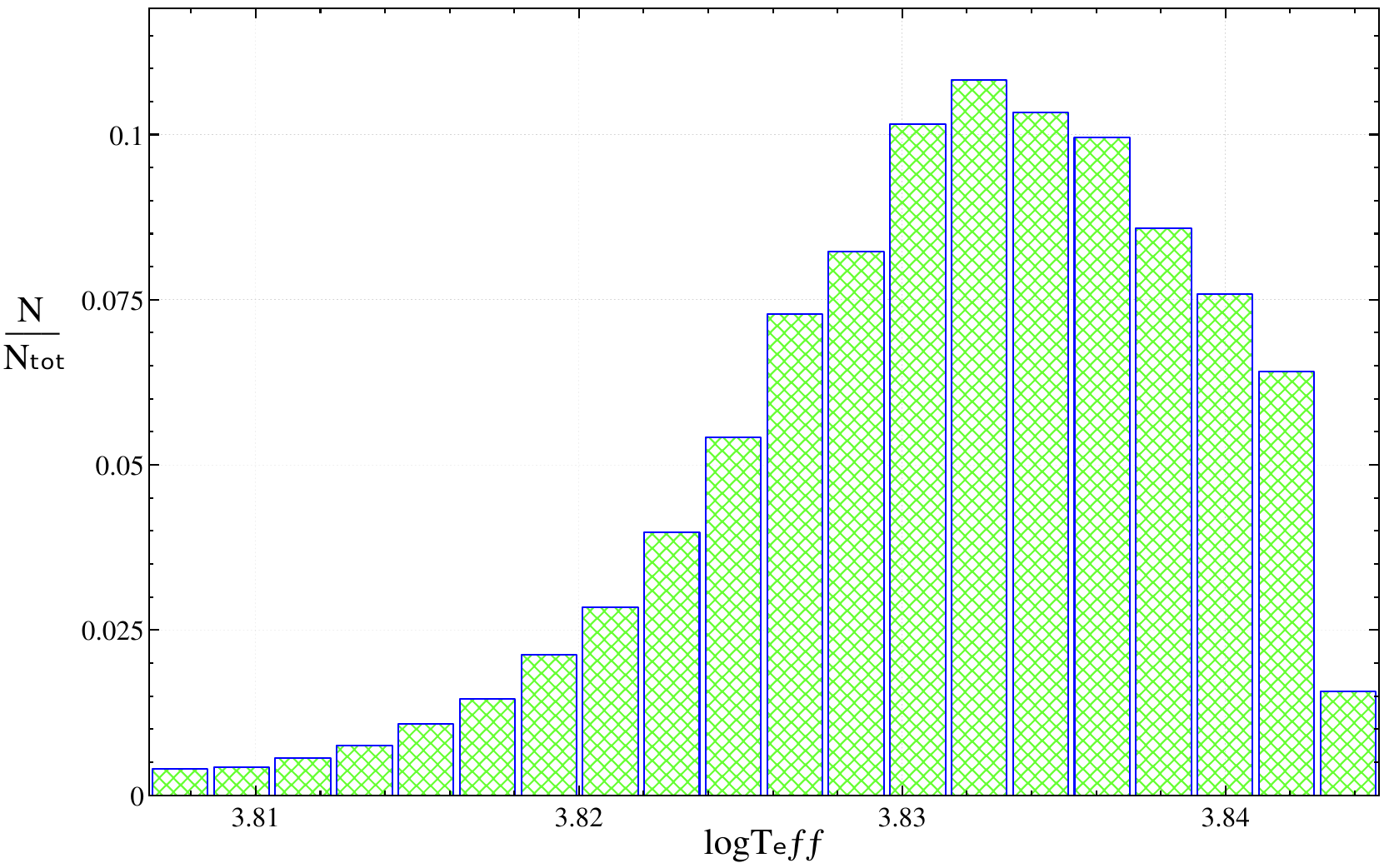}
	\includegraphics[clip,width=0.495\linewidth,height=60mm]{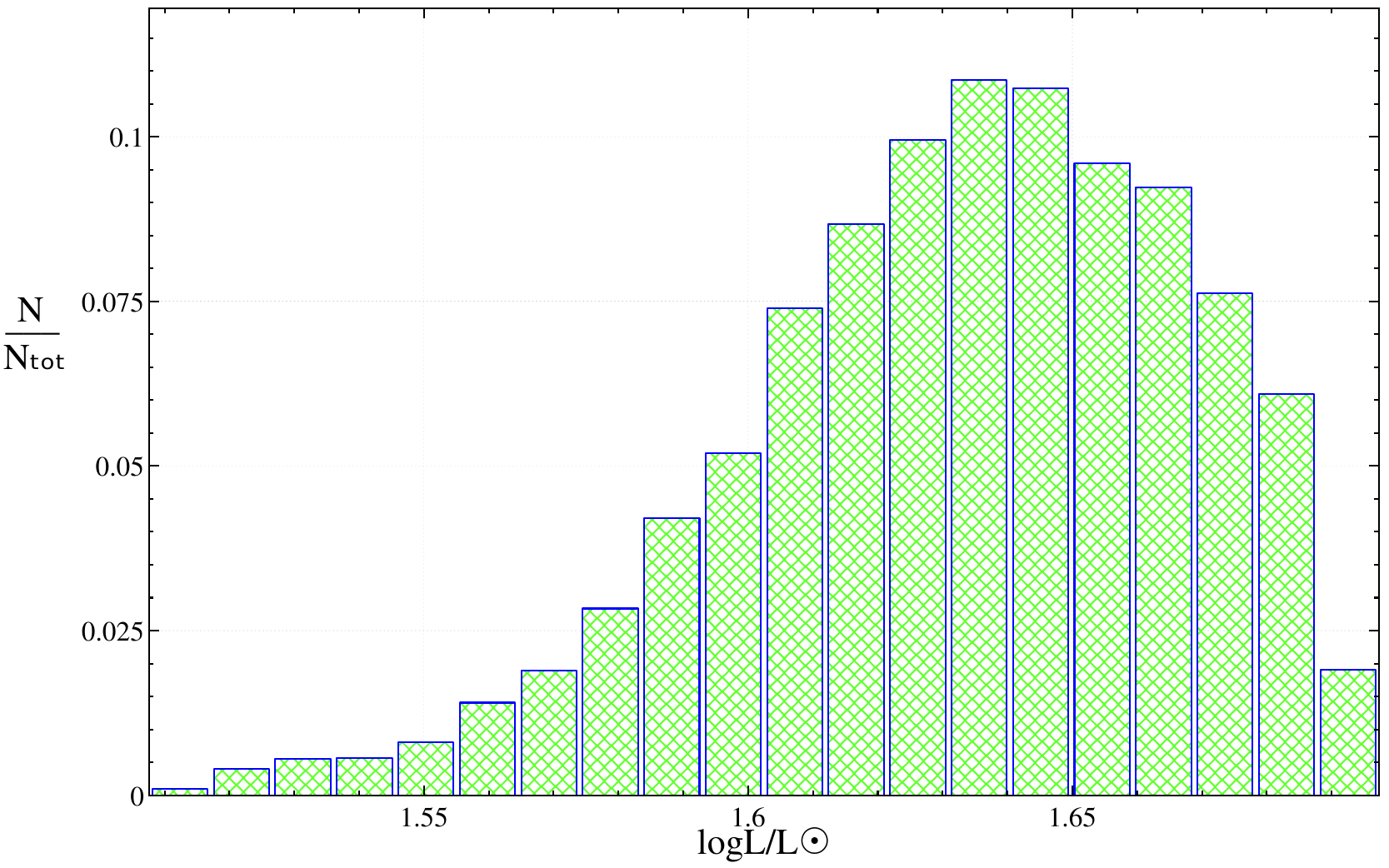}
	\includegraphics[clip,width=0.495\linewidth,height=60mm]{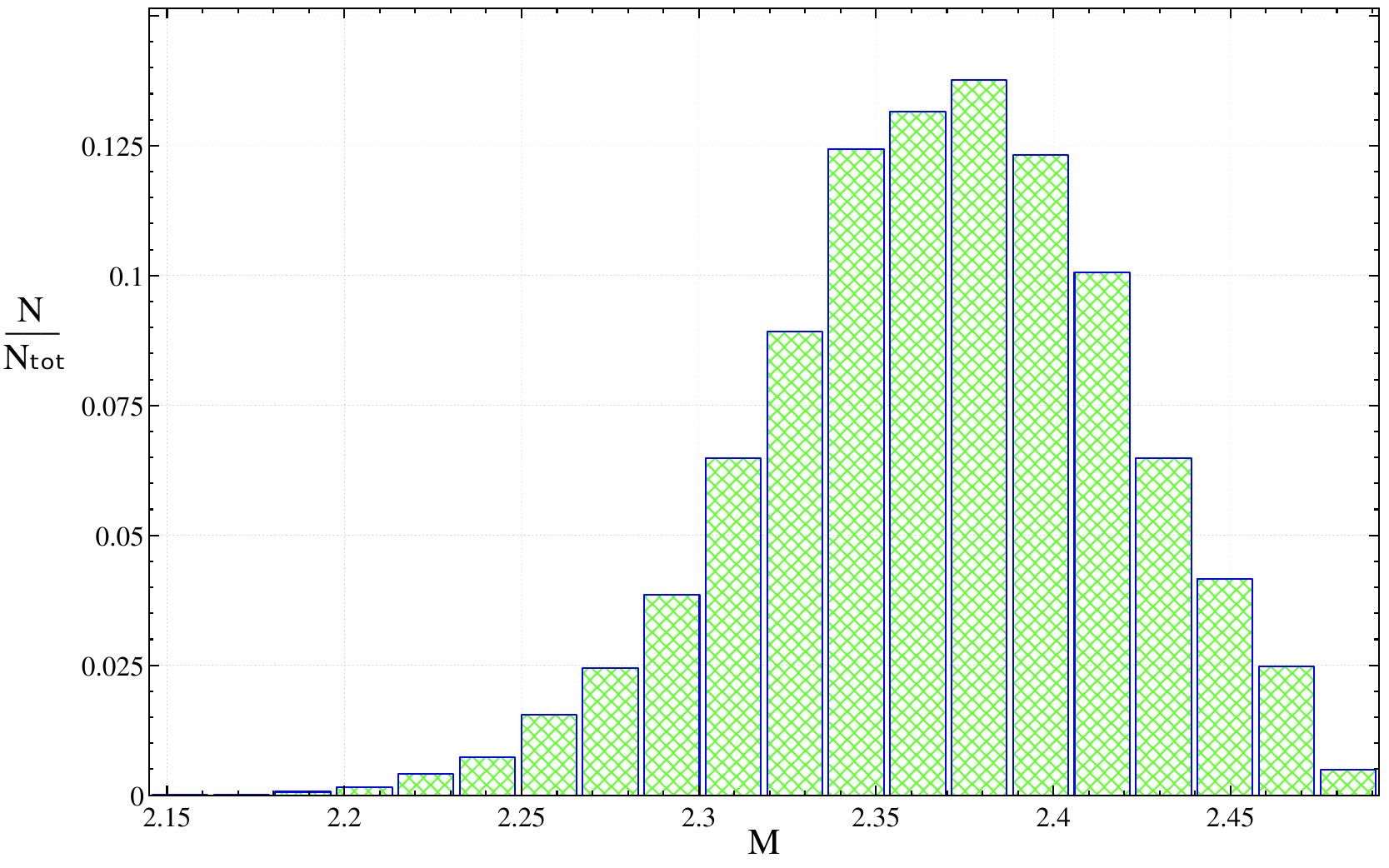}
	\includegraphics[clip,width=0.495\linewidth,height=60mm]{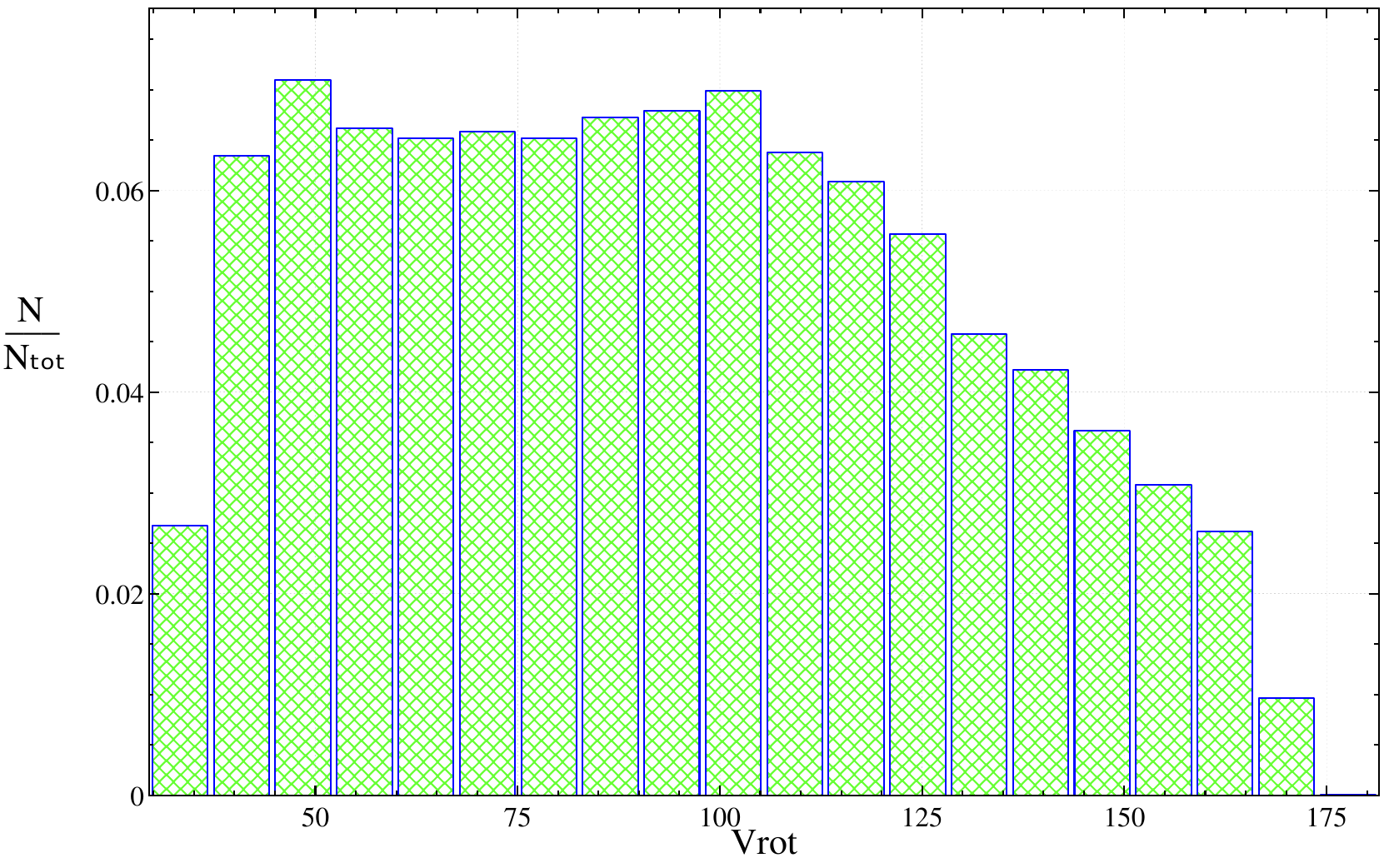}
	\includegraphics[clip,width=0.495\linewidth,height=60mm]{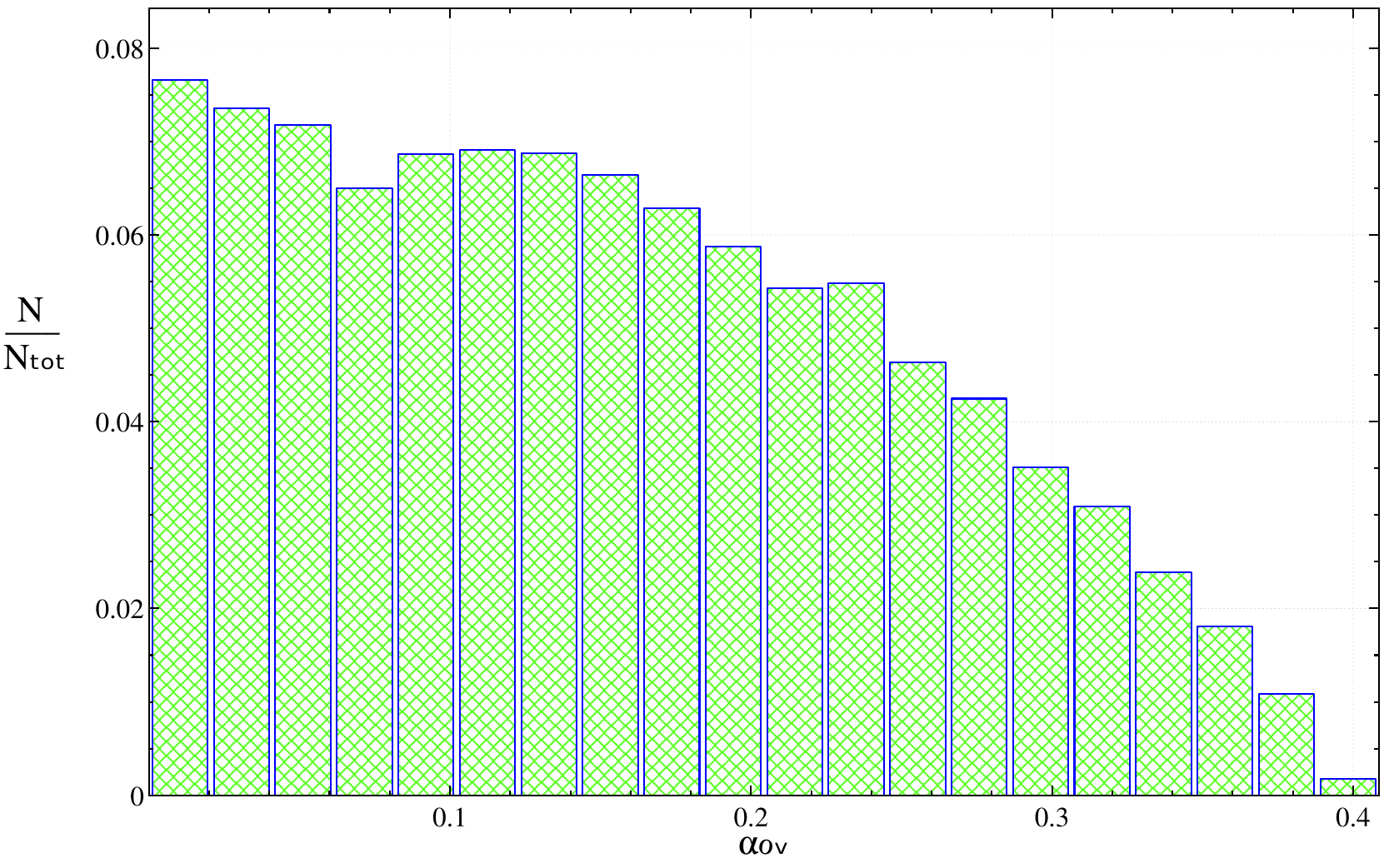}
	\includegraphics[clip,width=0.495\linewidth,height=60mm]{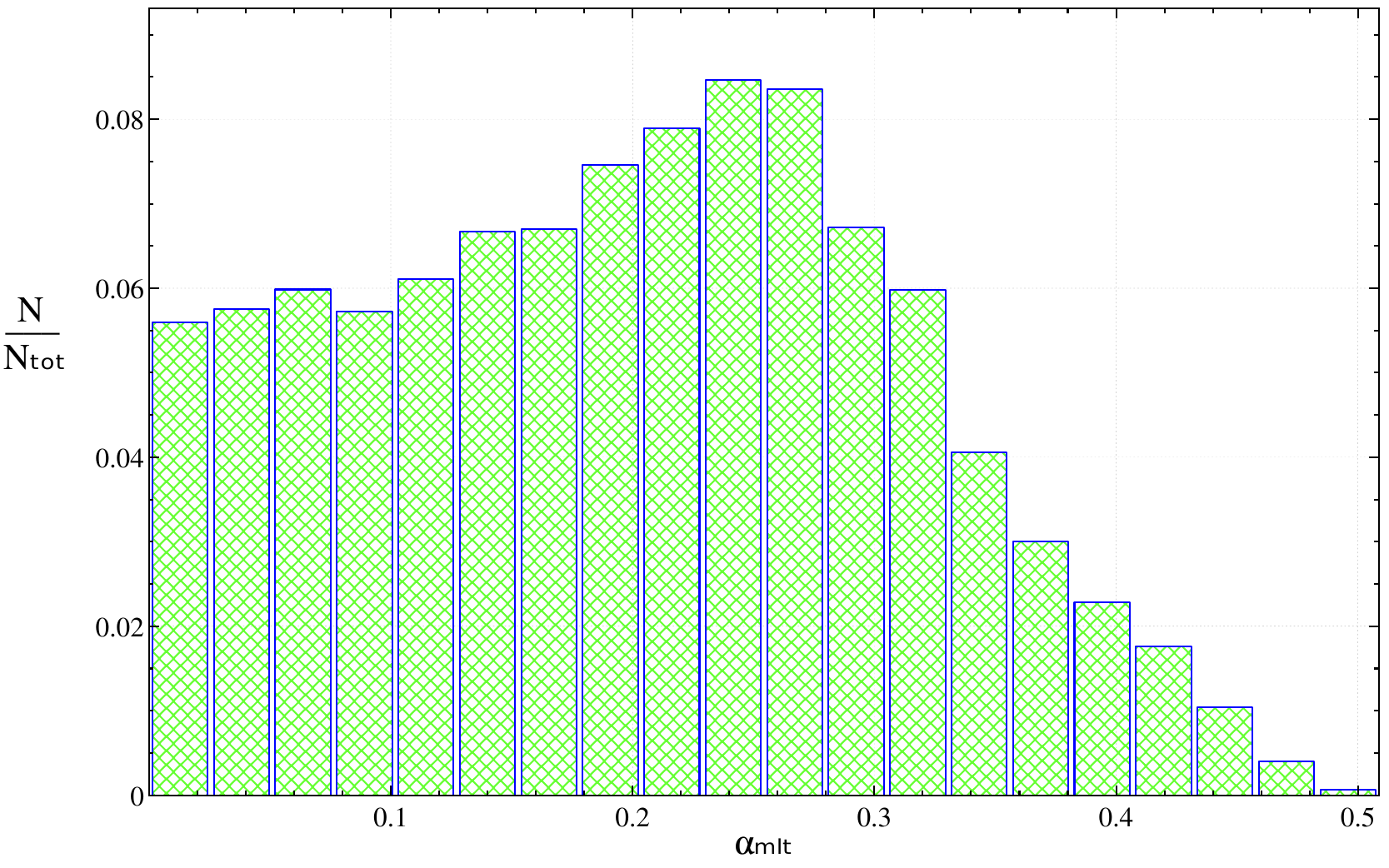}
	\caption{The same as in Fig.\,\ref{seismic_HSB_Z020} but for $Z=0.030$.}
	\label{seismic_HSB_Z030}
\end{figure*}

\begin{figure*}
	\centering
	\includegraphics[clip,width=0.495\linewidth,height=60mm]{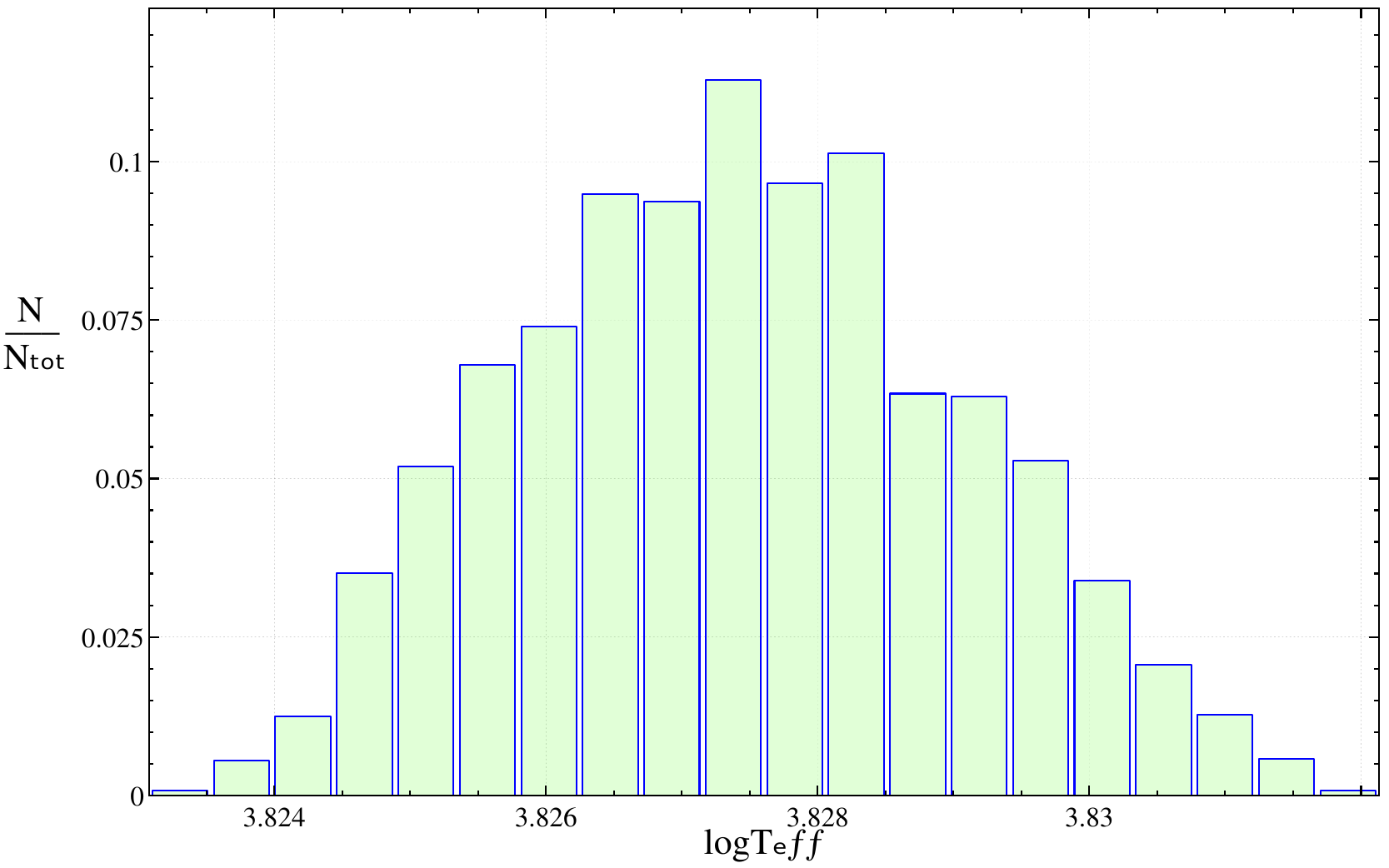}
	\includegraphics[clip,width=0.495\linewidth,height=60mm]{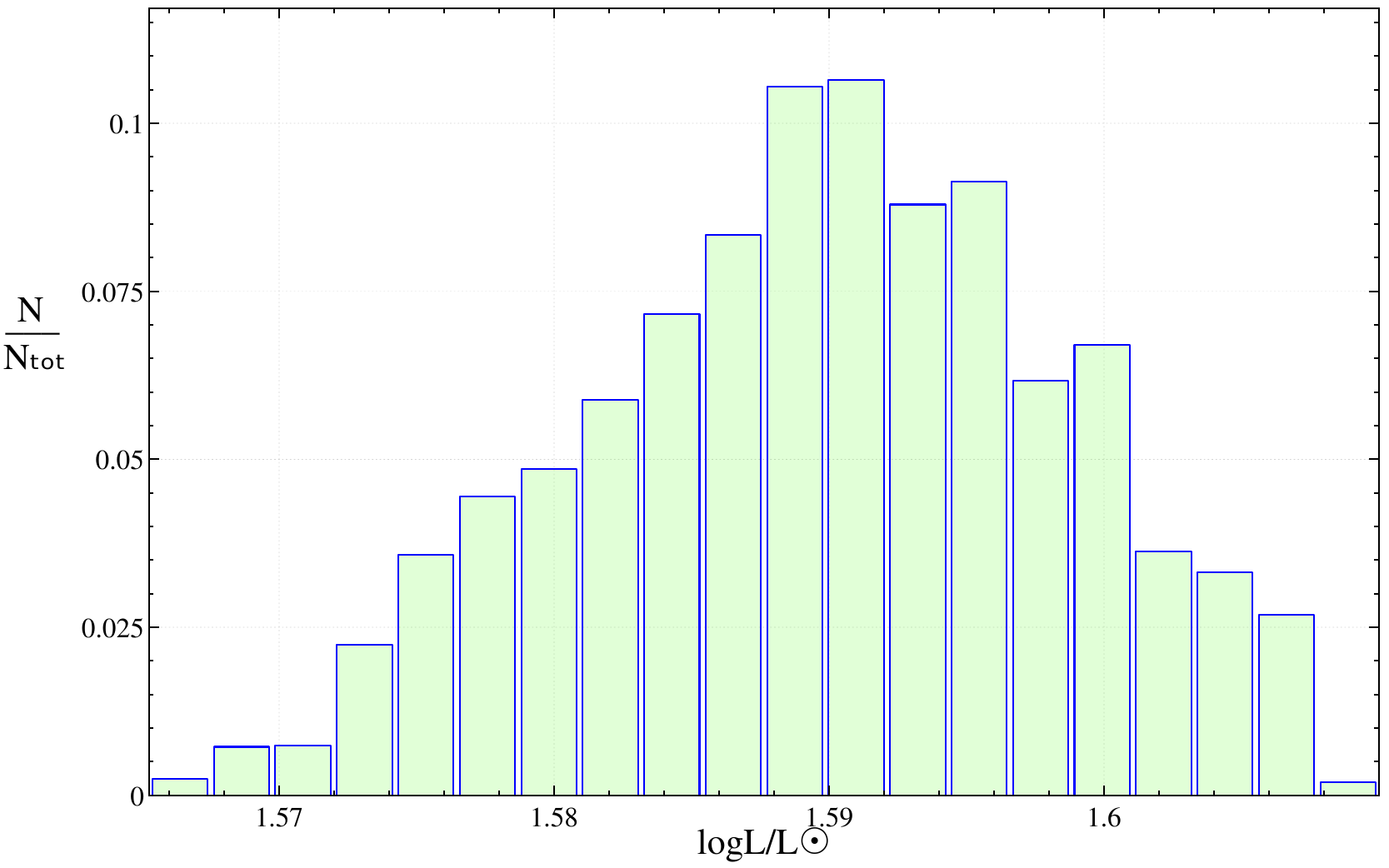}
	\includegraphics[clip,width=0.495\linewidth,height=60mm]{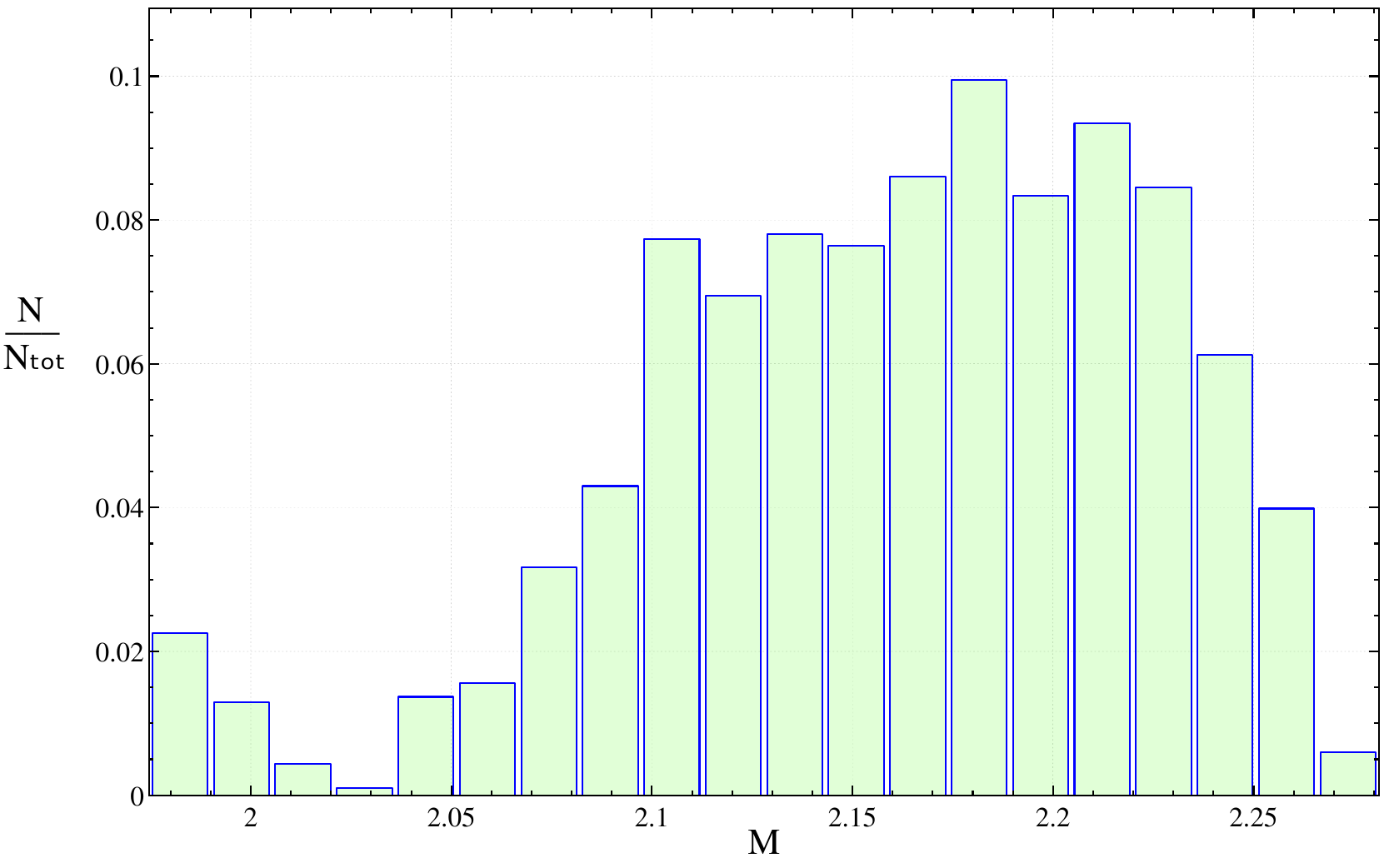}
	\includegraphics[clip,width=0.495\linewidth,height=60mm]{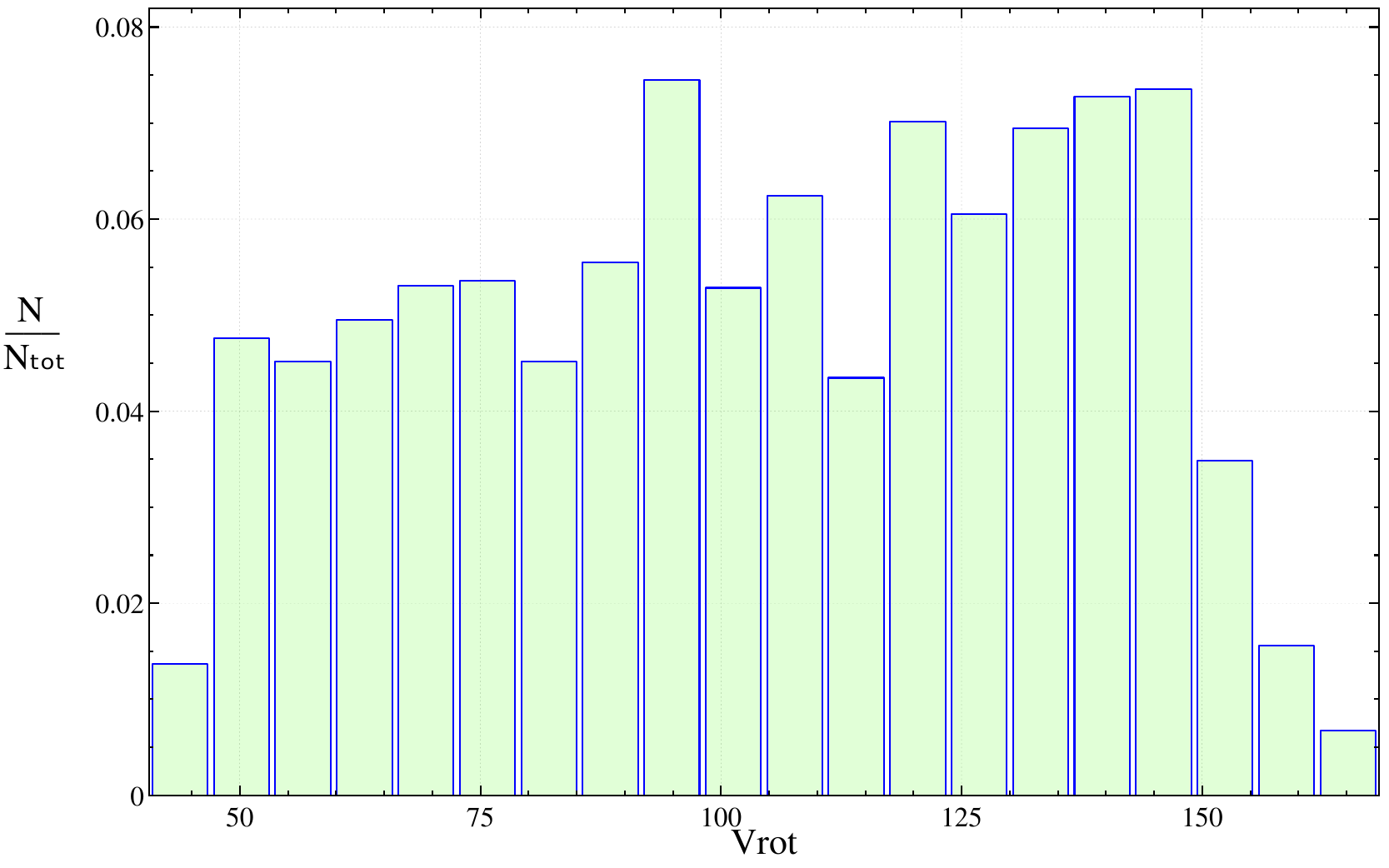}
	\includegraphics[clip,width=0.495\linewidth,height=60mm]{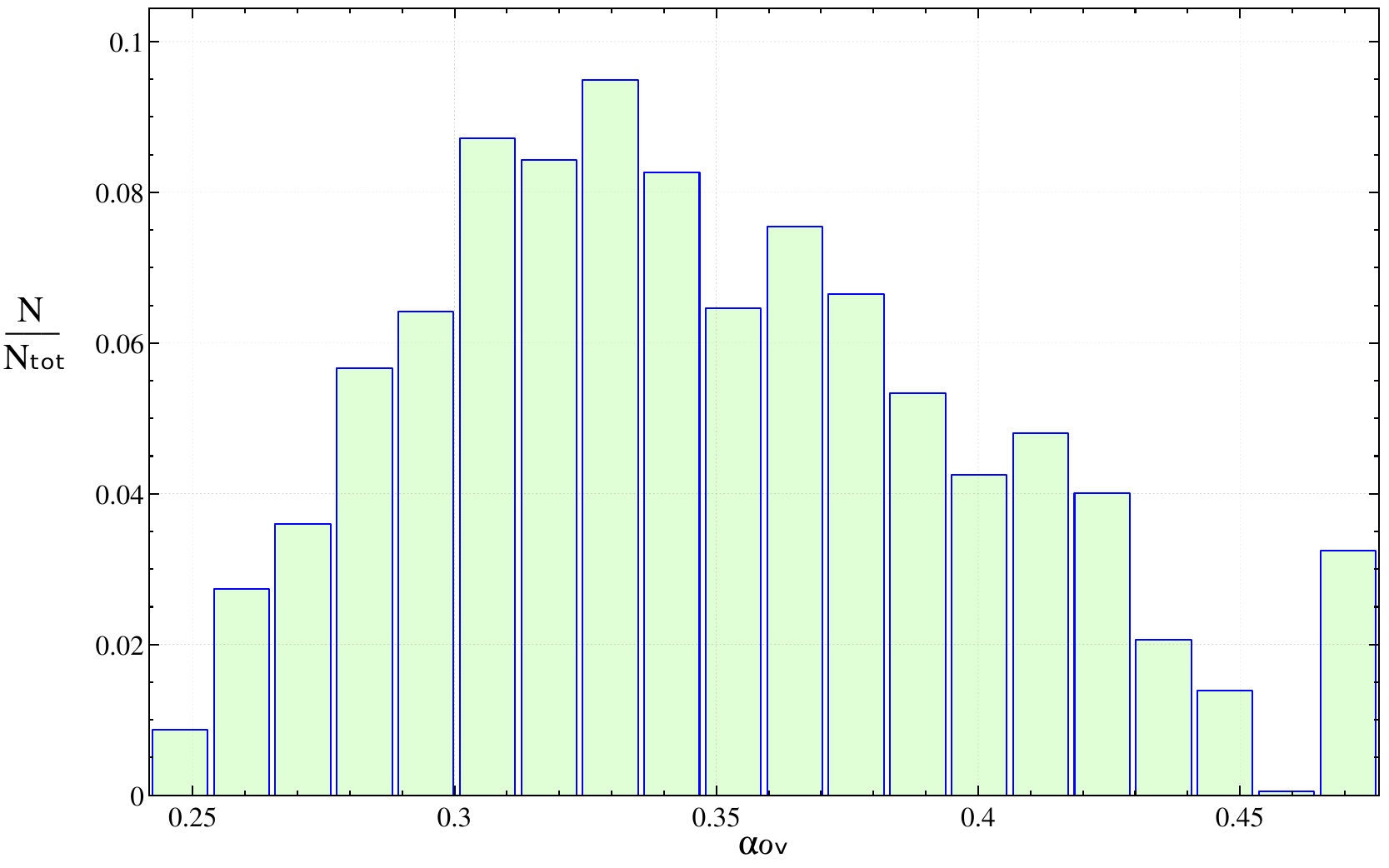}
	\includegraphics[clip,width=0.495\linewidth,height=60mm]{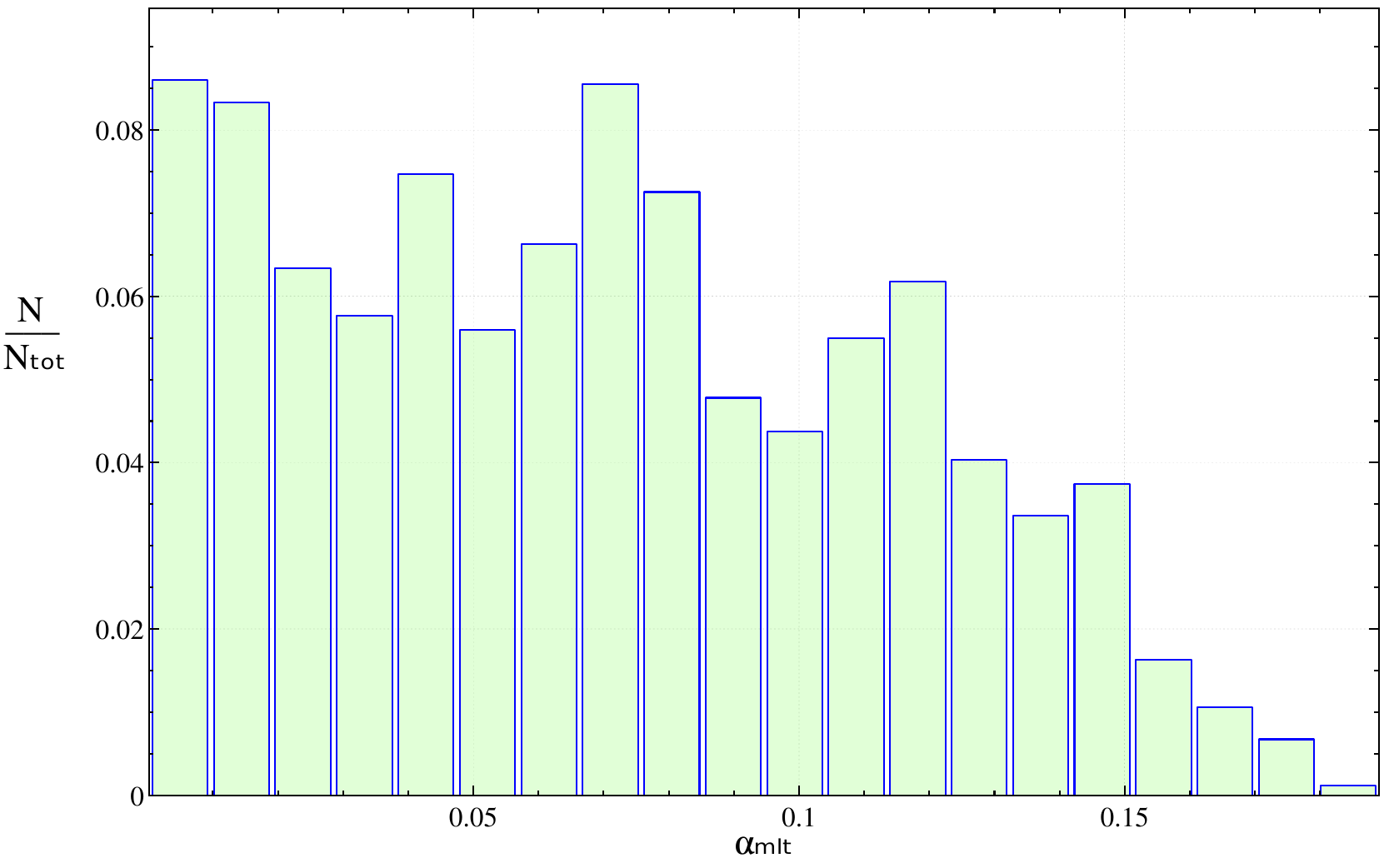}
	\caption{The normalized histograms for the parameters of the OC seismic models of V2367\,Cyg, calculated assuming 
		that the observed frequencies $\nu_1$ and $\nu_2$ are the first and second radial overtone, respectively 
		(the $p_2\& p_3$ hypothesis).  The OPAL opacities were adopted and  $X_0=0.70,~Z=0.020$ were fixed.}
	\label{seismic_OC_Z020}
\end{figure*}

\begin{figure*}
	\centering
	\includegraphics[clip,width=0.495\linewidth,height=60mm]{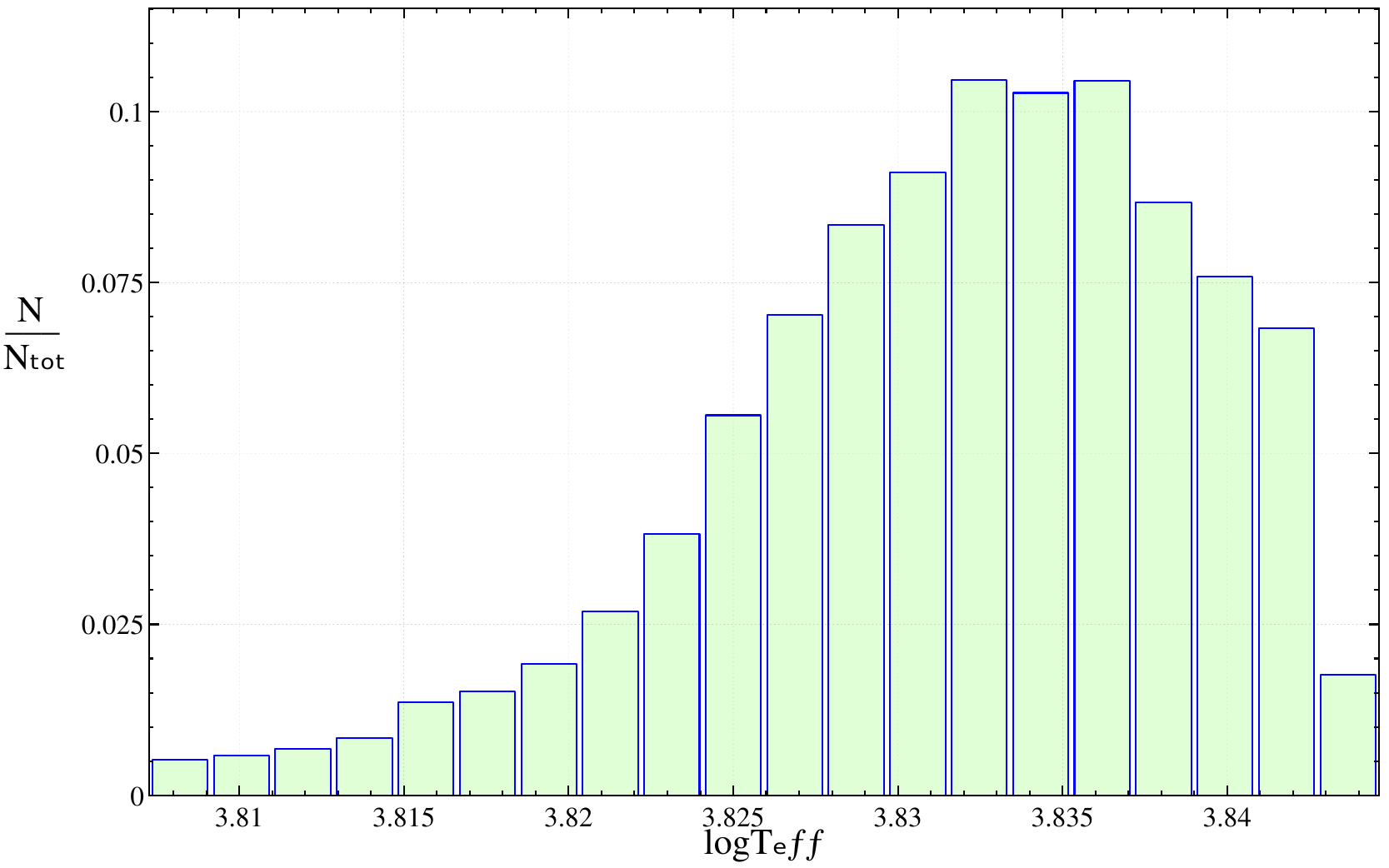}
	\includegraphics[clip,width=0.495\linewidth,height=60mm]{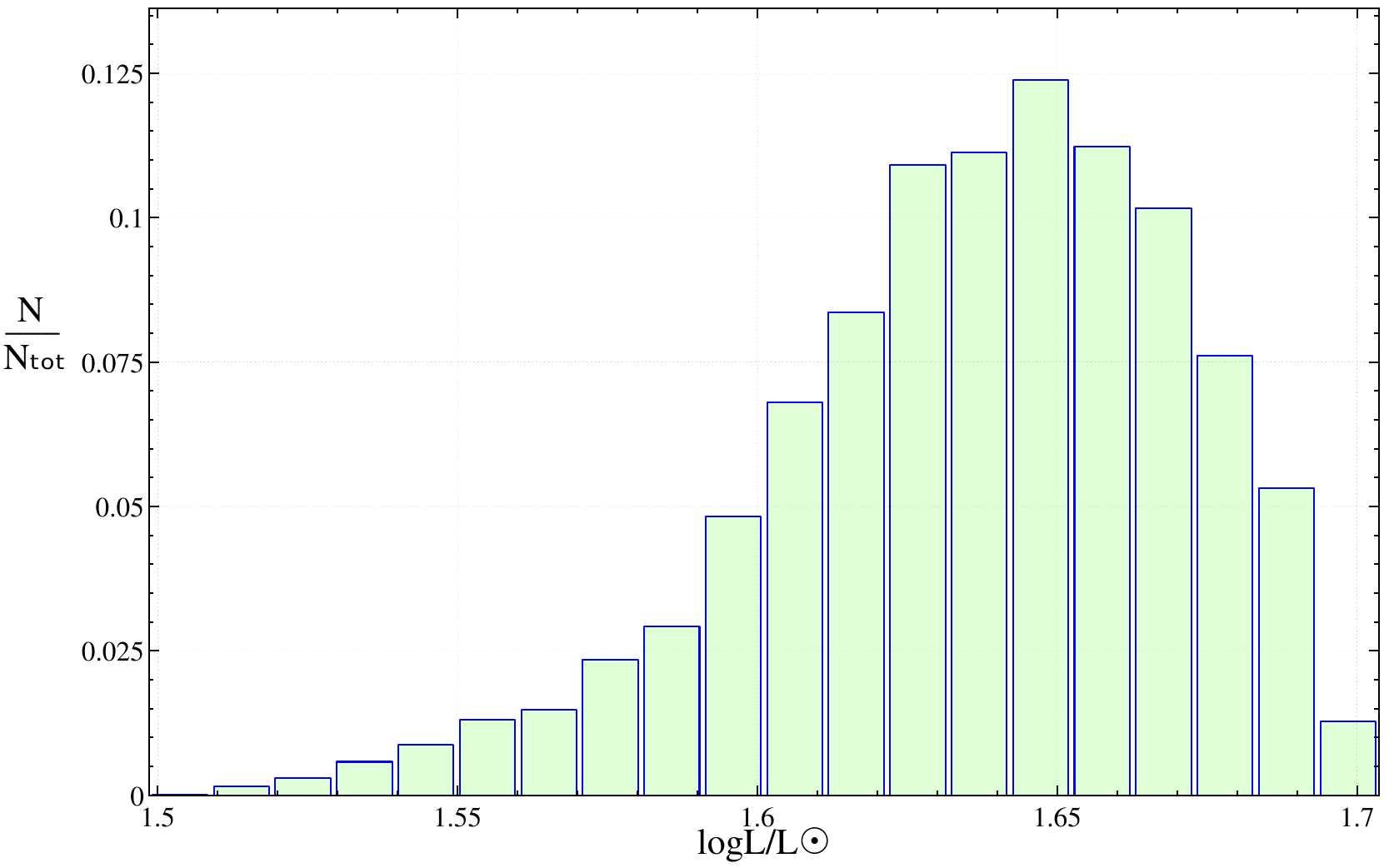}
	\includegraphics[clip,width=0.495\linewidth,height=60mm]{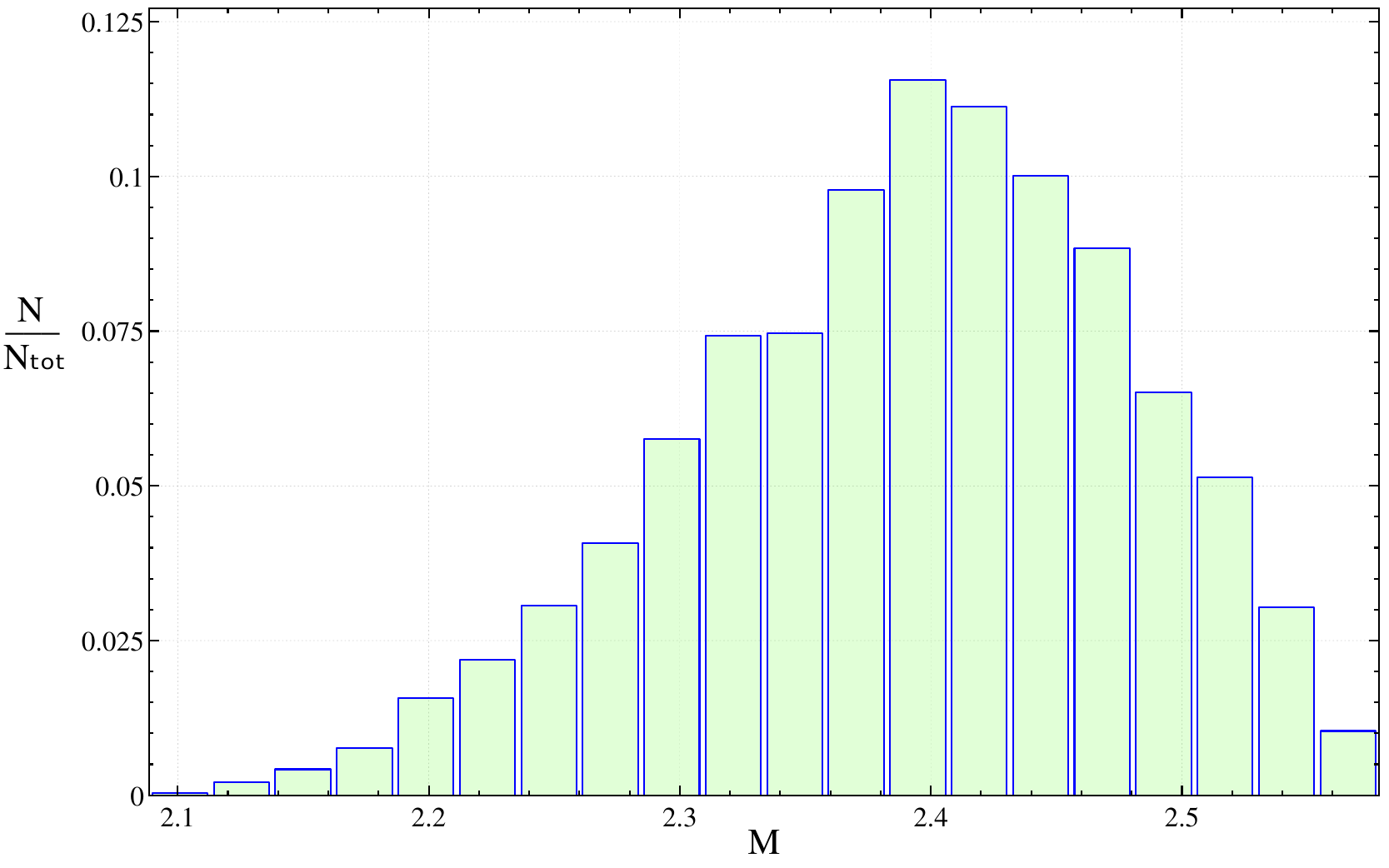}
	\includegraphics[clip,width=0.495\linewidth,height=60mm]{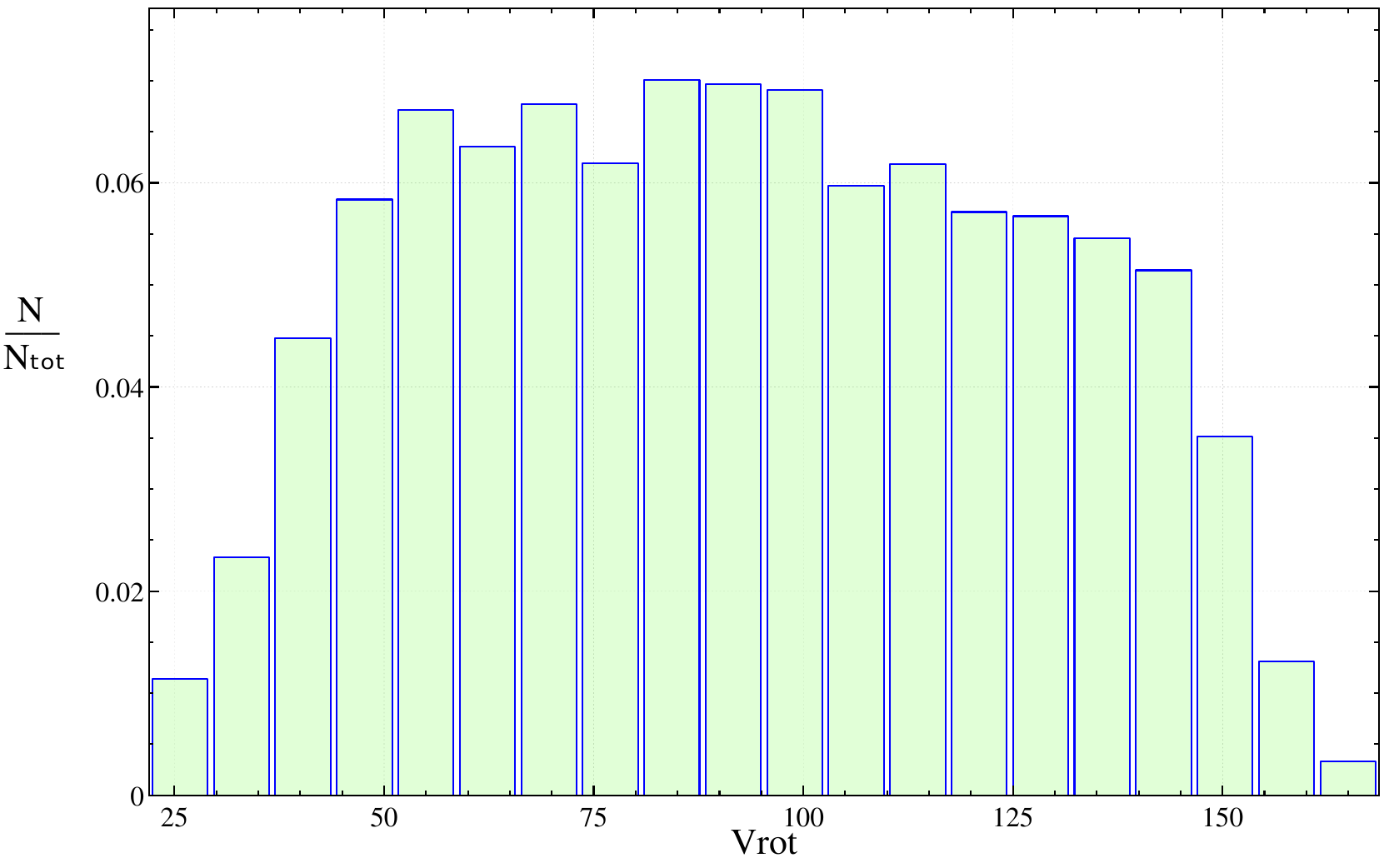}
	\includegraphics[clip,width=0.495\linewidth,height=60mm]{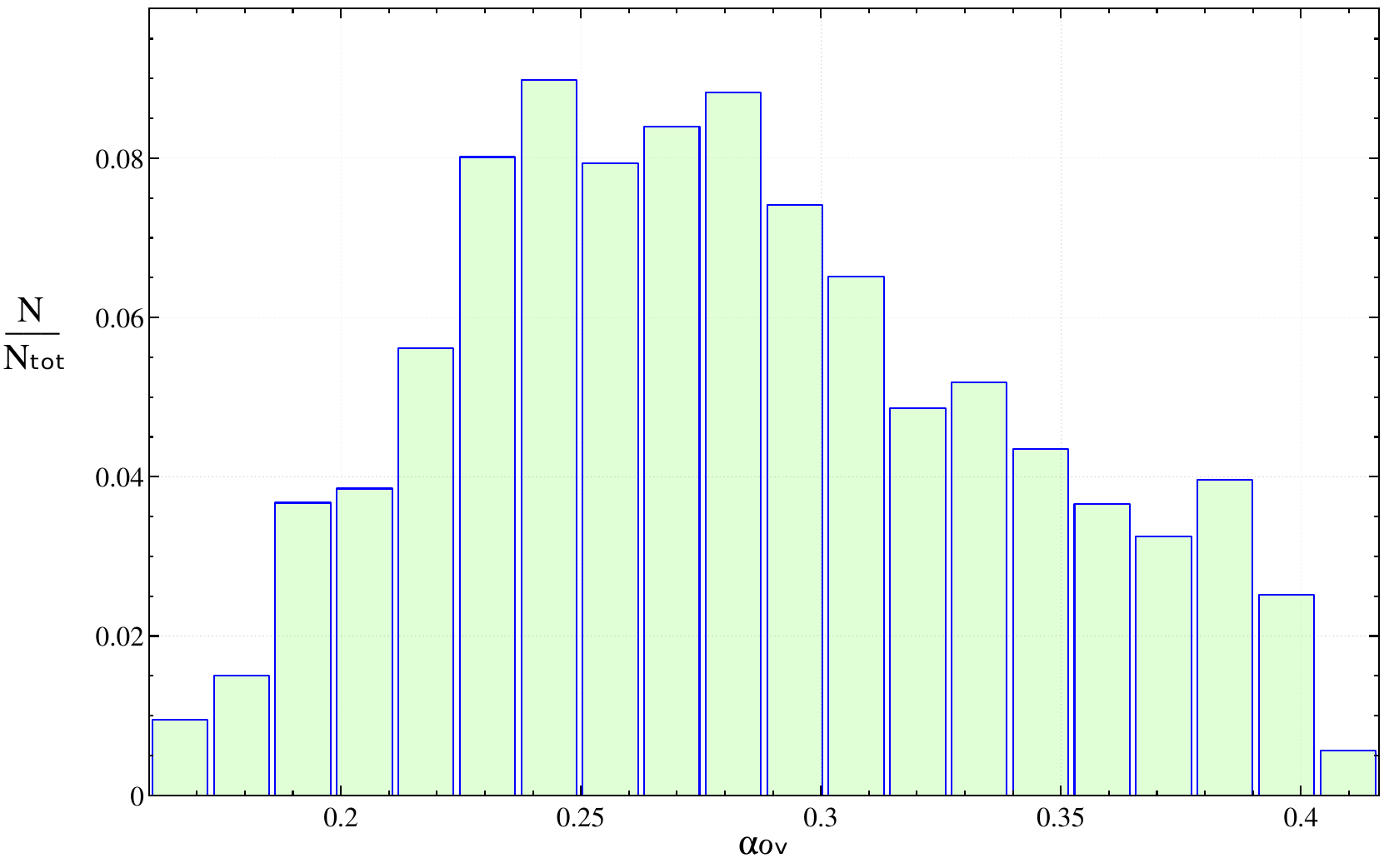}
	\includegraphics[clip,width=0.495\linewidth,height=60mm]{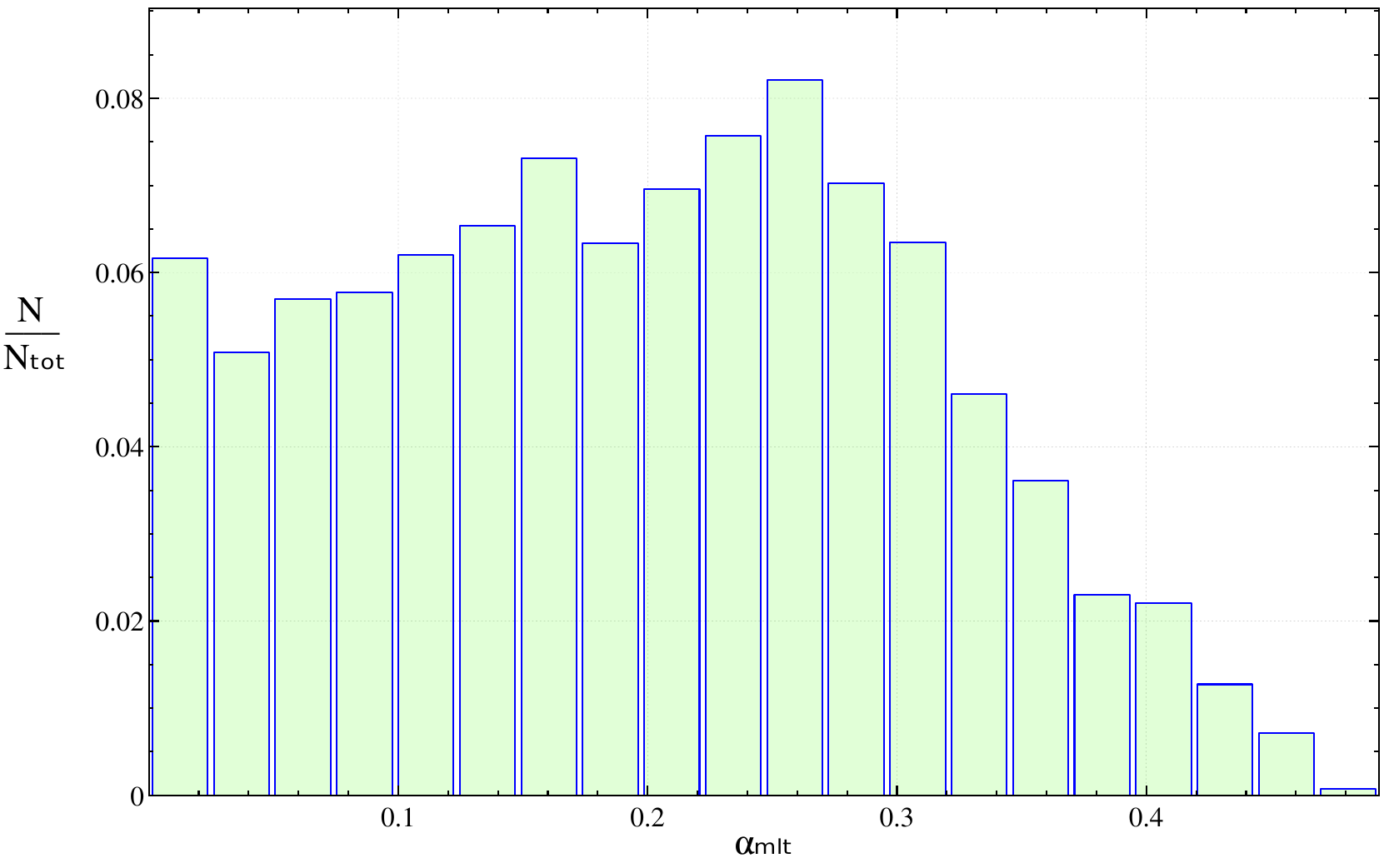}
	\caption{The same as in Fig.\,\ref{seismic_OC_Z020} but for $Z=0.030$.}
	\label{seismic_OC_Z030}
\end{figure*}

\begin{figure*}
	\centering
	\includegraphics[clip,width=0.495\linewidth,height=60mm]{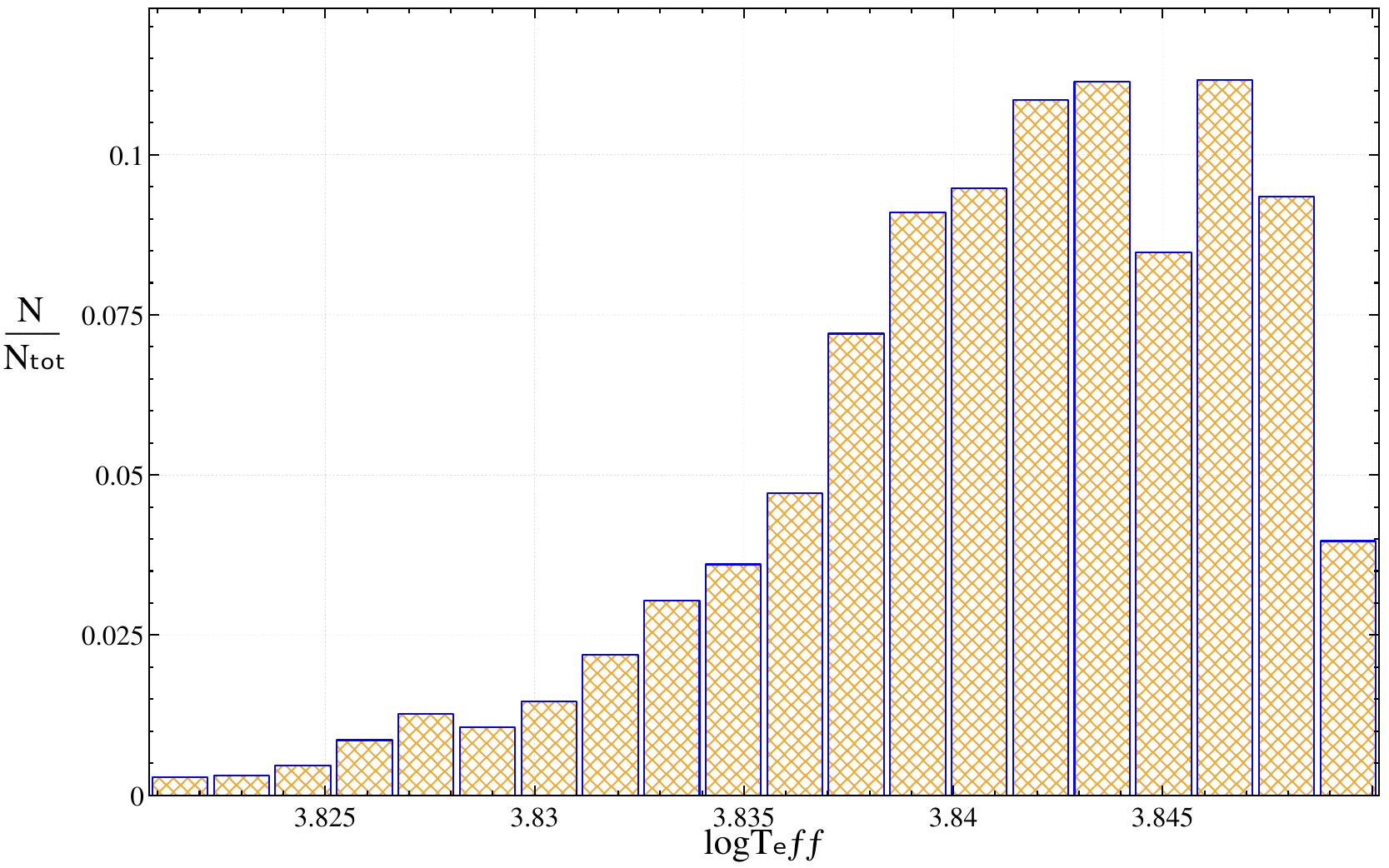}
	\includegraphics[clip,width=0.495\linewidth,height=60mm]{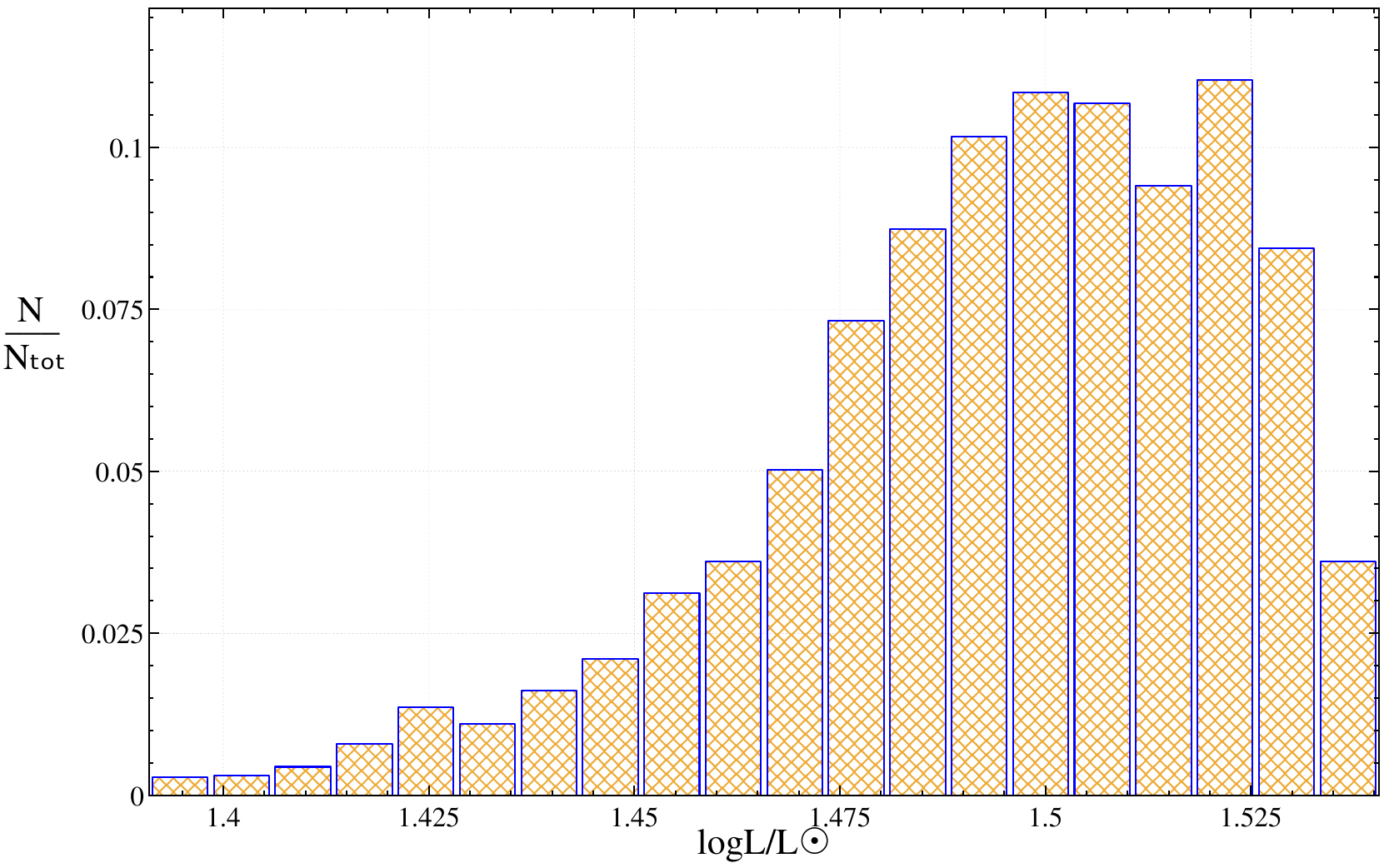}
	\includegraphics[clip,width=0.495\linewidth,height=60mm]{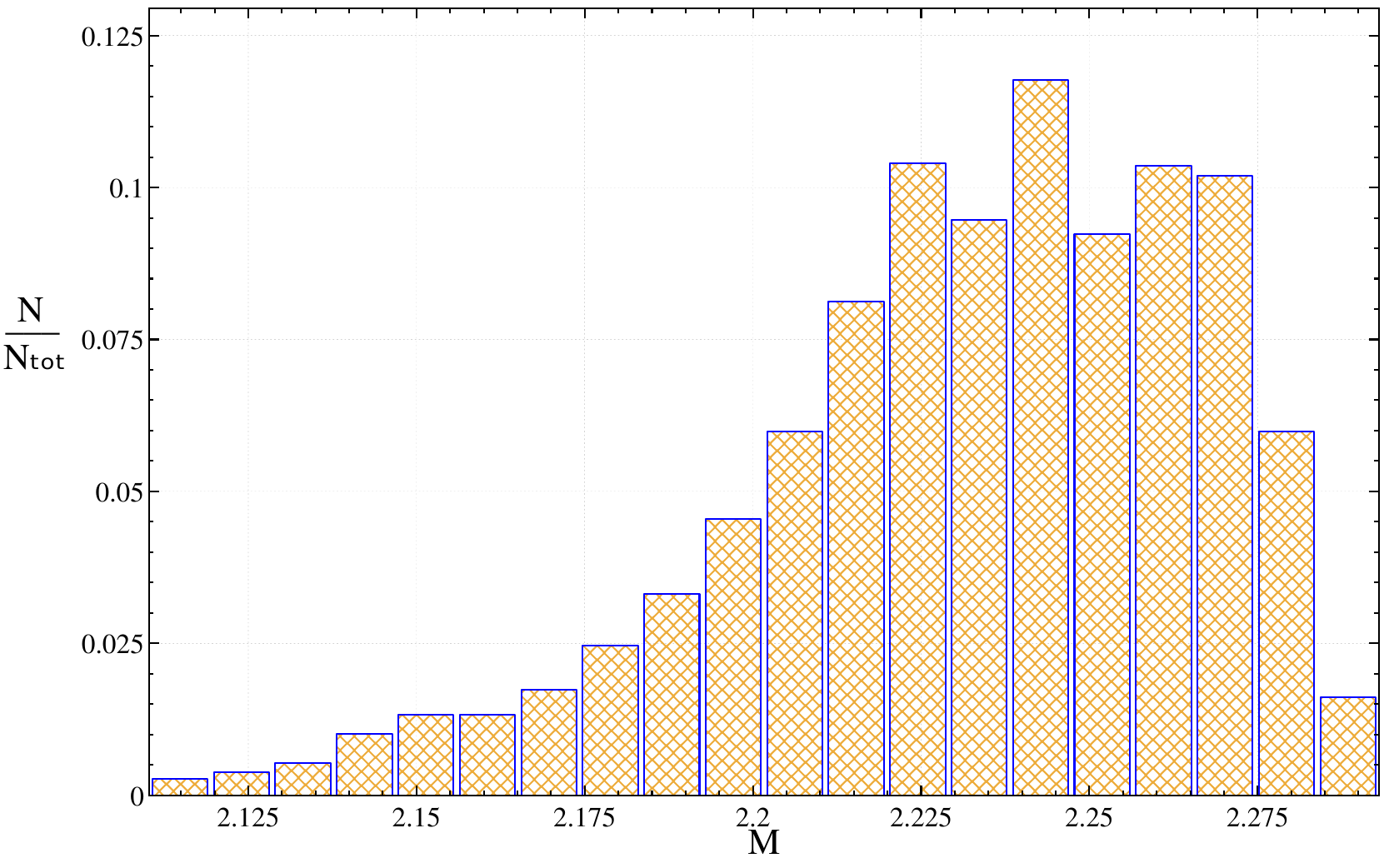}
	\includegraphics[clip,width=0.495\linewidth,height=60mm]{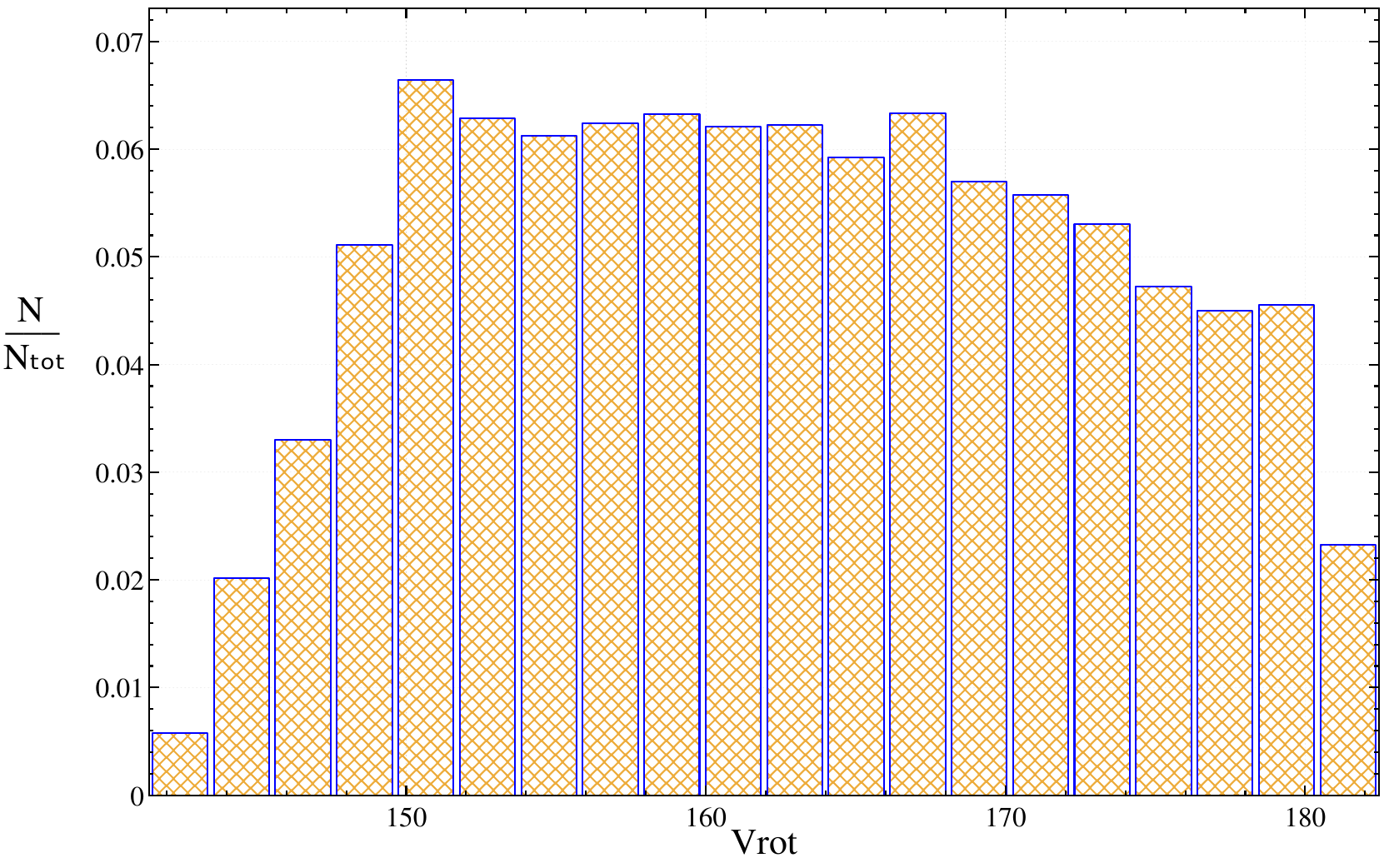}
	\includegraphics[clip,width=0.495\linewidth,height=60mm]{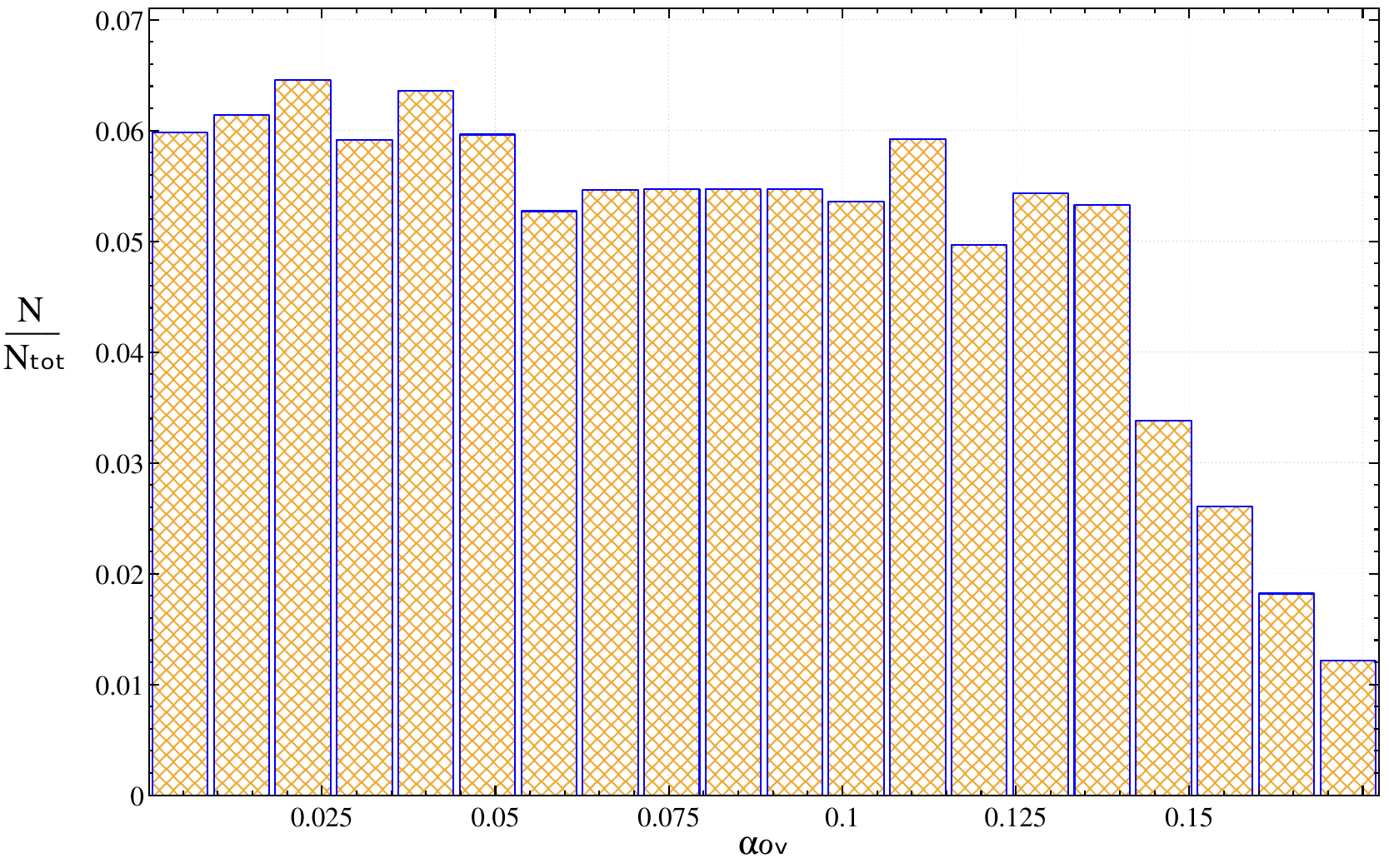}
	\includegraphics[clip,width=0.495\linewidth,height=60mm]{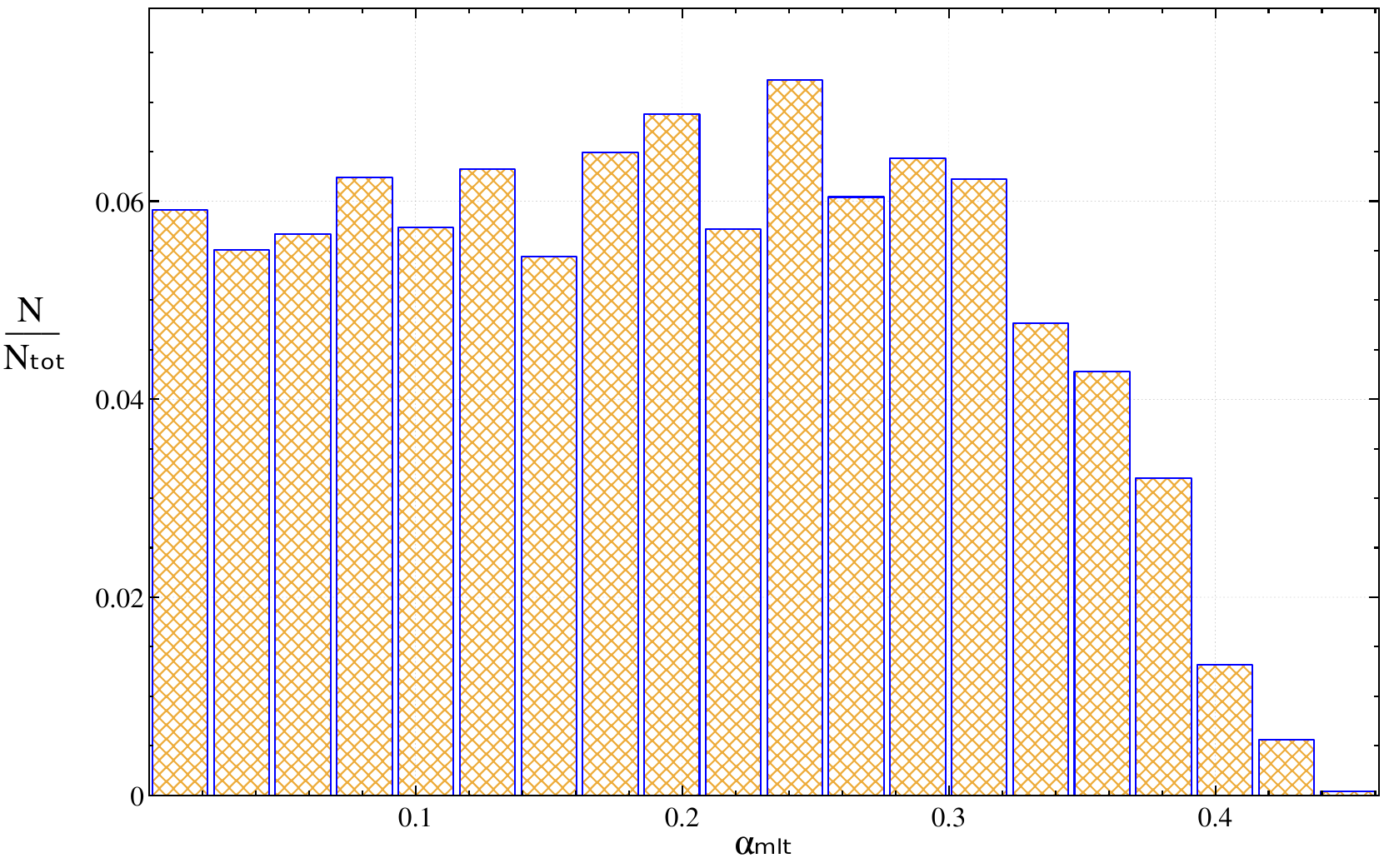}
	\caption{The normalized histograms for the parameters of the HSB seismic models of V2367\,Cyg, calculated assuming 
		that the observed frequencies $\nu_1$ and $\nu_2$ are the fundamental and first overtone radial mode, respectively 
		(the $p_1\& p_2$ hypothesis).  The OPAL opacities were adopted and  $X_0=0.70,~Z=0.030$ were fixed.}
	\label{seismic_HSB_Z030_p1}
\end{figure*}

\begin{figure*}
	\centering
	\includegraphics[clip,width=0.495\linewidth,height=60mm]{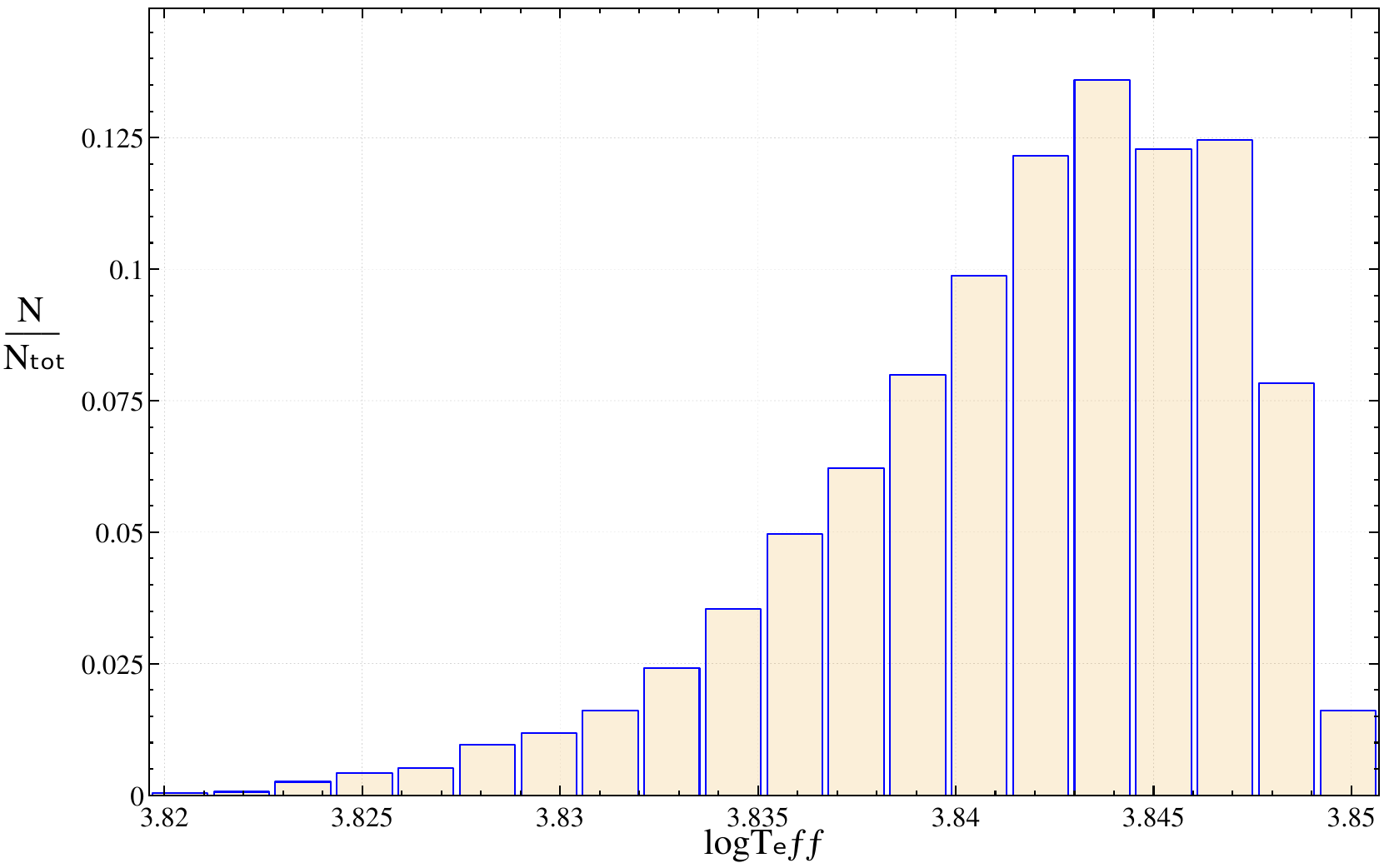}
	\includegraphics[clip,width=0.495\linewidth,height=60mm]{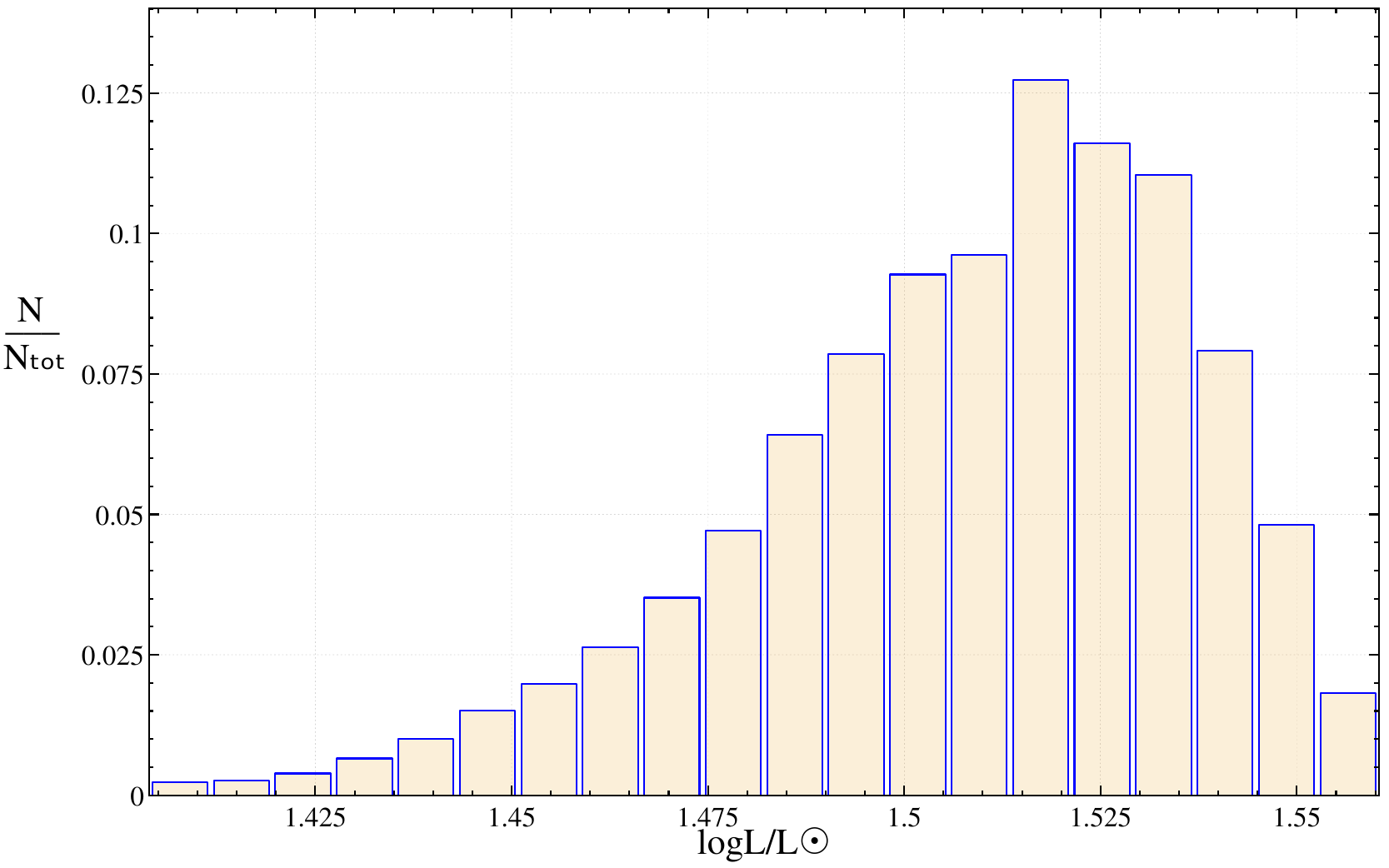}
	\includegraphics[clip,width=0.495\linewidth,height=60mm]{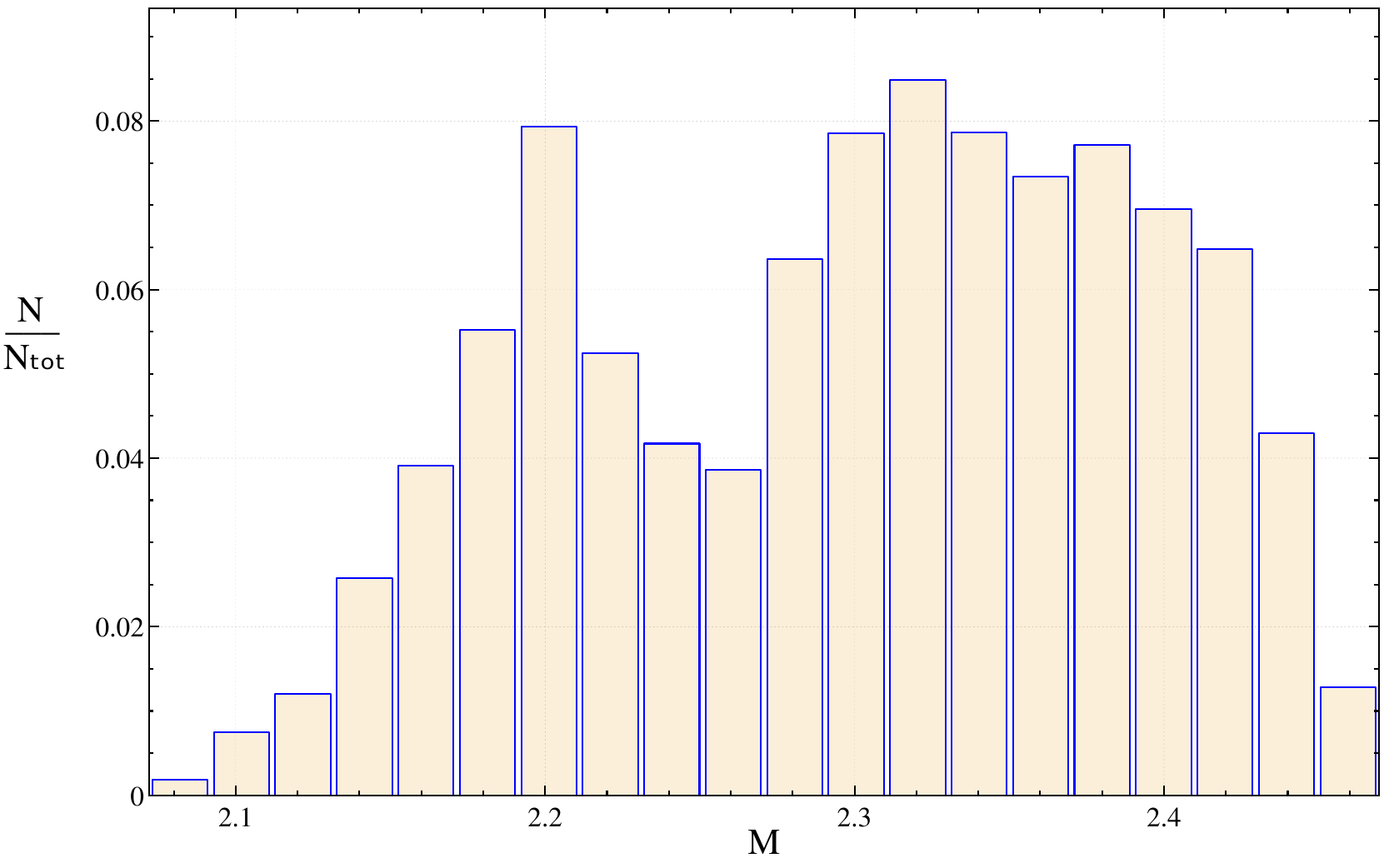}
	\includegraphics[clip,width=0.495\linewidth,height=60mm]{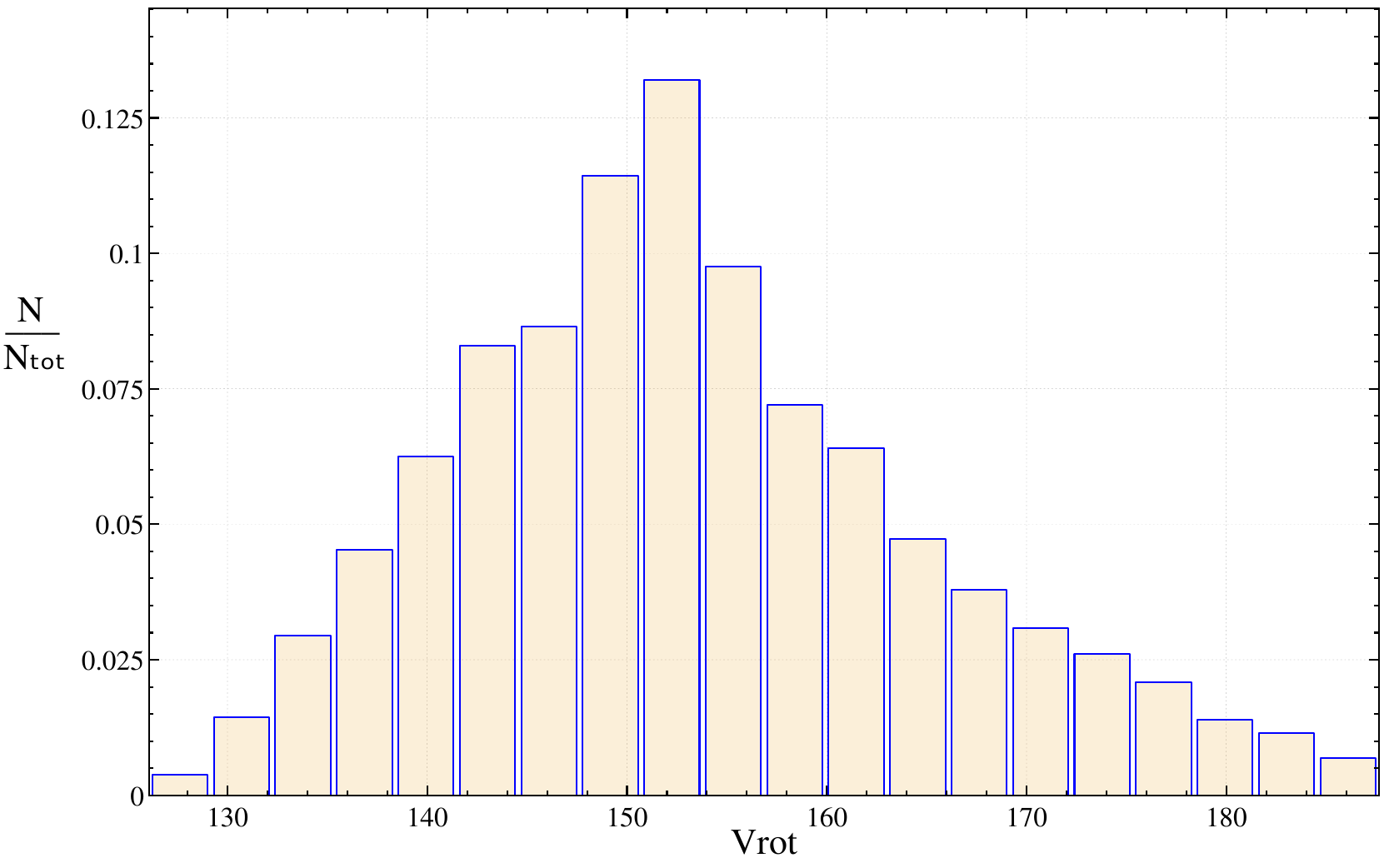}
	\includegraphics[clip,width=0.495\linewidth,height=60mm]{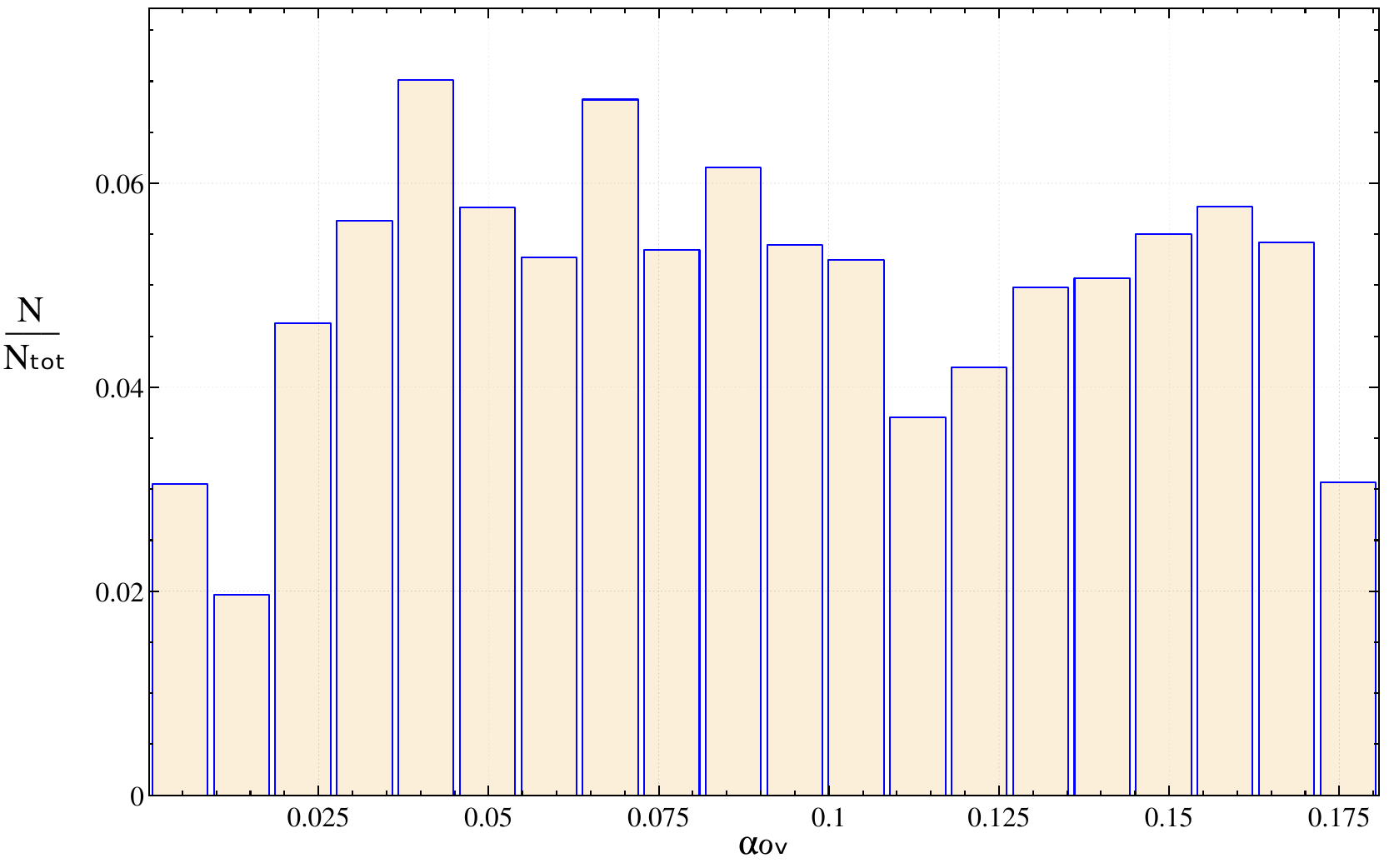}
	\includegraphics[clip,width=0.495\linewidth,height=60mm]{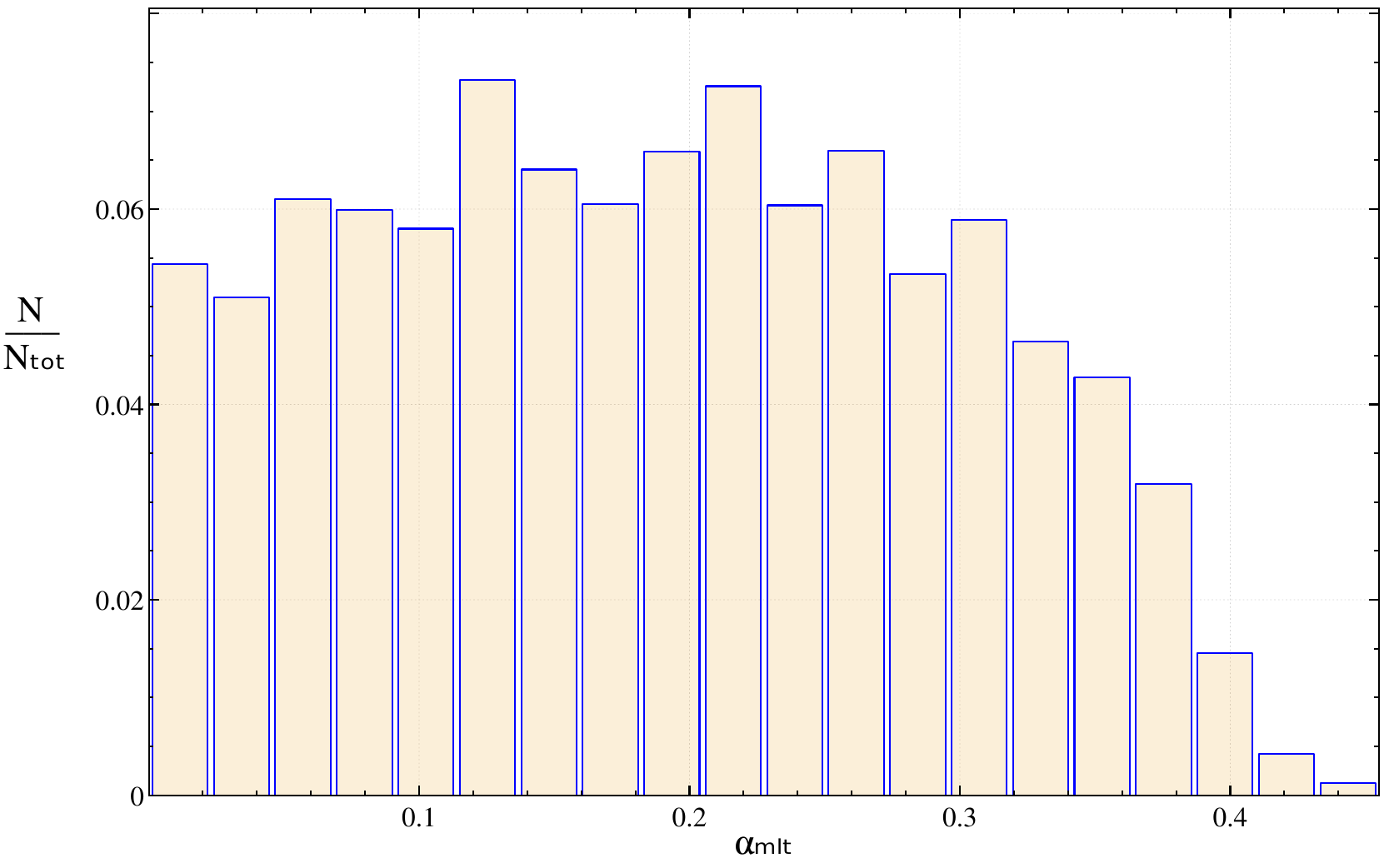}
	\caption{The same as in Fig.\,\ref{seismic_HSB_Z030_p1} but for parameters of the OC seismic models.}
	\label{seismic_OC_Z030_p1}
\end{figure*}

The results of our seismic modelling of V2367\,Cyg with the MC-based Bayesian analysis are presented as histograms.
In Fig.\,\ref{seismic_HSB_Z020}, we show histograms for the parameters $(\log T_{\rm eff},~\log L/{\rm L}_\odot,~ M, ~
V_{\rm rot},~\alpha_{\rm ov}$ and $\alpha_{\rm MLT})$ of the HSB seismic models for the $p_2\& p_3$ hypothesis and $Z=0.020$.
Fig.\,\ref{seismic_HSB_Z030} shows the same but for the seismic models with the metallicity $Z=0.030$.
The histograms for parameters of the OC seismic models are presented in Fig.\,\ref{seismic_OC_Z020} for $Z=0.020$ and in Fig.\,\ref{seismic_OC_Z030} for $Z=0.030$. In case of the $p_1\& p_2$ hypothesis, only models with the metallicity $Z=0.030$
had the real and imaginary part of the parameter $f$ consistent within the 1-$\sigma$ error with the empirical counterparts. 
Histograms  for these models are shown in Fig.\,\ref{seismic_HSB_Z030_p1}  and Fig.\,\ref{seismic_OC_Z030_p1}
for the parameters of  the HSB and OC models, respectively.

% Don't change these lines
\bsp	% typesetting comment
\label{lastpage}
\end{document}